\documentclass[11pt]{article}
\pdfoutput=1
\usepackage{cite}
\usepackage{hyperref}
\usepackage[dvipsnames]{xcolor}

\usepackage{lineno}
\usepackage{amsmath}
\usepackage{amssymb,amsfonts,amsthm}
\usepackage{epsfig,graphicx}
\usepackage{subcaption}
\usepackage{url} %url
\usepackage{bm} %letras griegas en negrita (uso: \bm{\alpha})
\usepackage{bbm}
\usepackage{slashed}
\usepackage{cite}
\usepackage[normalem]{ulem}
\usepackage{color}
\usepackage{pstricks}

\usepackage{fancyhdr}
\usepackage{textcomp}
\newcommand{\unit}{\leavevmode\hbox{\small1\kern-3.6pt\normalsize1}}

\makeatletter
\newcommand*{\rom}[1]{\expandafter\@slowromancap\romannumeral #1@}
\makeatother

\def\pslash  {{p\mkern-8mu/}}
\def\qslash  {{q\mkern-9mu/}}
\def\xslash  {{x\mkern-9mu/}}

\allowdisplaybreaks

\begin{document}

\thispagestyle{empty}
\begin{flushright}

 {\small IPPP-17-101}\\
  {\small DCTP-17-202}\\
{\small TUM-HEP-1126/18}\\

 \vspace*{2.mm}{January 10, 2018} %\color{red}{Modified: \today}}
\end{flushright}

\begin{center}
  {\bf {\LARGE $B+L$ violation at colliders and new physics}}

\renewcommand*{\thefootnote}{\fnsymbol{footnote}}
\setcounter{footnote}{3}

  \vspace{0.5cm}
  {\large
  David G. Cerde\~{n}o$^a$, Peter Reimitz$^b$, Kazuki Sakurai$^c$, Carlos Tamarit$^d$}
  \\[0.2cm] 

  {\footnotesize{$^a$ Institute for Particle Physics Phenomenology , Department of Physics\\
	Durham University, Durham DH1 3LE, United Kingdom
    
     $^b$ 
Institut f\"ur Theoretische Physik, Universit\"at Heidelberg\\Philosophenweg 16, 69120 Heidelberg, Germany

    $^c$ Institute of Theoretical Physics, Faculty of Physics\\ 
        University of Warsaw, ul. Pasteura 5, PLâ02â093 Warsaw, Poland
        
    $^d$ Physik Department T70, Technische Universit\"at M\"unchen\\
James Franck Stra\ss e 1, 85748 Garching, Germany
      }
    }

\vspace*{0.7cm}

  \begin{abstract}
  Chiral electroweak anomalies predict baryon ($B$) and lepton ($L$) violating fermion interactions, which can be dressed with large numbers of
 Higgs and gauge bosons. The estimation of the total $B+L$-violating rate from an initial two-particle state --potentially observable at colliders-- has been the subject of an intense discussion, mainly centered on the resummation of boson emission, which is believed to contribute to the cross-section with an exponential function of the energy, yet with an exponent (the ``holy-grail" function) which is not fully known in the energy range of interest. In this article we focus instead on the effect of fermions beyond the Standard-Model (SM) in the polynomial contributions to the rate. It is shown that $B+L$ processes involving the new fermions have a polynomial contribution that can be several orders of magnitude greater than in the SM, for high centre-of-mass energies and light enough masses. We also present calculations that hint at a simple dependence of the holy grail function on the heavy fermion masses. Thus, if anomalous $B+L$ violating interactions are ever detected at high-energy colliders, they could be associated with new physics.
\end{abstract}
\end{center}
\newpage
\tableofcontents
\newpage

%%%%%%%%%%%%%%%%%
%%%%%%%%%%%%%%%%%
%%%%%%%%%%%%%%%%%
%%%%%%%%%%%%%%%%%

%%%%%%%%%%%%%%%%%%%%%%%%%%%%%%%%%%%%%%%%%%%%%%%%%%%%%%%
%%%%%%%%%%%%%%%%%%%%%%%%%%%%%%%%%%%%%%%%%%%%%%%%%%%%%%%
%%%%%%%%%%%%%%%%%%%%%%%%%%%%%%%%%%%%%%%%%%%%%%%%%%%%%%%
%%%%%%%%%%%%%%%%%%%%%%%%%%%%%%%%%%%%%%%%%%%%%%%%%%%%%%%
%%%%%%%%%%%%%%%%%%%%%%%%%%%%%%%%%%%%%%%%%%%%%%%%%%%%%%%
\section{Introduction}
\label{sec:introduction}
%%%%%%%%%%%%%%%%%%%%%%%%%%%%%%%%%%%%%%%%%%%%%%%%%%%%%%%
%%%%%%%%%%%%%%%%%%%%%%%%%%%%%%%%%%%%%%%%%%%%%%%%%%%%%%%
%%%%%%%%%%%%%%%%%%%%%%%%%%%%%%%%%%%%%%%%%%%%%%%%%%%%%%%
%%%%%%%%%%%%%%%%%%%%%%%%%%%%%%%%%%%%%%%%%%%%%%%%%%%%%%%
%%%%%%%%%%%%%%%%%%%%%%%%%%%%%%%%%%%%%%%%%%%%%%%%%%%%%%%

The Standard-Model (SM) has accidental baryon ($B$) and lepton ($L$) symmetries at the classical level, which however become broken by quantum anomalies \cite{Adler:1969gk,Bell:1969ts,Bardeen:1969md}. Such effects can have a strong impact in the physics of the early Universe, as they may play a crucial role in the generation of the baryon asymmetry, for example in electroweak baryogenesis \cite{Kuzmin:1985mm}--in which anomalous processes convert a CP asymmetry into baryon number-- and leptogenesis \cite{Fukugita:1986hr}--in which the anomalous interactions convert a lepton asymmetry into a baryon asymmetry (for reviews, see \cite{Trodden:1998ym,Morrissey:2012db,Buchmuller:2004nz,Davidson:2008bu} and references therein). Although both types of mechanisms require physics beyond the Standard Model (BSM), the new fields (e.g. bosonic fields guaranteeing a strong first-order electroweak phase transition, as required for electroweak baryogenesis, or new right-handed neutrinos whose out-of-equilibrium decays trigger leptogenesis)  typically don't affect the anomalous interactions, which remain SM-like. Aside from these cosmological effects, the $B+L$ violating interactions could be potentially observable at colliders, which would involve striking signatures arising from hard processes with at least twelve SM fermions. This follows because $B+L$ anomalies impose
\begin{align}
\label{eq:deltaBL}
\Delta B=\Delta L=3n_{\rm top},
\end{align}
where $n_{\rm top}$ is the integer topological charge of a given gauge-field background; this gives 12 SM fermions in a background with $n_{\rm top}=1$.

 The interactions sourced by the $B+L$ anomaly are of a non-perturbative nature, and can be understood as transitions between vacua of the electroweak theory, classified by integer Chern-Simons number. The minimum energy barrier between the vacua is known as the sphaleron energy --where the sphaleron is the unstable field configuration at the top of the energy barrier \cite{Manton:1983nd}-- and is of the order of $E_{\rm sph}=9$ TeV in the SM \cite{Dashen:1974ck,Klinkhamer:1984di} as well as its extensions \cite{Kastening:1991nw,Moreno:1996zm,Ahriche:2007jp,Spannowsky:2016ile}. At finite temperature, as in the early Universe, thermal excitations of the plasma can overcome the sphaleron barrier, and the transition rate after the phase transition can be estimated in terms of the sphaleron energy and the temperature \cite{Linde:1981zj}. At nearly zero temperature, as in a particle collider or cosmic ray event, the transition between vacua is a tunneling process, and can be estimated with semiclassical instanton techniques \cite{Belavin:1975fg,tHooft:1976snw}. The transition rate is then determined by the Euclidean action of the $SU(2)$ instanton, going as
\begin{align}
\Gamma_{B+L}\propto e^{-\frac{4\pi}{\alpha_W}}= O(10^{-164}),
\end{align}
 where $\alpha_W=g_2^2/(4\pi)$, with $g_2$ the coupling constant of the weak interactions. Although the situation would seem rather hopeless \cite{tHooft:1976rip,Ellis:1987qi}, the fact that rates can be unsuppressed in a thermal plasma suggests that one could get observable $B+L$-violating rates at a collider if the initial state has an energy comparable to the sphaleron barrier \cite{Klinkhamer:1984di,Arnold:1987zg}. The first quantitative estimates in ref. \cite{Ringwald:1989ee} and \cite{Espinosa:1989qn} offered striking results: although the rate involving the minimum of 12 SM fermions remains exponentially small, amplitudes involving the emission of gauge bosons raise with powers of the centre-of-mass energy, and in fact the inclusive rate involves an exponential function of the energy, which becomes unsuppressed for $\sqrt{\hat{s}}\sim E_{\rm sph}$. Higher-order corrections have also been shown to exponentiate \cite{Arnold:1990va,Khlebnikov:1990ue,Arnold:1991cx,Mueller:1991fa,Khoze:1990bm} and this is believed to happen to arbitrary order for final-state  bosons (including Higgses \cite{Khoze:1990bm}, although their effect is subdominant \cite{Khlebnikov:1990ue,Mueller:1991fa,Arnold:1991dv}), which was interpreted as a hint for the existence of an alternative semiclassical expansion that should resum the perturbative corrections in the instanton background \cite{Arnold:1990va,Khlebnikov:1990ue}.  The problem is that the usual instanton perturbation theory only allows to estimate the exponential function as a series expansion in powers of $\sqrt{\hat{s}}/E_{0}$, with $E_0=\sqrt{6}\pi m_W/\alpha_W\sim 18$ TeV of the order of the sphaleron energy. Thus, instanton calculations at a fixed order lose predictive power in the regime of interest $\sqrt{\hat{s}}\gtrsim E_{\rm sph}$, as had been anticipated in \cite{Klinkhamer:1984di,Arnold:1987zg}. The total $B+L$ violating cross-section has then the structure
\begin{align}
\label{eq:sigmaF}
\sigma_{B+L}^{2\rightarrow {\rm any}}= f(\hat{s})e^{-\frac{4\pi}{\alpha_W}F[\sqrt{\hat{s}}/E_{0}]}\ ,
\end{align}
where the ``holy-grail" function $F[\sqrt{\hat{s}}/E_{0}]$ has an expansion in $(\sqrt{\hat{s}}/E_0)^{2/3}$ of the following form \cite{Arnold:1990va,Khlebnikov:1990ue,Arnold:1991cx,Mueller:1991fa,Khoze:1990bm}
\begin{align}
\label{eq:holymoly}
F\left[\frac{\sqrt{\hat{s}}}{E_{0}}\right]=1-\frac{9}{8}\left(\frac{\sqrt{\hat{s}}}{E_0}\right)^{4/3}+\frac{9}{16}\left(\frac{\sqrt{\hat{s}}}{E_0}\right)^{2}+{\cal O}\left(\frac{\sqrt{\hat{s}}}{E_0}\right)^{8/3}.
\end{align}
In order to gain more information about the holy grail function, one may use unitarity bounds \cite{Zakharov:1991rp,Maggiore:1991vi,Veneziano:1992rp}, calculations based on dispersion relations \cite{Zakharov:1990dj,Porrati:1990rk,Khoze:1990bm,Khoze:1991mx,Ringwald:2002sw}, as well evaluations based on the coherent-state approach to S-matrix elements, in which estimates for rates with many-initial particles --which can be calculated semi-classically-- are extrapolated down to two initial states \cite{Rubakov:1991fb,Tinyakov:1991fn}. Unitarity requires $F[\sqrt{\hat{s}}/E_0] \geq 0$, and does not rule out the possibility of $F$ becoming small enough at high-energies, so as to lead to observable rates. Estimates from dispersion relations based on instanton-anti-instanton interactions hint at $F$ becoming small at high-energies \cite{Khoze:1991mx }, and give a value of the prefactor going as \cite{Khoze:1990bm}
\begin{align}
\label{eq:prefactor}
f_{\rm SM}(\hat{s})=\frac{1}{m^2_W}\left(\frac{2\pi}{\alpha_W}\right)^{7/2}.
\end{align}
With the semiclassical calculations in the coherent state approach, lower bounds for $F$ were obtained allowing for $F < 0.1$ for $\sqrt{\hat{s}}\gtrsim 45$ TeV \cite{Bezrukov:2003er,Bezrukov:2003qm}; however,  direct estimates for spherically symmetric saddle points give a value of $F$ that flattens out at $F\sim 0.5$ . The aforementioned lower bound was used in \cite{Ringwald:2003ns} to estimate rates of the order of $10^{-3}$ fb at $\sqrt{\hat{s}}\sim 30$ TeV, which would be potentially observable.
For more in-depth reviews of the calculations summarised above, see \cite{Mattis:1991bj,Tinyakov:1992dr,Guida:1993qy,Rubakov:1996vz}.

Aside from the previous estimates relying on semiclassical techniques, a new approach was advocated in reference \cite{Tye:2015tva}, which proposed that the tunneling problem in the space of bosonic fields is equivalent to a one-dimensional effective quantum dynamics in terms of the Chern-Simons number, whose potential is periodic. Tunneling becomes then unsuppressed for energies within conducting energy bands of the Bloch wave-functions; this gives rates at $\sqrt{\hat{s}}\sim E_{\rm sph}$ which are quoted to be 70 orders of magnitude above those corresponding to $F=0.5$, which, using \eqref{eq:sigmaF} and \eqref{eq:prefactor}, would give a cross-section of the order of $100$ fb at the sphaleron energy. This result seems to contradict the exponential suppression in instanton calculations. The existence of exponential suppression has also been justified with general arguments based on the idea that an initial two-particle state has an exponentially suppressed overlap with the configurations which dominate tunneling (see e.g. \cite{Funakubo:2016xgd}, which argues that such overlap factors were not accounted for in \cite{Tye:2017hfv}). Some counter-arguments were given recently in \cite{Tye:2017hfv}; the debate is still open, and it has revived the interest in the possibility of observing $B+L$ violating interactions at colliders or in high-energy cosmic ray events; for recent studies see for example \cite{Ellis:2016ast,Brooijmans:2016lfv,Ellis:2016dgb}.

Most of the previous discussion in the literature was mainly concerned with the enhancement from gauge boson emission. Equation \eqref{eq:deltaBL} prevents interactions with arbitrary number of fermion emissions within each topological sector, while the exponential dependence in \eqref{eq:sigmaF} is due  to the emissions of large number of bosonic particles. In fact, it can be shown formally that fermions only contribute to the polynomial factor $f(\hat{s})$ in \eqref{eq:sigmaF} \cite{Mattis:1991bj,Espinosa:1991vq}. In this paper we address the question of whether fermions beyond the Standard Model can enhance this prefactor, and thus play a role in the potential for observation of $B+L$ violating processes at colliders. Since these processes are due to $SU(2)$ anomalies, only new fermions in non-trivial $SU(2)$ representations can have an impact. If the new fermions are chiral, heavy masses require large Yukawa couplings to electroweak scalars like the Higgs; to allow for large masses while avoiding strong coupling, we will focus instead in vector representations. In this case, chiral anomalies in backgrounds with $n_{\rm top}=1$  predict not only the SM-like 12 fermion interaction, but also interactions involving the exotic fermions. If these are heavy enough, the decoupling theorem \cite{Appelquist:1974tg} implies that the SM-like processes will have SM-like rates. However, the polynomial factor in the cross section can still be enhanced with respect to the SM for the interactions involving the BSM fermions. Regarding the exponential energy-dependence accounting for gauge-boson emission, the arguments leading to equation \eqref{eq:sigmaF} still apply for the processes with BSM particles, but the associated holy  grail function could be sensitive to the masses of the exotic fermions. Our main goal will be to study ratios of cross sections for BSM processes over SM-like processes; doing so, we expect to eliminate sensitivity with respect to uncertainties in the overall normalisation. We will work in the sector with $n_{\rm top}=1$, and restrict to partial amplitudes with fixed numbers of gauge bosons, estimated with instanton techniques as in \cite{tHooft:1976rip,tHooft:1976snw,Shifman:1979uw,Ringwald:1989ee,Espinosa:1989qn}. The novelty of our approach lies in the emphasis put in the selection rules enforced by chiral anomalies in the presence of fermion masses, and the use of decoupling arguments to arrive to an instanton density valid for instanton scales both above and below these masses. Ratios for processes with zero  bosons will allow to estimate the impact of BSM fermions in the polynomial contributions to the cross section. On the other hand, estimates of the enhancement of the rates with the number of bosons will allow us to qualitatively infer whether $F$ is sensitive to the BSM fermions. This is because the first energy-dependent term  in the expansion \eqref{eq:holymoly} of the holy grail function is known to capture the sum of the cross sections with fixed numbers of gauge bosons, at leading order in the semiclassical expansion around a single instanton. A dependence of the holy grail function on the mass of heavy fermions is  of course expected from the fact that, for a fixed centre of mass energy $s$, having to produce more exotic fermions reduces the energy available for boson emission. Nevertheless, for energies much above the masses of the heavy fermions one should expect to recover the SM function in \eqref{eq:holymoly}.

To illustrate the impact of exotic fermions, we will focus  in simple anomaly-free extensions of the SM including a pair of Weyl fermions in the fundamental of $SU(2)_L$, or a new Weyl fermion in the adjoint. Such particles can be found in BSM extensions with Supersymmetry (SUSY), such as the two Higgsinos and the electroweak gaugino in the Minimal Supersymmetric Standard Model (MSSM). Given this, we will also study scenarios including both types of BSM fermions, either in simple MSSM realizations with degenerate superparticles, or in scenarios in which all of them are decoupled except for the Higgsinos and 
which are motivated by dark-matter and gauge-coupling unification.

The paper is organised as follows. In section\,\ref{sec:anomalies} we review how the $SU(2)_L$ anomalies corresponding to chiral rotations of the fermions, as well as 
$B$ and $L$ transformations,
predict anomalous processes that violate $B+L$ and which, in the presence of exotic fermions, include not only SM-like interactions, but also reactions involving the BSM fermions.  The  quantitative estimate of the rates of $B+L$ reactions will be the subject of section \ref{sec:instantons}, paying particular attention to anomaly selection rules and decoupling. The formalism will be applied to concrete scenarios of new physics in sections \ref{sec:heavy_fermions}, before the final discussion. We include two appendices, one summarising our Euclidean-space conventions, and another giving details on the fermionic zero modes in the fundamental and adjoint representations, and their associated instanton form-factors.

%%%%%%%%%%%%%%%%%%%%%%%%%%%%%%%%%%%%%%%%%%%%%%%%%%%%%%%
%%%%%%%%%%%%%%%%%%%%%%%%%%%%%%%%%%%%%%%%%%%%%%%%%%%%%%%
%%%%%%%%%%%%%%%%%%%%%%%%%%%%%%%%%%%%%%%%%%%%%%%%%%%%%%%
%%%%%%%%%%%%%%%%%%%%%%%%%%%%%%%%%%%%%%%%%%%%%%%%%%%%%%%
%%%%%%%%%%%%%%%%%%%%%%%%%%%%%%%%%%%%%%%%%%%%%%%%%%%%%%%
\section{$B+L$ violating processes from $SU(2)_L$ anomalies}
\label{sec:anomalies}
%%%%%%%%%%%%%%%%%%%%%%%%%%%%%%%%%%%%%%%%%%%%%%%%%%%%%%%
%%%%%%%%%%%%%%%%%%%%%%%%%%%%%%%%%%%%%%%%%%%%%%%%%%%%%%%
%%%%%%%%%%%%%%%%%%%%%%%%%%%%%%%%%%%%%%%%%%%%%%%%%%%%%%%
%%%%%%%%%%%%%%%%%%%%%%%%%%%%%%%%%%%%%%%%%%%%%%%%%%%%%%%
%%%%%%%%%%%%%%%%%%%%%%%%%%%%%%%%%%%%%%%%%%%%%%%%%%%%%%%

Consider a theory with $N_F$ Weyl fermions, $\psi_k$, in representations $r_k$ of a gauge group with coupling $g$, assumed to be semisimple --as is the case
for $SU(2)$, on which this paper will focus. Each representation $r_k$
has generators $T^a,\,a=1,\dots,\dim(r_k)$, and an associated Dynkin index $T_k$, defined from the relation
\begin{align}
\label{eq:TR}
\text{Tr}_{r_k} T^a T^b=T_k\,\delta^{ab}.
\end{align}
The fermions may also be charged with charges $q^S_k$ under a global $U(1)$ symmetry $S$ with transformations\footnote{$S$ may be the baryon or lepton number of the $SU(2)$ doublets.  In the former case, the doublets of quark Weyl spinors have $q_k^S =1/3$, whereas in the latter case the lepton doublets have $q_k^S =1$.}
\begin{align}
 \psi_k\rightarrow e^{i q^S_k \alpha}\psi_k.
\end{align}
Quantum effects will in general give rise to an anomalous conservation law for the associated current $J^\mu_S = \sum_k q_k^S \psi_k^\dagger \overline\sigma^\mu \psi_k$:
\begin{equation}\label{eq:anomaly}\begin{aligned}
 \int d^4x\,\partial_\mu J_S^\mu=&\,N_S n_{\rm top},\\
 %%%%
 N_S=&\,2\sum_k q^S_k T_k,\\
 %%%% 
 n_{\rm top}=&\,\int d^4x\frac{g^2}{16\pi^2} \,{\rm Tr}\,\tilde F_{\mu\nu} F^{\mu\nu},\quad \tilde F_{\mu\nu}=\frac{1}{2}\epsilon_{\mu\nu\rho\sigma} F^{\rho\sigma}.
\end{aligned}\end{equation}
In the above equation, $F_{\mu\nu}=\partial_\mu A_\nu-\partial_\nu A_\mu-ig[A_\mu,A_\nu]$ is the field strength --with $A_\mu=A_\mu^aT^a$ the gauge potential-- and $\epsilon_{\mu\nu\rho\sigma}$ is the Levi-Civita tensor with $\epsilon_{0123}=1$. The notation
$n_{\rm top}$ reflects the fact that $n_{\rm top}$ is a topological invariant --the integral of a total derivative, and thus determined by boundary terms-- known to take integer values. For field configurations with finite energies,
the gauge potential must approach a pure gauge configuration,
$A_\mu =i g^{-1}\mathcal{U}\partial_\mu \mathcal{U}^\dagger$, 
at  space-time infinity,
which defines a map from the 3-sphere at the space-time infinity
to the gauge group, forming an equivalence class labelled by $n_{\rm top}$. The 
anomaly implies a violation of the conservation of the charge $Q_S=\int d^3 x J_S^0$ associated with the global symmetry:
\begin{align}
\label{eq:charge}
Q_S(t=\infty)-Q_S(t=-\infty)=\int d^4 x \frac{dJ_{S}^0}{dt}=\int d^4 x\,\partial_\mu J_S^\mu = N_S n_{\rm top}\ ,
\end{align}
where we assume the current is not flowing-in or -out at the boundary of the spacial infinity, $\oint_S \vec{J_S} \cdot d \vec{s} = 0$.
The anomalous processes
predicted by the relation \eqref{eq:charge} will be associated with effective interaction vertices arising from nonperturbative dynamics, as reviewed in the next section. The nonperturbative character
of the 
anomalous effects can be understood from the fact that they appear in association with the topological charge
$n_{\rm top}$, which, being the integral of a total derivative, does not generate any perturbative vertices. 

Of particular importance are the chiral symmetries $C_k$ --present when there are no mass terms that couple pairs of fermions charged under the gauge group--  which rotate the Weyl fermions in a given nontrivial representations $k$ of the group, with $q_k=1$. This implies that 
\begin{align}
\label{eq:chiral_anomaly_k}
 \Delta Q^k_{\rm chiral}= 2 T_k n_{\rm top}\equiv N^{0,k}_F n_{\rm top}.
\end{align}
One can define as well combinations of the above flavoured chiral rotations, in particular that in which all Weyl fermions are rotated with the same phase. This leads to the following relation for the total chiral charge, $Q_{\rm chiral}$,
\begin{align}
\label{eq:chiral_anomaly}
 \Delta Q_{\rm chiral}= \sum_k N^{0,k}_F n_{\rm top}\equiv N^{0}_Fn_{\rm top}
\end{align}

For fermions in the fundamental of $SU(2)$, $T(\rm fund)=1/2$, while for fermions in the adjoint, $T(\rm adj)=2$. Assuming classical invariance under chiral rotations, the anomaly \eqref{eq:chiral_anomaly} predicts that the processes with minimal violation of $Q_{
\rm chiral}$ correspond to gauge-field backgrounds
with $n_{\rm top}=1$, with every fundamental fermion contributing one unit to $\Delta Q_{\rm chiral}$, every
adjoint fermion contributing 4 units, etcetera. This means that the corresponding effective interaction vertex consistent with the anomaly of $\Delta Q_{\rm chiral}$ under $SU(2)$, and with minimal 
charge violation,
will involve one field insertion for every Weyl fermion in the fundamental, 4 fermion insertions for 
every Weyl fermion in the adjoint, and $2T_k$ insertions for any other representation $r_k$. Such anomalous interactions must involve all the Weyl fermions present in the theory that transform nontrivially under $SU(2)$, as enforced by the anomalous conservation laws of the flavoured chiral symmetries \eqref{eq:chiral_anomaly_k}.

In the presence of mass terms that couple pairs of fermions charged under the gauge group, the chiral symmetry is explicitly broken.\footnote{

Note that this does not apply to 
the fermion mass terms or the Yukawa terms
in the Standard Model, since
the right-handed quarks and leptons are
$SU(2)$ singlets. One can still define a classical chiral symmetry by complementing the rotations of the doublets with compensating transformations of the singlets.}

Even with a broken symmetry, one may still treat the masses as spurions with an associated chiral charge which would render the mass-terms invariant. Then one may still use equation \eqref{eq:chiral_anomaly} to constrain the effective Lagrangian, but with the understanding that mass insertions also count towards $\Delta Q$; in this sense, \eqref{eq:chiral_anomaly} becomes a selection rule. In this way, one gets not just the previous effective vertices involving all the Weyl fermions in nontrivial representations of $SU(2)$, but also additional lower-dimensional operators, in which pairs of fermion fields are traded for the conjugate of their corresponding mass (note that, if ${\cal L}\supset -m \psi_k \psi_l+c.c.$, then $m^*$ carries the same spurious chiral charge as the product of two Weyl spinors. The  maximum number of insertions of a given mass is the one that saturates the contribution of the associated fermions to $\Delta Q_{\rm chiral}$. This follows from considering alternative chiral symmetries which do not involve rotations of the massive fermions, and thus remain classically exact; the associated $\Delta Q$ give the minimum amount of chiral violation in the anomalous interactions).

In the SM, every generation, $k$, has 3 quark doublets $q_k^{cw}$ --where $c=1,\dots,3$ is a colour index, and $w=1,2$  a weak index-- and a lepton doublet $l^w$, all in the fundamental of SU(2). Despite the presence of Yukawa couplings, there is still an exact classical chiral symmetry under which the left-handed doublets have unit charge, and the $SU(2)_L$ singlets transform with compensating phases that leave the Yukawa terms invariant. Eq.~\eqref{eq:chiral_anomaly} then predicts anomalous interactions with $n_{\rm top}=1$ involving 12
fermion fields, of the form
\begin{align}
\label{eq:Lanom}
 \Delta{\cal L}\sim y_{ \{c\};\{w\}} \prod_{i=1}^3   q^{c^1_i w^1_i}_i q^{c^2_i w^2_i}_i q^{c^3_i w^3_i}_i l^{w^4_i}_i,
\end{align}
where the index $i$ is not summed over, and the $y_{\{c\};\{w\}} $ must be compatible with gauge invariance under the SM gauge group. The number and type of fermions per generation in the interaction vertex follow from considering chiral symmetries --and their anomalies-- in which only some of the fields are charged. The result is a determinant-like interaction, involving one fermion of each type; such anomalous vertices where discovered by 't Hooft in the QCD context \cite{tHooft:1976rip}. Crucially, the interaction breaks baryon and lepton number, while preserving $B-L$, as it is clear from the fact that all quark doublets $q_k$ carry $B=1/3$, while the  lepton doublets $l_k$ carry $L=1$. This, of course, fits with the anomalous identities for $B$ and $L$ that follow from \eqref{eq:anomaly}, which implies $\Delta B=\Delta L=3n_{\rm top}$. 

Having reviewed the situation in the SM, one may wonder if new physics with massive fermions can have an effect on the $B+L$ violating interactions. Naturally, in order to partake in weak anomalous processes, the new massive fermions should be charged under $SU(2)$. A priori there is no reason that these particles carry $B$ or $L$, yet they could have some anomalous fermion number. But even if the fermion number is non-anomalous, the new particles will still partake in anomalous interactions. This is because, as seen before, the anomaly under chiral rotations enforces interactions that involve all the new Weyl fermions charged under $SU(2)$. If the new fermions are chiral --i.e., if one  can find a classical symmetry under which all $SU(2)$ Weyl fermions have unit charge-- then all anomalous interactions must involve the new fields, and there is no limit in which one recovers the SM interactions of the form of equation \eqref{eq:Lanom}. This might seem puzzling given the decoupling theorem, which would appear to warrant an SM-like limit if the new particles become heavy. However, new chiral fermions can only become  heavy by coupling strongly to the Higgs, so that the decoupling theorem does not apply.\footnote{Decoupling holds in the limit in which particles are made heavy, while keeping their couplings constant.
This possibility is however strongly constrained by the measurements of the Higgs productions and decays at the LHC.} For nonchiral new fermions --that is, with masses incompatible with classical chiral symmetries-- then, aside from the interactions involving all $SU(2)$ Weyl fermions, there will be additional vertices in which pairs of new fermion fields are traded for  insertions of their associated mass. In this case one predicts SM-like vertices as in \eqref{eq:Lanom}. Now the decoupling theorem applies, and in the limit of heavy new particles one expects to recover identical rates as in the SM case for the SM-like processes.

Regarding the possibilities for new weakly charged fermions, it should be noted that they are restricted by the Witten and gauge anomalies. The Witten anomaly \cite{Witten:1982fp} requires an even number of fermions in representations with half-integer Dynkin index --such as the fundamental, but not the adjoint. In regards to the gauge anomalies, they are of no concern for $SU(2)$, as its anomaly is determined by the invariant symmetric tensors $d_\rho^{abc}={\rm Tr}_\rho\{T^a,T^b\}T^c$, which vanish in $SU(2)$. However, if the new particles carry representations under other gauge groups --such as $SU(3)$, $U(1)_Y$ or a hidden gauge group-- there will be additional constraints.

To finish this discussion and pave the way for the last part of the paper, we will consider four example scenarios with nonchiral fermions, which will be analyzed in section \ref{sec:heavy_fermions}.

{\bf Dirac fermion in the fundamental}

Such a Dirac fermion, 
$\Psi_F=\{\psi_{F,\alpha},{\tilde\psi}_F^{\dagger,\dot{\alpha}}\}$,
involves two left-handed Weyl fermions, $\psi_F$ and $\tilde \psi_F$,
in the fundamental and antifundamental  representations, respectively.\footnote{Note that the fundamental and antifundamental representations of $SU(2)$ are related by a similarity transformation involving the antisymmetric matrix $\epsilon=i\sigma^2$, with $\sigma^2$ the second Pauli matrix.} There is no Witten anomaly, and one may write a Dirac mass ${\cal L}\supset m_F \overline{\Psi}_F \Psi_F = m_F\tilde\psi_F\psi_F+c.c.$  implies $\Delta{Q_{\rm chiral}}=14$, and one predicts then two types of vertices:
\begin{itemize}
\item Vertex with 14 Weyl fermions, of the form
\begin{align}
\label{eq:Lanomf}
 \Delta{\cal L}_F \sim 
 y_{ \{c\}; \{ w \} }
 \,\left(\prod_{i=1}^3  q^{c^1_i w^1_i}_i q^{c^2_i w^2_i}_i q^{c^3_i w^3_i}_i l^{w^4_i}_i\right)\psi_F^{w^5} \tilde \psi_F^{w^6};
\end{align}
\item SM-like vertex with 12 fermions as in \eqref{eq:Lanom}, which still satisfies  $\Delta_{Q_{\rm chiral}}=14$ by involving an insertion of $m_F^*$ in substitution of $ \psi_F\tilde \psi_F$.
\end{itemize}

The fermion number is defined such that it is 
associated with the phase rotation of 
the Dirac fermion $\Psi_F=\{\psi_{F,\alpha},{\tilde\psi}_F^{\dagger,\dot{\alpha}}\}$
and $\psi_F$ and $\tilde \psi_F$ carry opposite charges.
The vertex in \eqref{eq:Lanomf} thus preserves the number of the new species; instead, the anomalous interactions will involve for example the creation or annihilation of particle-antiparticle pairs of the new fermion.\footnote{This is the reason that models of asymmetric dark matter in which the $B+L$ asymmetry is related to a dark-sector fermion number through  $SU(2)$ anomalies require dark fermions to be chiral.} 

{\bf Weyl fermion in the adjoint}

We may write the adjoint Weyl fermion as $\psi_A^a T^a$.
Again, there is no Witten anomaly, and one can write down a gauge-invariant mass term, $m_A \psi_A^a \psi_A^a$, incompatible with chiral symmetries, and which can assign $m_A$ a spurious chiral charge of $-2$. The anomaly equation is now  $\Delta{Q_{\rm chiral}}=16$, and one predicts three types of vertices:
\begin{itemize}
\item Vertex with 16 Weyl fermions, of the form
\begin{align}
\label{eq:LanomA}
 \Delta{\cal L}_A\sim y_{ \{c\} ;\{w \}; \{a\} } \,\left(\prod_{i=1}^3q^{c^1_i w^1_i}_i q^{c^2_i w^2_i}_i q^{c^3_i w^3_i}_i l^{w^4_i}_i\right)\psi^{a_1}_A\psi^{a_2}_A\psi^{a_3}_A \psi^{a_4}_A,
\end{align}
\item Vertex with 14 fermions (involving an insertion of $m_A^*$),
\begin{align}
\label{eq:LanomAp}
 \Delta{\cal L}_A\sim y_{ \{c\} ;\{w \}; \{a\} } m_A^* \,\left(\prod_{i=1}^3
q^{c^1_i w^1_i}_i q^{c^2_i w^2_i}_i q^{c^3_i w^3_i}_i l^{w^4_i}_i\right)\psi^{a_1}_A\psi^{a_2}_A ,
\end{align}
\item SM-like vertex with 12 fermions as in \eqref{eq:Lanom}, which can be understood from
an insertion of $(m^*_A)^2$.

\end{itemize}

We note that  adjoint fermions appear in supersymmetric gauge theories, and their nonperturbative, anomalous interactions have been intensively studied (see e.g. \cite{Terning:2003th}). This motivates us to also consider the situation in the Minimal Supersymmetric Standard Model (MSSM) or related models.

{\bf SUSY inspired models}

The MSSM involves the following additional fermions charged under $SU(2)$: two Weyl fermions in the (anti) fundamental --the Higgsinos $\psi_{H_u}$ and $\psi_{H_d}$, which can be grouped into a Dirac fermion in the fundamental-- and a Weyl spinor in the adjoint, the $SU(2)$ gaugino $\lambda_2$. The Higgsinos are coupled through a supersymmetric mass term, $\mu$, while the gauginos have a Supersymmetry-breaking mass, $M_2$. Then we are in a situation which combines the previous two scenarios. The allowed vertices are: 
\begin{itemize}
\item Vertex with 18 Weyl fermions, of the form
\begin{align}
\label{eq:LanomMSSM1}
 \Delta{\cal L}_{\rm MSSM}\sim y_{ \{c\} ;\{w \}; \{a\} } \,\left(\prod_{i=1}^3q^{c^1_i w^1_i}_i q^{c^2_i w^2_i}_i q^{c^3_i w^3_i}_i l^{w^4_i}_i\right)\lambda^{a_1}_2\lambda^{a_2}_2\lambda^{a_3}_2 \lambda^{a_4}_2 \psi_{H_u}^{w^5}  \psi_{H_d}^{w^6},
\end{align}
\item Vertex with 16 fermions, involving an insertion of $M_2^*$). This gives an interaction of the form
\begin{align}
\label{eq:LanomMSSM2}
 \Delta{\cal L}_{\rm MSSM}\sim y_{ \{c\} ;\{w \}; \{a\} } M_2^*\,\left(\prod_{i=1}^3q^{c^1_i w^1_i}_i q^{c^2_i w^2_i}_i q^{c^3_i w^3_i}_i l^{w^4_i}_i\right)\lambda^{a_1}_2\lambda^{a_2}_2 \psi_{H_u}^{w^5}  \psi_{H_d}^{w^6},
\end{align}
\item Vertex with 16 fermions (involving an insertion of $\mu^*$). This gives an interaction similar to \eqref{eq:LanomA}.

\item Vertex with 14 fermions, involving two insertions of $M_2^*$, giving an interaction as in \eqref{eq:Lanomf}.

\item Vertex with 14 fermions, involving an insertion of $M_2^*\mu^*$. This gives an interaction as in \eqref{eq:LanomAp}.

\item 12 fermion SM-like vertex as in \eqref{eq:Lanom}, arising from an insertion of $(M^*)^2\mu^*$.
\end{itemize}

The previous vertices will exist in any realization of the MSSM in which the Higgsinos and weak gauginos are not decoupled, regardless of the masses of the rest of the BSM particles.  For concreteness, in this article we will consider a Split-SUSY inspired scenario, in which the Higgsinos and weak gauginos are the only dynamical BSM fields, as well as a degenerate MSSM limit in which all BSM scalars and fermions of the MSSM are assumed to have approximately degenerate masses.
%%%%%%%%%%%%%%%%%%%%%%%%%%%%%%%%%%%%%%%%%%%%%%%%%%%%%%%
%%%%%%%%%%%%%%%%%%%%%%%%%%%%%%%%%%%%%%%%%%%%%%%%%%%%%%%
%%%%%%%%%%%%%%%%%%%%%%%%%%%%%%%%%%%%%%%%%%%%%%%%%%%%%%%
%%%%%%%%%%%%%%%%%%%%%%%%%%%%%%%%%%%%%%%%%%%%%%%%%%%%%%%
%%%%%%%%%%%%%%%%%%%%%%%%%%%%%%%%%%%%%%%%%%%%%%%%%%%%%%%
\section{$B+L$ violating rates from instanton effective Lagrangians}
\label{sec:instantons}
%%%%%%%%%%%%%%%%%%%%%%%%%%%%%%%%%%%%%%%%%%%%%%%%%%%%%%%
%%%%%%%%%%%%%%%%%%%%%%%%%%%%%%%%%%%%%%%%%%%%%%%%%%%%%%%
%%%%%%%%%%%%%%%%%%%%%%%%%%%%%%%%%%%%%%%%%%%%%%%%%%%%%%%
%%%%%%%%%%%%%%%%%%%%%%%%%%%%%%%%%%%%%%%%%%%%%%%%%%%%%%%
%%%%%%%%%%%%%%%%%%%%%%%%%%%%%%%%%%%%%%%%%%%%%%%%%%%%%%%
In this section we review how the nonperturbative anomalous interactions can be recovered by means of instanton techniques. Although, as said in the introduction, perturbations around instanton backgrounds are known not to be able to accurately capture
the effect of gauge boson emission --since in the SM the leading instanton results only give the first energy-dependent term in the expansion \eqref{eq:holymoly} of the holy-grail function-- they will suffice to estimate the behaviour of the polynomial function $f(\hat{s})$ in \eqref{eq:sigmaF}. We will nevertheless estimate corrections from gauge boson emission in the instanton background in order to assess how the masses of heavy fermions affect the $(\sqrt{\hat{s}}/E_0)^{4/3}$  contribution to the holy grail function. 

We will start by reviewing the usual construction of effective Lagrangians for fermions in an instanton background, emphasising how the selection rule \eqref{eq:chiral_anomaly} can be used to understand features related to the properties of the fermionic zero modes and the insertions of fermion masses. Next, we will review the effect of adding a scalar field that breaks the gauge symmetry, and the construction of anomalous effective interactions including both fermions and bosons.
We will improve upon the usual expressions by accounting for decoupling effects, and we will provide approximate formulae for the effective
interactions that will be used in the next section to estimate relative cross sections. 

%%%%%%%%%%%%%%%%%%%%%%%%%%%%%%%%%%%%%%%%%%%%%%%%%%%%%%%
%%%%%%%%%%%%%%%%%%%%%%%%%%%%%%%%%%%%%%%%%%%%%%%%%%%%%%%
%%%%%%%%%%%%%%%%%%%%%%%%%%%%%%%%%%%%%%%%%%%%%%%%%%%%%%%
\subsection{\label{subsec:instanton_fermions}Instanton effective Lagrangians in gauge theories with fermions}
Instanton calculations rely on semiclassical (saddle-point) approximations to the path integral \cite{tHooft:1976snw} (see e.g. \cite{Vainshtein:1981wh} for a pedagogical introduction).  Consider an SU$(N)$ gauge theory with gauge fields $A_\mu^a$ and fermions $\psi_k$ --the effect of a Higgs scalar will be considered later.
The gauge fields fall into equivalence classes of configurations with different values of the integer topological charge $n_{\rm top}$ in \eqref{eq:anomaly}, so that the path integration over gauge fields can be expressed as a sum of path integrals over the different topological sectors.   Crucially, equation \eqref{eq:chiral_anomaly} for the chiral anomaly implies that in any $n_{\rm top}\neq0$ background the chiral charge must be altered, and thus the saddle point approximations to the path integrals for the sectors with nonzero $n_{\rm top}$ must generate the anomalous interactions of section \ref{sec:anomalies}. 

In this theory, the classical vacua 
are given by the pure gauge configuration
$A_\mu(x) = ig^{-1}{\mathcal U}^{\dagger}(x) \partial_\mu {\mathcal U(x)}$.
For a given time slice,
the classical vacua are given by 
maps from three-dimensional space, with coordinates ${\bf x}$,
to group elements ${\tt g} \in SU(2)\simeq S^3$.
In the topological gauge, where 
$A_0(x) = 0$ for any $x$
and
$A_\mu({\bf x}) \to 0$
for $|{\bf x}|\to \infty$, the latter condition allows to identify spatial infinity with a single point,
and
the space ${\mathbb R}^3$ is compactified into $S^3$.
One can then see that the classical vacua 
are classified by the 
Chern-Simons number, $N_{\rm CS}$,
corresponding to the winding number $\pi_3(S^3) = {\mathbb Z}$.
It can be shown that $n_{\rm top}=N_{\rm CS}(t=\infty)-N_{\rm CS}(t=-\infty)$. We expect then the existence of mutually orthogonal ``perturbative'' vacuum states $|n\rangle$ with associated integer $N_{\rm CS}=n$, and a true vacuum state $|0\rangle$ given by a linear combination of the former. Since $|0\rangle$ has to be gauge invariant, and time-independent gauge-transformations change $N_{\rm CS}$ by integer amounts, this forces $|0\rangle=\sum_n e^{in\theta}|n\rangle$ for an arbitrary $\theta$. Then, after a rotation to Euclidean space (see appendix \ref{app:Euclidean} for our conventions) we may write the generating functional, given by the amplitude $_{\rm out}\langle 0|0\rangle_{\rm in}=\sum_{mn} {}_{\rm out}\langle m|n\rangle_{\rm in}e^{i(n-m)\theta}$, as:
\begin{equation}
\label{eq:Z}
\begin{aligned}
Z=&\,\sum_{mn} {}_{\rm out}\langle m|n\rangle_{\rm in}e^{i(n-m)\theta}=\,\sum_N\int [dA_N]\prod_k[d\psi_k][d\psi^\dagger_k]\exp\left(-S\right), \\
%%%%%
S=&\,S_g+S_\theta+S_F+S_{g.f.},\\
%%%%%
S_g=&\,\int d^4x \,\frac{1}{2}\,{\rm Tr}\,F_{\mu\nu} F_{\mu\nu},\\
%%%%
S_\theta=&\, i\theta n_{\rm top}=\int d^4x\frac{i\theta g^2}{16\pi^2} \,{\rm Tr}\,\tilde F_{\mu\nu} F_{\mu\nu},\\
%%%%
S_F=&\, \int d^4x\, -\psi_k^\dagger i\overline{\sigma}_\mu D_\mu \psi_k+\left(\frac{1}{2}M_{kl} \psi_k \psi_l+c.c\right); \quad \overline{\sigma}_\mu =(-\vec{\sigma},i),\\
%%%%%
S_{g.f.}=&\,\int d^4x \,\frac{1}{2\xi}F^a[A_\mu]F^a[A_\mu]+\bar c^a \frac{\delta F^a[A_\mu]}{\delta A_\mu^b} (D_\mu c)^b.
\end{aligned}
\end{equation}
In the above equations, $[dA_N]$ represents path integration of gauge fields over the sector with topological charge $n_{\rm top} = N$. Note that we identified $n_{\rm top}$ in \eqref{eq:Z} with the difference $n-m$ of the Chern-Simons numbers of the in and out vacua, as anticipated before. Although the relation between $N_{\rm CS}$ and $n_{\rm top}$ only works in the topological gauge, 
the same applies to the identification of vacua with static pure gauge configurations; for arbitrary gauges we then take equation \eqref{eq:Z} as the definition of the partition function of the true vacuum.
The inclusion of the $\theta$-term, which we motivated by demanding gauge-invariance of the vacuum,  can also be argued by demanding that the Lagrangian includes the most general renormalisable and gauge-invariant interactions. The covariant derivative of the fermions is $D_\mu=\partial_\mu-igA_\mu^aT^a$. $S_{g.f.}$ is the contribution responsible for gauge-fixing, involving gauge-fixing functions $F[A]^a$, a gauge-fixing parameter
$\xi$, and  Grassmannian ghost fields $c^a$ in the adjoint representation of the gauge group, with $(D_\mu c)^a=\partial_\mu c^a+gf^{abc}A^b_\mu c^c$. In the SM, the $\theta$ angle for $SU(3)$ is constrained as $|\theta_3|< 10^{-10}$ by the non-observation of the neutron dipole moment \cite{Baker:2006ts}, while for $SU(2)$ it is unobservable, as it can be rotated away by a combination of $B$ and $L$ transformations \cite{Anselm:1992yz,Perez:2014fja}.\footnote{Strictly speaking, it is not $\theta_3$ that is constrained, but a combination of $\theta_3$ with the phases in the fermion mass matrix that remains invariant under chiral rotations. We will elaborate on a related subtlety concerning $\theta$, chiral rotations and fermion decoupling in section \ref{subsec:decoupling}.} 

As said before, instanton calculations rely in approximating the path integration within each topological sector by performing a saddle point expansion around configurations
which extremise $S=S_g+S_\theta+S_F+S_{g.f}$. The sector of zero topological charge corresponds to ordinary perturbation theory around a background with $A_\mu=0$, so that the first  nontrivial corrections correspond to $n_{\rm top}=\pm1$ saddle points: the single instanton and anti-instanton.  Saddle-points with higher $n_{\rm top}$  (multi-instantons) are expected to be approximately equivalent to a dilute gas
of instantons, and so their classical action is $n$ times the single instanton action, so that one expects  a higher exponential suppression in $\exp(-S)$ --see however \cite{Tye:2015tva,Tye:2017hfv}, which suggest otherwise. Here we will limit ourselves to $|n_{\rm top}|=1$.  As was argued in section~\ref{sec:anomalies},  $|n_{\rm top}|=1$  corresponds to anomalous  interactions with a minimal amount of nontrivial violation of the chiral charge; we will focus on $n_{\rm top}=1$, which yields the interactions vertices we wrote for the SM or its extensions in equations \eqref{eq:Lanom}, \eqref{eq:Lanomf}, \eqref{eq:LanomA}, \eqref{eq:LanomAp}. The $n_{\rm top}=-1$ case is similar, yielding interactions with the opposite violation of the charge (see \eqref{eq:chiral_anomaly}), and captured by taking the complex conjugate of the $n_{\rm top}=1$ interaction vertices.

The anomalous effective interactions for fermions can be recovered with instanton techniques by using the following procedure \cite{tHooft:1976snw,tHooft:1976rip,Shifman:1979uw,Ringwald:1989ee,Espinosa:1989qn}. First, one computes Green functions involving fermions in the $n$-th topological sector by performing a saddle-point expansion around an $n$-instanton configuration. Then, one defines the effective Lagrangian in the $n$-th sector as the one which gives rise to tree-level vertices which reproduce the previous Green functions.

%%%%%%%%%%%%%%%%%%%%%%%%%%%%%%%%%%%%%%%%%%%%%%%%%%%%%%%
%%%%%%%%%%%%%%%%%%%%%%%%%%%%%%%%%%%%%%%%%%%%%%%%%%%%%%%
%%%%%%%%%%%%%%%%%%%%%%%%%%%%%%%%%%%%%%%%%%%%%%%%%%%%%%%
\subsubsection{The $n=1$ instanton configuration}

The $n=1$ instanton-configuration is an extremum of the Euclidean action $S$ in  \eqref{eq:Z}, with associated topological charge $n_{\rm top}=1$. Instantons for arbitrary simple gauge groups can be constructed from the $SU(2)$ solution found in \cite{Belavin:1975fg}, which can be embedded into the $SU(2)$ subalgebras of larger groups. The fermion fields in this BPST instanton solution can be set to zero, while, for a choice of gauge enforcing  $\partial_\mu A_\mu=0$ in the absence of a scalar or fermion background, the gauge fields go as
\begin{align}
\label{eq:inst}
A_\mu=U^\dagger \,\frac{2}{g}\frac{\eta_{a\mu\nu}(x-x_0)_\nu}{(x-x_0)^2+\rho^2}\tau^a\, U.
\end{align}
In the previous equation, $\tau^a$, represent the generators of an arbitrary $SU(2)$ subalgebra, with $a=1,\,2,\,3$, and $U$ denotes a rigid (space-time independent) rotation in the full gauge group (as opposed to the SU(2) subgroup).
The previous gauge-field configuration is localised in space-time (hence the name "instanton") around an arbitrary point $x=x_0$, with a spread controlled by an arbitrary scale $\rho$. $\eta_{a\mu\nu}$, with $a=1,\,2,\,3$, $\mu,\nu=1,..,4$  are the so-called 't~Hooft symbols, which, under the convention that $x_4$ designates Euclidean time, satisfy
\begin{equation}\label{eq:etas}\begin{aligned}
\eta_{a\mu\nu}=&\,\epsilon_{a\mu\nu},\,\,\mu,\nu\in\{1,2,3\},\\
%%%%
\eta_{a4\nu}=&\,-\eta_{a\nu4}=-\delta_{4\nu},\,\,\nu\neq4,\\
%%%%%
\eta_{a44}=&\,0.
\end{aligned}\end{equation}
The BPST instanton can be seen to have topological charge one, and its  Euclidean action is given by
\begin{align}
\label{eq:Sinst}
S_{\rm inst}=\frac{8\pi^2}{g^2} .
\end{align}
When studying quantum fluctuations around the instanton solution, it is convenient to pick a gauge  such that  $A_\mu^a$ vanishes at infinity like  $A_\mu^a\sim r^{-n}$ with $n\geq2$. This simplifies the treatment of the instanton's zero energy bosonic fluctuations \cite{Bernard:1979qt}. Such behaviour is not satisfied by the BPST instanton of equation \eqref{eq:inst}, but one may remedy this by performing a gauge transformation with a group element ${\mathcal U}(x)$ within the $SU(2)$ subalgebra. In our conventions, the gauge transformations of $A_\mu$ go as $A_\mu=A_\mu^a T^a\rightarrow {\mathcal U}^\dagger A_\mu {\mathcal U}+i g^{-1} \mathcal{U}^\dagger\partial_\mu \mathcal{U}$.
Choosing 
\begin{align}
\mathcal{U}(x)=\tilde U^\dagger\frac{i\overline\sigma_\mu(x-x_0)_\mu}{\sqrt{(x-x_0)^2}}\tilde U,
\end{align}
where $\tilde U$ is a new arbitrary rigid rotation, the resulting instanton configuration is
\begin{align}
\label{eq:inst_sing}
A_\mu=\tilde U^\dagger\,\frac{2}{g}\overline{\eta}_{a\mu\nu}(x-x_0)_\nu\frac{\rho^2}{(x-x_0)^2[(x-x_0)^2+\rho^2]}\tau^a\,\tilde U,
\end{align}
where the $\overline{\eta}_{a\mu\nu}$ symbols are obtained from the relations in  \eqref{eq:etas} by changing the sign of $\delta$. The previous equation gives the instanton in the so-called ``singular gauge", given that it blows up at $x=x_0$; however, the action, being gauge-invariant, remains as in equation \eqref{eq:Sinst}, and the singularity is merely a gauge artifact.

%%%%%%%%%%%%%%%%%%%%%%%%%%%%%%%%%%%%%%%%%%%%%%%%%%%%%%%
%%%%%%%%%%%%%%%%%%%%%%%%%%%%%%%%%%%%%%%%%%%%%%%%%%%%%%%
%%%%%%%%%%%%%%%%%%%%%%%%%%%%%%%%%%%%%%%%%%%%%%%%%%%%%%%
\subsubsection{Zero modes, fluctuations, and effective Lagrangians for small instantons}

The fields $\varphi\equiv\{A^a_\mu,\psi_k\}$ can be expanded around the instanton configuration $\varphi_{\rm inst}$ of equation 
\eqref{eq:inst_sing}, $\varphi(x)=\varphi_{\rm inst}(x)+\tilde\varphi(x)$. Then, using the fact that the instanton configuration extremises the Euclidean action and has unit topological charge, one may write
\begin{align}
S[\varphi]=\frac{8\pi^2}{g^2}+i\theta+\int d^4x d^4y\left.\frac{\delta^2S[\varphi]}{\delta \varphi(x)\delta\varphi(y)}\right|_{\varphi_{\rm inst}}\tilde\varphi(x)\tilde\varphi(y)+O(\tilde\varphi^3).
\end{align}
The fluctuations $\tilde\varphi(y)$ can be expanded in eigenvalues of the fluctuation operator $\delta^2S[\varphi]/\delta \varphi(x)\delta\varphi(y)$. 
The integration over modes with nonzero eigenvalues is Gaussian, and gives determinant factors. On the other 
hand, the integration over bosonic zero modes can be recasted into an integration over collective coordinates --arbitrary parameters in the instanton solution-- for which one needs to calculate the associated Jacobians.

The $SU(N)$ instanton in a theory with gauge fields and fermions has the following zero modes \cite{tHooft:1976snw,Bernard:1979qt}:
\begin{itemize}
\item Four translation zero modes, associated with shifts in the collective coordinate $x_0$ in equations \eqref{eq:inst}, \eqref{eq:inst_sing}.
\item A dilatation zero mode, associated with shifts in the scale $\rho$ of the instanton.
\item $4N-5$ zero modes related to gauge transformations, related to shifts in the parameters of the rigid rotation $\tilde U$ in \eqref{eq:inst}. This gives 3 zero modes in $SU(2)$, coinciding with the dimension of the Lie group. For $N>2$ the number of gauge modes $4N-5$ does not coincide with the dimension of $SU(N)$ because some transformations $\tilde U$ leave the instanton solution invariant. The instanton background preserves a $U(N-2)$ symmetry, so that $4N-5$ is the number of generators broken in the instanton solution.
\item $N^0_F=2\sum_k T_k$ fermionic zero modes (in the absence of fermion masses).
\end{itemize}
The bosonic zero modes can be expressed as derivatives of the instanton solution with respect to the collective coordinates $\gamma=\{x_{0\mu},\rho,\tilde U\}$, supplemented by compensating gauge transformations that bring the configuration back to the chosen gauge slice. The singular gauge is convenient because all the zero modes can be treated in the same footing when calculating their associated Jacobian.

The number of fermionic zero modes can be understood from the index theorem \cite{Atiyah:1968mp} of the Dirac operator in the instanton background \cite{Jackiw:1977pu,Fujikawa:1980eg}.
Moreover it coincides with the maximal number of fermion fields in the anomalous interactions discussed in section \ref{sec:anomalies}. This connection becomes
clearer when constructing the fermion effective Lagrangian induced by the instanton. To do so, one has to consider fermionic Green functions in the instanton background.
First, the Green function without any fermion fields --the one-instanton contribution to the vacuum-to-vacuum amplitude-- is given as anticipated above by the contribution from the instanton action \eqref{eq:Sinst}, times a contribution over the bosonic zero modes --involving an integration over collective coordinates $\gamma=\{x_0,\rho,\tilde U \}$-- times determinant factors:
\begin{align}
\label{eq:kkp1}
_{\rm out}\langle n|n+1\rangle_{\rm in}=\exp\left[-\frac{8\pi^2}{g^2(\mu)}-i\theta\right]\int d\gamma J (\gamma) {\det}' \,{\mathfrak M}_A^{-1/2}  \,{\det}' \, {\mathfrak M}_{\rm gh} \,{\det} \,{\mathfrak M}_\Psi.
\end{align}
Above, $\mu$ is the reference renormalisation scale, and $J(\gamma)$ is a zero-mode Jacobian. ${\mathfrak M}_A$, ${\mathfrak M}_{\rm gh}$ and ${\mathfrak M}_\Psi$ denote the fluctuation operators of gauge fields, ghosts and fermions, respectively, while ${\rm det}'$ denotes the determinant with the  zero modes omitted. In the case of fermions,  the determinant may include the zero modes, though the eigenvalues with smallest magnitude are determined by the fermion masses and they can be nonzero. 

The lowest fermion modes are especially relevant, since, as will be seen, they determine the leading contributions to the effective Lagrangian in the instanton background in the limit $\rho M_{kl}\ll1$. A treatment of the fermion fluctuation operator is simplified when using four-component spinors. Consider a basis in which Weyl fermions interact through mass terms that pair each fermion with at most one other. If a mass term links a pair of fermions $(k,l)$, with $k\neq l$ (so that, in order to ensure gauge invariance, $k,l$ are  conjugate representations, then one can group the pair into a Dirac fermion $\Psi_{kl}$.  Weyl fermions in self-conjugate representations, such as the adjoint, can have mass terms $M_{mm}$ with themselves, so one can construct Dirac fermions $\Psi_{mm}$ satisfying a (Minkowski) Majorana condition:\footnote{One cannot define a Majorana condition in Euclidean space compatible with the $SO(4)$ symmetry --see e.g.\cite{Ramond:1981pw}. We define the spinor fields and the partition function by analytic continuation from Minkowksi space \cite{Vainshtein:1982ic,Leutwyler:1992yt}.}
\begin{align}\label{eq:Dirac}
\Psi_{kl}\equiv\left[
\begin{array}{c}
\psi_{k,\alpha}\\
%%%%
\psi_l^{\dagger,\dot{\alpha}}=\epsilon^{\dot{\alpha}\dot{\beta}}\psi^\dagger_{l,\dot\beta}
\end{array}
\right],
\quad
\Psi_{kk}\equiv\left[
\begin{array}{c}
\psi_{k,\alpha}\\
%%%%
\psi_k^{\dagger,\dot{\alpha}}
\end{array}
\right],
\end{align}
where we used dotted indices to distinguish Weyl fermions from their conjugates, and used the $2\times2$ antisymmetric tensor $\epsilon^{\dot{\alpha}\dot{\beta}}$, with $\epsilon^{\dot{1}\dot{2}}=1$, to raise dotted indices. In this way we can always group all our fermions into massive or massless Dirac or Majorana four-component spinors.
Defining Euclidean gamma matrices, left and right projectors, as well as a Euclidean Dirac adjoint as detailed in appendix \ref{app:Euclidean}, we may write $S_F$ in \eqref{eq:Z} as
  \begin{equation}\begin{aligned}
S_F=\sum_{[kl]} S_{kl}+\sum_{m} S_{m}, \,\,S_{kl}=&\int d^4x\, \overline\Psi_{kl}(-i\gamma_\mu D_\mu 
+{\cal M}_{kl})\Psi_{kl},\\
%%%%%
S_{mm}=&\frac{1}{2}\int d^4x\, \overline\Psi_{mm}(-i\gamma_\mu D_\mu 
+{\cal M}_{mm})\Psi_{mm}.
\end{aligned}\end{equation}
We use $[kl]$ to denote a sum with $k\neq l$ over the Dirac fermions $\Psi_{kl}$ constructed from distinct Weyl fermions $\psi_k,\psi_l$ linked by a mass term $M_{kl}$. The  4-component mass matrices are (including the case $k=l$):
\begin{align}
\label{eq:M}
{\cal M}_{kl}=M_{kl}P_L+M^*_{kl}P_R.
\end{align}
To account for the distinction between Dirac and Majorana fermions, we then write the determinant of fermionic fluctuation operator ${\mathfrak M}_\Psi$ in \eqref{eq:kkp1} as
\begin{align}
\label{eq:detDirac}
{\rm det} {\mathfrak M}_\Psi=\prod_{[kl]}{\rm det}\,{\mathfrak M}_{kl}\prod_{m}\,({\rm det}\,{\mathfrak M}_{mm})^{1/2},
\end{align}
with ${\mathfrak M}_{kl}=-i\gamma_\mu D_\mu 
+{\cal M}_{kl}$ the fermionic fluctuation operator for the 4-component spinor $\Psi_{kl}$. If $k=l$, the Majorana constraint is taken care by the square root, so that ${\rm det}\,{\mathfrak M}_{mm}$ should be thought of as acting on unconstrained Dirac spinors (see e.g. \cite{Vainshtein:1982ic,Witten:1982fp,Leutwyler:1992yt}).\footnote{Problems with the square root of the determinant are at the heart of Witten's anomaly; however, we only use Majorana spinors for Weyl fermions in the self-conjugate representations such as the adjoint, for which there is no problem.}
It turns out that in the  instanton background with $n_{\rm top} = 1$, 
when acting on the Dirac spinor $\Psi_{kl}$, $\gamma_\mu D_\mu$  has 
\begin{align}
\label{eq:N0mn}
N^0_{kl}=T_k+T_l
\end{align}
right-handed zero modes. 
This can be understood from the Atiyah-Singer index theorem \cite{Atiyah:1968mp,Jackiw:1977pu,Fujikawa:1980eg}, which relates the number of zero modes $n_\pm$ with positive and negative chiralities for a given Dirac fermion to the topological charge:
\begin{align}
n_+-n_-=N^0_{kl} n_{\rm top}.
\end{align}
The instanton background, with $n_{\rm top}=1$, satisfies a self-duality condition which can be seen to imply that $n_-=0$ \cite{Jackiw:1977pu}, which gives then $N^0_{kl}$ right-handed zero modes. Indeed, using the anticommutation property $\{\gamma_\mu,\gamma_\nu\}=2\delta_{\mu\nu}$, the definition $\gamma_{\mu\nu}=\frac{i}{4}[\gamma_\mu,\gamma_\nu]$ and the property $[D_\mu,D_\nu]=-i F_{\mu\nu}$, it follows that
\begin{equation}
\label{eq:zeroFeq}
\begin{aligned}
\gamma_\mu D_\mu\Psi=0\Rightarrow \gamma_\nu\gamma_\mu D_\nu\ D_\mu\Psi=D^2\Psi-\gamma_{\mu\nu} F_{\mu\nu}\Psi=0.
\end{aligned}
\end{equation}
In the instanton background one has the advertised self-duality property
\begin{align}
F_{\mu\nu}=\frac{1}{2}\epsilon_{\mu\nu\rho\sigma}F_{\rho\sigma},
\end{align}
while the matrices $\gamma^{\mu\nu}$ satisfy
\begin{align}
\gamma_{\mu\nu}P_L=-\frac{1}{2}\epsilon_{\mu\nu\rho\sigma}\gamma_{\rho\sigma}P_L,\quad \gamma_{\mu\nu}P_R=\frac{1}{2}\epsilon_{\mu\nu\rho\sigma}\gamma_{\rho\sigma}P_R.
\end{align}
From this it follows that  \eqref{eq:zeroFeq} separates into the following equations for the left and right-handed components:
\begin{align}
D^2\Psi_L=0,\quad D^2\Psi_R-\gamma_{\mu\nu} F_{\mu\nu}\Psi_R=0.
\end{align}
The operator appearing in the equation for $\Psi_L$ is positive definite when acting on normalisable spinors, so that there is no zero mode solution for $\Psi_L$. As will be commented later, the number $n_+-n_-$ can also be related to the selection rule \eqref{eq:chiral_anomaly}.  A typical example is given by a Dirac fermion in the fundamental, with $T(\rm fund)=1/2$, for which there is one right-handed mode, as in the original calculation by 't Hooft \cite{tHooft:1976snw}. Dirac fermions in the adjoint have $T(\rm fund)=2$ and four zero modes \cite{Chadha:1977mh,Jackiw:1977pu}. The previous result means that, for small $M_{kl}$, the eigenvalue of ${\mathfrak M}_{\Psi_{kl}}$ with minimal modulus is ${\cal M}_{kl}P_R=M^*_{kl}$. Therefore, from the determinants in \eqref{eq:detDirac}
we expect a factor of $\prod_{[k,l]}(M^*_{kl})^{T_k+T_l}\prod_{m}(M^*_{mm})^{T_m}$ to the vacuum amplitude, coming from the contributions of the lowest modes. After accounting for the bosonic zero modes and the leading logarithmic contributions to the remaining determinants in the limit $M_{kl}\rho\ll1$, the result is \cite{tHooft:1976snw,Bernard:1979qt}
\begin{equation}
\label{eq:instanton_vacuum}\begin{aligned}
&_{\rm out}\langle n|n+1\rangle_{\rm in}=\, \int d^4x\frac{d\rho}{\rho^5}\,C_M(\rho) C_I(\rho),\\
%%%%
&C_I(\rho)=\,c\left[\frac{8\pi^2}{g^2(\rho)}\right]^{2N}\exp\left[-\frac{8\pi^2}{g^2(\rho)}-i\theta\right],\\
%%%%
&C_M(\rho)=\prod_{[k,l]}(M^*_{kl}\rho)^{T_k+T_l}\prod_{m}(M^*_{mm}\rho)^{T_m}.
\end{aligned}\end{equation}
In the equation above, $c$ is a constant, and the integrals over $x$ and $\rho$ are associated with the translation and dilatation zero modes, respectively. The factor of $c\rho^{-5}(8\pi^2/g^2)^{2N}$ is, up to a power of $\rho$,  the Jacobian of the $4N$ bosonic zero modes.
We note that the presence of the $M^*_{kl}$ with their corresponding power follows the selection rule of equation \eqref{eq:chiral_anomaly}: as discussed in section \ref{sec:anomalies}, \eqref{eq:chiral_anomaly} can still be used in the presence of masses if they are assigned a chiral charge of $-2$ (so that the $M^*_{kl}$ have charge $+2$). Since all the Weyl spinors are assumed to belong to either one of the Dirac or Majorana 4-component spinors, it follows that the instanton induced contribution in \eqref{eq:instanton_vacuum} satisfies indeed \eqref{eq:chiral_anomaly}. Conversely, one could use \eqref{eq:chiral_anomaly} to justify the existence of $T_k+T_l$ fermion zero modes for each Dirac fermion $\Psi_{kl}$, and the fact that these modes must be right-handed.\footnote{The anomaly equation \eqref{eq:chiral_anomaly} requires a positive violation of chiral charge in the $n_{\rm top} = 1$ instanton background. Then the leading contribution to the instanton amplitude for small fermion masses must involve powers of $M^*_{kl}$, with positive chiral charge. The lowest fermion modes must then have eigenvalues set by $M^*_{kl}$ instead of $M_{kl}$, which implies right-handedness (see \eqref{eq:M}). The number of zero modes $T_k+T_l$ then follows from the required amount of charge violation enforced by \eqref{eq:chiral_anomaly} or the related identities for other chiral symmetries.} Finally, the factor of $\exp[-{8\pi^2}/{g^2(\rho)}]$  incorporates the action of the semiclassical instanton solution, $\exp[-{8\pi^2}/{g^2(\mu)}]$,  plus leading logarithmic corrections from the fluctuations beyond the zero modes 
(i.e.~the leading contribution from ${\det}' \,{\mathfrak M}_A^{-1/2}$ and ${\det}' \, {\mathfrak M}_{\rm gh}$ factors in Eq.~\eqref{eq:kkp1}). 
When the dominant mass scale is $1/\rho$ (as in the limit $\rho M_{kl} \ll 1$), the $\mu$ dependence in $g(\mu)$ is cancelled --as must happen for physical observables-- by factors of $\log{\rho\mu}$, with coefficients fixed by the beta function of $g(\mu)$. The corrections then resum into the  coupling $g(\rho)$, as is most clear if one chooses $\mu=\rho$, which cancels all $\log(\rho\mu)$ corrections (for the two-loop, RG-improved version of \eqref{eq:instanton_vacuum}, see \cite{Morris:1984zi,Ringwald:1999ze}). The coupling $g(\rho)$ runs with the inverse of the instanton size with the usual beta function. For an $SU(N)$ theory with fermions and scalars, at one-loop order one has
\begin{align}
\label{eq:bdef}
\rho\frac{\partial}{\partial\rho}\left[\frac{8\pi^2}{g^2(\rho)}\right]=-b,\quad b=\frac{11}{3}N-\frac{2}{3}\sum_m T_m-\frac{1}{3}\sum_s T_s,
\end{align}
where the sums in $m$ and $s$ run over representations of Weyl fermions and complex scalars, respectively.
Since the vacuum-to-vacuum transition $_{\rm out}\langle n|n+1\rangle_{\rm in}$ has to be independent of the unphysical renormalisation scale, the determinant corrections must cancel the logarithmic $\mu$ dependence. 

Beyond the vacuum-to-vacuum amplitude, one may also construct Green functions with fermion fields. In the background of a single instanton --before integrating over the location, size and rigid group rotations-- the Green function is given by the single-instanton vacuum-to-vacuum amplitude (the integrand in \eqref{eq:instanton_vacuum}) times a product of fermion propagators in the instanton background. The latter are defined as the inverse of the fermion kinetic terms. Ignoring the phases of the $M_{kl}$ for simplicity,  the propagator for a Dirac fermion $\Psi_{kl}$ can be written in terms of the orthonormal eigenmodes $\Psi^r_{kl}$ of $\gamma_\mu D_\mu$ in the instanton background, with eigenvalues $\lambda_r/\rho$:
\begin{align}
\label{eq:Diracprop}
(-\gamma_\mu D_\mu+M_{kl})^{-1}=&\,\sum_r \frac{\rho\Psi^r_{kl} \Psi^{r\dagger}_{kl}}{\lambda_r+ \rho M_{kl}},\quad
%%%
\gamma_\mu D_\mu\Psi^r_{kl}= \frac{\lambda_r}{\rho} \Psi^r_{kl}.
\end{align}
In the limit $\rho M_{kl}\ll1$, the sum is dominated by the zero mode contributions. As a consequence of this, Green functions involving pairs of fermions $\Psi\overline\Psi$ in the instanton background involve the instanton density $C_M(\rho)C_I(\rho)$ in equation \eqref{eq:instanton_vacuum}, times insertions of the product of fermion zero modes $\Psi^0 \Psi^{0\dagger}$:
\begin{equation}
\label{eq:GreenF}
\begin{aligned}
&\langle \Psi_{kl}(x)\overline \Psi_{kl}(y)\dots \Psi_{pq}(z)\overline \Psi_{pq}(w)\rangle=\\
%%%%
& \int d^4x \frac{d\rho}{\rho^5}\,d\tilde U C_M(\rho)C_I(\rho)\frac{\sum_i\Psi^{0(i)}_{kl}(x) \Psi^{0(i)\dagger}_{kl}(y)}{M^*_{kl}}\dots \frac{\sum_j\Psi^{0(j)}_{pq}(z) \Psi_{pq}^{0(j)\dagger}(w)}{M^*_{pq}},
\end{aligned}
\end{equation}
where we recovered the appropriate phase of the mass matrices corresponding to the zero modes, and introduced sums over the possible zero modes, labelled by indices $(i)$, $(j)$, etc. When considering Green functions with external Majorana spinors, the propagator is again given by the inverse of the Dirac operator; the Majorana constraint would only influence vertex Feynman rules and the combinatorics of contractions \cite{Denner:1992me}. For $k\neq l$, Green functions with $N^0_{kl}$ insertions of $\Psi_{kl}\overline\Psi_{kl}$ --the same as the number of zero modes in the $\Psi_{kl}$ sector-- the inverse factors of $M^*_{kl}$ in \eqref{eq:GreenF} cancel the factors of $M^*_{kl}$ in the instanton density $C_M(\rho)C_I(\rho)$, giving no net power dependence on the mass $M_{kl}$. Similarly, for a Majorana spinor $\Psi_{mm}$, Green functions with $N^0_{mm}/2$ fermion pairs --half the number of zero modes-- have no dependence on $M^*_{mm}$. For a smaller number of fermion-field insertions, some powers of the masses in $C_M(\rho)C_I(\rho)$ remain uncancelled, but the selection rule \eqref{eq:chiral_anomaly} is always respected. Green functions with more than $N^0_{kl}$ insertions of  $\Psi_{kl}\overline\Psi_{kl}$ (or $N^0_{mm}$/2 insertions of $\Psi_{mm}\overline\Psi_{mm}$) are forbidden by the equivalent of \eqref{eq:chiral_anomaly} for chiral rotations that only affect $\Psi_{kl}$ ($\Psi_{mm}$), together with the requirement of a well-defined limit for massless fermions. 

For example, the selection rule for the "flavoured" chiral rotation of $\Psi_{kl}$ implies violations of the chiral charge by $2N^0_{kl}$ units. Then, effective interactions with more than $N^0_{kl}$ pairs of $\Psi_{kl}\overline\Psi_{kl}$ would require compensating negative powers of $M^*_{kl}$, which would diverge in the massless limit. 
Since such limit is physical and cannot be divergent, the corresponding interactions should can not be generated.

The effective instanton Lagrangian is constructed such that it mimics the correlators \eqref{eq:GreenF}, but in terms of fermions with ordinary propagators. In order to estimate physical observables, it suffices to construct an on-shell effective Lagrangian, which assumes on-shell conditions for the momenta involved in the Fourier transform of the propagators in the instanton background. The interactions in this on-shell effective Lagrangian are obtained by going to momentum space and amputating the Green-functions \eqref{eq:GreenF} with the usual propagators. This requires to evaluate complicated integrals over the rigid rotations $\tilde U$ appearing in the zero modes. However, one can estimate the result as the product of the group averaging over each individual propagator.\footnote{For calculations in which the group averaging is done in full detail, see for example \cite{tHooft:1976snw,Shifman:1979uw,Moch:1996bs}).} It turns out that each propagator average, when evaluated for on-shell momenta, gives an instanton ``form-factor" times a left-handed projector.
\begin{align}\label{eq:zeroamp}
& \sum_{i}\int d\tilde U(\pslash+M_{kl})\Psi^{0(i)}_{kl}(p)\Psi^{0(i)\dagger}_{kl}(q)(\qslash+M) {\big |}_{\rm o.s.}\equiv \rho^{-1}{\cal F}_{kl}P_L.
\end{align}
Note how the $P_L$ factor ensures that the on-shell effective Lagrangian only includes the undotted Weyl spinors $\psi_m$ (see \eqref{eq:Dirac}). 
This is as expected from the violation of chiral charge in the instanton background, as discussed in \ref{sec:anomalies}.
The zero modes and their associated form factors for fermions  in the fundamental and adjoint representations are discussed in appendix \ref{app:zeromodes}. The form factors depend on $\rho$ and the physical masses $|M_{kl}|$. For adjoint fermions there is a subleading dependence on scalar products $p\cdot q$, which arise as higher-order corrections in a $\rho |M_{kl}|$ expansion, but also vanish in the soft limit. In our estimates we will keep the full $\rho |M_{kl}|$ dependence --as appropriate for considering new massive fermions-- but still assume a soft limit. In this way the form-factors are scalar functions of $\rho$ and $|M_{kl}|$. Denoting $u\equiv \rho |M_{kl}|$, we consider four types of form factors:\\

\noindent Massless fermion in the fundamental of $SU(2)$:
\begin{align}
\label{eq:FF0}
{\cal F}^F_0(\rho)=&\,2 \pi ^2\rho^3,
\end{align}
\noindent Massive fermion in the fundamental of $SU(2)$:
\begin{align}
\label{eq:FFF}
{\cal F}^F_M(\rho)=&\,8 \pi ^2 \rho^3 \left(u I_0\left(\frac{u}{2}\right) K_1\left(\frac{u}{2}\right)-I_1\left(\frac{u}{2}\right) \left(u K_0\left(\frac{u}{2}\right)+2 K_1\left(\frac{u}{2}\right)\right)\right){}^2,
%%%
\end{align}
\noindent Massive fermion in the adjoint of $SU(2)$:
\begin{align}
\label{eq:FFA}
{\cal F}^A_M(\rho)=&\,16\pi^2\rho^3[u K_1(u)-2K_0(u)]^2\\
%%%
\nonumber&+\frac{32\pi^2}{M^{6}\rho^3}[-16+u(u(8+u^2)K_0(u)+4(4+u^2)K_1(u))]^2.
\end{align}
In the above equations, $I_i$ are modified Bessel functions of the first kind, and $K_i$ are Bessel functions of the second kind.
The small and large $\rho M$ expansions of the form factors are as follows:
\begin{equation}\label{eq:Fexpansions}\begin{aligned}
{\cal F}^F_M(\rho)=&\,8\pi^2\rho^3+{\cal O}(\rho M)^2,& {\cal F}^F_M(\rho)=&\,\frac{2601\pi^2\rho^3}{32(\rho M)^6}+{\cal O}(\rho M)^{-8},\\
{\cal F}^A_M(\rho)=&\,16 \pi ^2 \rho^2 \left[2 \log \frac{M \rho}{2}+2 \gamma_E +1\right]^2\!\!\!\!+{\cal O}(\rho M)^2,& {\cal F}^A_M(\rho)=&\,24\pi^3\rho^3e^{-2\rho M}(\rho M+{\cal O}(\rho M)^{0}),
\end{aligned}\end{equation}
where $\gamma_E$ is Euler's constant.
The form factor for massless fermions in the fundamental is appropriate for SM fermions. It was used in the original computations in \cite{Shifman:1979uw}, as well as well as in the $B+L$ estimates in \cite{Ringwald:1989ee,Gibbs:1994cw}. To account for electroweak symmetry breaking, \cite{Espinosa:1989qn} used the constrained instanton method \cite{Affleck:1980mp}, ending with results reproducing \eqref{eq:FF0} in the limit of small fermion masses. Our treatment of Higgs effects will be less sophisticated (see section \ref{subsec:bosonic}) but we will account for the full mass-dependence of the heavy fermions through the form factors \eqref{eq:FFF}, and \eqref{eq:FFA}. In particular, it should be noted how the form factors for massive fermions go to zero as $\rho M$ grows.

The result in \eqref{eq:zeroamp}, together with \eqref{eq:GreenF} implies that the on-shell effective Lagrangian for ordinary fermions  involves the integral in $\rho$ of $\rho^{-5}C_M(\rho)C_I(\rho)$ times  factors of ${\cal F}_{kl}$ for every fermion pair. For every fermion pair in the effective Lagrangian, the factor $\rho M^*_{kl}$ inside $C_M(\rho)$ in \eqref{eq:instanton_vacuum} is cancelled by the $1/M^*_{kl}$ in the propagators in \eqref{eq:GreenF}, and the $\rho^{-1}$ in the definition of the form factors in \eqref{eq:zeroamp}. The resulting effective Lagrangians have the following schematic form, in terms of the original Weyl spinors:
\begin{equation}\label{eq:Lschematic}
\begin{aligned}
 {\Delta L}\supset&\,\int\frac{d\rho}{\rho^5}\,C_I(\rho)
\,\times \prod_{[k,l]}\sum_{j=0}^{N^0_{kl}} \left({\cal F}_{kl}\right)^{j}(\psi_k\psi_l)^{j}(\rho M^*_{kl})^{N^0_{kl}-j}\\
%%%%%
&\times\prod_{m}\sum_{i=0}^{1/2N^0_{mm}} \left({\cal F}_{mm}\right)^{i}(\psi_m\psi_m)^{i}(\rho M^*_{mm})^{1/2N^0_{mm}-i},\quad \rho |M_{mm}| \ll 1.
\end{aligned}\end{equation}
In the equation above, $C_I(\rho)$ is given in \eqref{eq:instanton_vacuum}, $N$ corresponds to the gauge group $SU(N)$, and $N^0_{kl}$ is given in \eqref{eq:N0mn}. For $SU(2)$ with fermions in the fundamental and adjoint, the form factors ${\cal F}_{kl}$ have to be chosen from formulae \eqref{eq:FF0} through \eqref{eq:FFA}. The Lagrangian is schematic because of the simplifying assumption on the integration over rigid rotations, which gives rise to a simplified Lorentz structure for the fermion contractions. When going beyond such approximations, different operators constructed in terms of gamma matrices may appear, but all the contributions have coefficients of the order of those in \eqref{eq:Lschematic}. 
Such effective Lagrangians were introduced by 't Hooft \cite{tHooft:1976snw,tHooft:1976snw} and computed explicitly for $SU(2)$ and $SU(3)$ in \cite{Shifman:1979uw}, in theories with up to three flavours of Dirac fermions, and with a proper treatment of the averaging over rigid rotations.

We note that as expected, the Lagrangian satisfies the selection rule \eqref{eq:chiral_anomaly}, as every contribution carries a chiral charge of $N^0_F$ units.
We also remind the reader again that the calculations that led to \eqref{eq:Lschematic} assumed $ \rho |M_{kl}| \ll 1$. Such assumption played a role when approximating fermion propagators by the zero mode contribution, and also in the calculation of the determinants of the nonzero modes. We will next consider the effects of adding a Higgs scalar, and review how the effective anomalous interactions \eqref{eq:Lschematic} can be extended to include bosonic fields.

To finish this section, let us summarise how the Weyl fermions in the SM fit into the above formalism. In the SM, one has the following undotted spinors: there are $SU(2)$ doublets from the quarks, with a multiplicity of 6 (three generations, times three colour indices). There are additionally three lepton doublets from the three generations.
This makes a total of 12 Weyl spinor doublets, which we may denote as $\psi_{k,i}$, $k=1,\dots,12$, where $i=1,2$ is the index of the fundamental representation. One can also define 12 undotted spinors in the antifundamental of $SU(2)$ as $\tilde \psi_{k,i}=\epsilon_{ij}\psi_{k,j}$, where $\epsilon_{ij}$ is the usual $2\times2$ antisymmetric matrix. We choose then a basis of fields given by 6 of the $\psi_k$, (e.g. with $k=1,\dots6$) and 6 of the $\tilde\psi_l$ (e.g. for $l=7,\dots,12$), which can be grouped into 6 Dirac spinors in the fundamental of $SU(2)$, which take the following form:
\begin{align*}
\Psi_{kl}\equiv\left[
\begin{array}{c}
(\psi_{k})_\alpha\\
%%%%
(\tilde \psi_l)^{\dagger\dot{\alpha}}
\end{array}
\right].
\end{align*}
Of course, in such contraction the $SU(3)$ and $U(1)_Y$ symmetries are not explicit, but the ensuing instanton interactions will be gauge invariant. This follows from the fact that the effective Lagrangian for the massless $SU(2)$ fermions involves an interaction of determinant type, involving one copy of each 
Weyl fermion charged under $SU(2)_L$\cite{tHooft:1976snw,tHooft:1976rip}. This  ensures invariance under $SU(3)$. Similarly, since the trace of hypercharge is zero in the SM, invariance under $U(1)_Y$ follows. 
%%%%%%%%%%%%%%%%%%%%%%%%%%%%%%%%%%%%%%%%%%%%%%%%%%%%%%%
%%%%%%%%%%%%%%%%%%%%%%%%%%%%%%%%%%%%%%%%%%%%%%%%%%%%%%%
%%%%%%%%%%%%%%%%%%%%%%%%%%%%%%%%%%%%%%%%%%%%%%%%%%%%%%%

\subsection{\label{subsec:bosonic}Adding bosonic fields.}

The previous instanton calculations apply to a theory with gauge fields and fermions, and rely on a saddle point expansion around the instanton configuration. However, in the presence of spontaneous symmetry breaking by means of a scalar field $H$ acquiring a vacuum expectation value $\langle H^\dagger H\rangle =v^2/2$, there are no instanton saddle points of the classical Euclidean action, except for $\rho\rightarrow0$. This can be understood from the fact that the mass term of the scalar field and the ensuing VEV break  the rescaling symmetry in the bosonic sector, for which the instanton parameter $\rho$ is a collective coordinate. However, one can still understand the usual instantons as saddle points of the action under a constraint that fixes the scale $\rho$; then the path integral can still be approximated by expanding around the constrained instantons and integrating over the constraint \cite{Affleck:1980mp}. In a constrained instanton, the Higgs field acquires a nontrivial profile that solves the Euclidean equation of motion in the instanton background,  with boundary conditions guaranteeing a finite energy. This requires the field to approach the minimum of its potential energy at infinity. For $\rho v\ll1 $, the solution in the singular gauge for an $SU(2)$ scalar doublet as the Higgs can be approximated as \cite{tHooft:1976snw}
\begin{align}
H=\left[\frac{(x-x_0)^2}{(x-x_0)^2+\rho^2}\right]^{1/2} \frac{v}{\sqrt{2}} \,\hat{h},
\end{align}
with $\hat{h}$ a constant doublet satisfying $h^\dagger h$=1.
The classical action of the instanton now picks up an additional contribution going as 
\begin{align}
\Delta S=\pi^2 v^2 \rho^2.
\end{align}
$\Delta S$ acts a cutoff on the $\rho$ integration in the effective Lagrangian \eqref{eq:Lschematic}, so that the only relevant instantons are those with scales $\rho\lesssim 1/v$ --that is, those scales for which the mass of the gauge bosons induced by the symmetry breaking becomes subdominant, so that one expects to recover the usual instanton behaviour. It can also be seen that a further effect of the Higgs is to introduce corrections involving logarithms of $\rho$ which modify the effective coupling $g(\rho)$ in the scalarless theory --appearing within $C_I(\rho)$ in \eqref{eq:Lschematic}, see also \eqref{eq:instanton_vacuum}-- so as to reproduce the running coupling in the higgsed theory \cite{Affleck:1980mp}.

Knowing the semiclassical bosonic configurations, one may also compute Green functions involving bosonic fields in the instanton background. By amputating these Green functions with ordinary propagators, one can identify the corresponding interactions in the effective Lagrangian which involve ordinary scalars and gauge bosons, aside from fermions. This can be done with the aid of the following identities, valid once one includes corrections in the constrained instanton formalism\cite{Ringwald:1989ee}:
\begin{equation}\label{eq:Fourierboson}\begin{aligned}
A^a_\mu(p)=&\,\frac{4\pi^2\rho^2 i}{g}\frac{\eta_{a\mu\nu}p_\nu}{p^2+m^2_W}+O(\rho p),\\
%%%%%%
H(p)=&\,-\frac{\sqrt{2}\pi^2\rho^2v}{p^2+m^2_h}+O(\rho p).
\end{aligned}\end{equation}
Since in the presence of the Higgs field the $\rho$ integral is cutoff at $\rho\sim 1/v$,  the relevant values of $\rho$ satisfy $\rho m_W<1$, $\rho m_h<1$. Then when evaluating the Fourier transforms in \eqref{eq:Fourierboson}
for on-shell momenta, the small $\rho m$ expansion is justified. This is in contrast to the case of heavy BSM fermions, for which we will use the form factors in equation \eqref{eq:FF0} through \eqref{eq:FFA} with the full $\rho M$ dependence.
With this we can now expand the effective Lagrangian of equation \eqref{eq:Lschematic} to account for bosonic interactions involving $n_h$ and $n_W$ Higgses and $W$ bosons (still valid for $\rho M_{kl}\ll1$):
\begin{equation}
\label{eq:Lschematic2}\begin{aligned}
 {\Delta L}\supset&\sum_{n_W,n_h}\int\frac{d\rho}{\rho^5}\,\tilde C_I(\rho)\,(-\sqrt{2}\pi^2 \rho^2 v h)^{n_h}\left(-\frac{4\pi^2\rho^2}{g}\eta_{a\mu\nu}\partial_\nu W^{a}_{\mu}\right)^{n_W}\\
 %%%%%
 &\times\prod_{[k,l]}\left\{\sum_{j=0}^{N^0_{kl}} \left({\cal F}_{kl}\right)^{j}(\psi_k\psi_l)^{j}(\rho M^*_{kl})^{N^0_{kl}-j}\right\}\\
%%%%%
&\times\prod_{m}\left\{\sum_{i=0}^{1/2N^0_{mm}} \left({\cal F}_{mm}\right)^{i}(\psi_m\psi_m)^{i}(\rho M^*_{mm})^{1/2N^0_{mm}-i}\right\},\quad  \rho |M_{kl}| \ll 1,
%%%%%%
\end{aligned}\end{equation}
where, accounting for the cutoff effect from the Higgs,
\begin{align}
\label{eq:CIfinal}
\tilde C_I(\rho)=&\,c\left[\frac{8\pi^2}{g^2(\rho)}\right]^{2N}\exp\left[-\frac{8\pi^2}{g^2(\rho)}-i\theta-\pi^2 v^2 \rho^2\right],
\end{align}
with $g(\rho)$ the running coupling as in equation \eqref{eq:bdef}.

%%%%%%%%%%%%%%%%%%%%%%%%%%%%%%%%%%%%%%%%%%%%%%%%%%%%%%%
%%%%%%%%%%%%%%%%%%%%%%%%%%%%%%%%%%%%%%%%%%%%%%%%%%%%%%%
%%%%%%%%%%%%%%%%%%%%%%%%%%%%%%%%%%%%%%%%%%%%%%%%%%%%%%%
\subsection{Effective Lagrangians accounting for decoupling effects\label{subsec:decoupling}}

As has been emphasised, the effective Lagrangians of equations \eqref{eq:Lschematic}
and \eqref{eq:Lschematic2} are only valid for $\rho M_{kl}\ll1$. Although the $\rho$ integral is cutoff by the Higgs profile, so that only $\rho\lesssim 1/v$ is relevant, 
the assumption for the fermion masses could be violated by new fermions beyond the Standard Model, which could be very heavy. 

We wish to obtain modified formulae that are also valid in the limit $\rho M_{kl}\gtrsim1$, so that we can perform a more reliable $\rho$ integration. For $\rho M_{kl}\ll1$, the powers of $\rho M_{kl}$ in the formulae for the effective Lagrangian came from the contributions of the zero mode, either in the fermion determinant or the fermion propagator. However, for a Dirac fermion $\Psi_{kl}$ with a large mass, we expect all the  eigenvalues of $-i\gamma_\mu D_\mu+{\cal M}_{kl}$ to be of the order of the mass, so that we expect a different power of $M_{kl}$ in the determinant than the one that follows simply from the zero modes. Note that, although we argued that the powers of $M_{kl}^*$ in the effective Lagrangian are exactly those needed to satisfy the selection rule \eqref{eq:chiral_anomaly}, one can still have corrections involving $|M|=(M^*M)^{1/2}$, which carry no chiral charge. 

The modified dependence on the masses of the fermion determinant can be estimated by imposing decoupling. The decoupling theorem \cite{Appelquist:1974tg} ensures that, if a particle can be made heavy while keeping its couplings fixed, then its physical effects become irrelevant, and the behaviour of the theory can be captured with an effective theory in which the heavy particle is absent. For a pair of Weyl fermions in mutually conjugate representations of the group, $\psi_p$, $\psi_q$, and coupled through a large mass $M_{pq}$ --or equivalently for a new massive Dirac fermion-- decoupling must happen for large $M_{pq}$. This is not the case of chiral fermions, e.g., those acquiring a mass term through a Yukawa coupling with a singlet fermion, as in the SM: in this case, if all other masses are kept fixed, a large fermion mass can only be achieved by increasing the Yukawa coupling, which prevents decoupling. As we have seen, the effective Lagrangian includes interactions with a varying number of Weyl fermions. The decoupling of a pair of Weyl fermions means that the effective interactions in the UV theory that do not involve the heavy fermions should be reproduced by the IR theory without such fermions, up to subleading corrections. In terms of the effective Lagrangian in equation  \eqref{eq:Lschematic2}, this would imply the following large $M_{pq}$ behaviour,
\begin{align}
\label{eq:check}
\tilde C^{\rm UV}_I(\rho)(\rho M_{pq}^*)^{N^0_{pq}}=\tilde C^{\rm IR}_I(\rho)+O\left(\frac{1}{\rho |M_{pq}|}\right),
\end{align}
where $\tilde C^{\rm UV}_I(\rho)$ and $\tilde C^{\rm IR}_I(\rho)$ correspond to the instanton densities (see \eqref{eq:CIfinal}) in the  theories with and without the pair of Weyl fermions $\psi_p,\psi_q$.

Using formula \eqref{eq:CIfinal}, we can check whether \eqref{eq:check} is satisfied. The difference in the beta function coefficient $b$ that determines the running coupling $g(\rho)$ in the UV and IR theories is
\begin{align}
\label{eq:deltab}
 b_{UV}-b_{IR}=-\frac{2}{3}(T_p+T_q)=-\frac{2}{3}N^0_{pq}.
\end{align}
We expect both running couplings to match at the scale of the mass of the heavy fermions (up to subleading threshold effects), which gives
\begin{align}
\label{eq:expmatching}
 \exp\left[-\frac{8\pi^2}{g^2_{UV}(\rho)}\right]=\exp\left[-\frac{8\pi^2}{g^2_{IR}(\rho)}\right](|M_{pq}|\rho)^{-\frac{2}{3}N^0_{pq}}.
\end{align}
The matching of the $\theta$ angles in the UV and IR theories is a bit more subtle, when the masses have nontrivial phases. Under an infinitesimal chiral transformation  that only affects the fermions $\psi_p$ and $\psi_q$, and with an associated parameter $\alpha$, the fermion mass $M_{pq}$ changes as
\begin{align}
\label{eq:deltapq}
\delta_\alpha {\rm Arg}\, M_{pq}= 2\alpha.
\end{align}
On the other hand, the $\theta$ parameter is also modified as a consequence of the anomalous conservation of the chiral current. Under the chiral transformation the effective action $\Gamma$ in Minkowski space changes as
\begin{align}
 \delta_\alpha \Gamma=-\alpha \int d^4x\, \partial_\mu J^\mu= -2N^0_{pq}\,\alpha\, n_{\rm top},
\end{align}
where we used equation \eqref{eq:anomaly} applied to the transformations at hand. Since the $\theta$ interaction is proportional to the topological charge, the above result implies that the chiral transformation induces an anomalous shift in $\theta$:
\begin{align}
\label{eq:deltathetapq}
\delta_\alpha\theta=-2N^0_{pq}\alpha.
\end{align}
As is clear from equations \eqref{eq:deltapq}, \eqref{eq:deltathetapq}, $\theta$ and $M_{pq}$ have correlated transformations, such that $\theta+N^0_{pq}{\rm Arg}\,M_{pq}$ remains invariant.
In the IR theory, there are no fermions $\psi_p,\psi_q$, and so the IR couplings must be invariants under the chiral transformations of the pair of Weyl fermions $\psi_p,\psi_q$. This means that the matching of $\theta$ goes as
\begin{align}
\label{eq:thetamatch}
\theta_{IR}=\theta_{UV}+N^0_{pq}{\rm Arg} \,M_{pq}=\theta_{UV}-iN^0_{pq}\log\frac{M_{pq}}{|M_{pq}|}.
\end{align}
The difference in the remaining factor 
$(8\pi^2/g^2(\rho))^{2N}$ in the formula \eqref{eq:CIfinal}
is only up to powers of
$\log(\rho|M|)$,
which are expected to be explained by 
loop corrections.
Ignoring this difference and substituting 
Eq.~\eqref{eq:expmatching} and Eq.~\eqref{eq:thetamatch}
into Eq.~\eqref{eq:CIfinal},
we find
\begin{align}
\label{eq:Cmismatch}
\tilde C^{\rm UV}_I(\rho)(\rho M_{pq}^*)^{N^0_{pq}}=\tilde C^{\rm IR}_I(\rho) (|M_{pq}|\rho)^{N^0_{pq}/3}.
\end{align}
This violates the requirement \eqref{eq:check} of decoupling, which should not be surprising:  the formula we used for $C_I(\rho)$ were obtained in the limit $|M_{pq}\rho|\ll1$, while  the condition \eqref{eq:check} applies in the large $\rho |M_{pq}|$ limit.\footnote{Note that the argument made near \eqref{eq:bdef}, justifying that quantum corrections from the determinants had the effect of substituting $g(\mu)$ by $g(\rho)$, assumed that $\rho$ was the dominant scale, so that all leading logarithms were of the form $\log\mu\rho$. For large fermion masses however $\rho |M|$ becomes large and one gets an additional dependence on the fermion masses, which we recover by imposing decoupling.
} Still, equation \eqref{eq:Cmismatch} offers a way out to implement decoupling: in the large $M_{pq}$ limit $M_{pq}\rho\gtrsim1$ the insertions of $\rho M_{pq}^*$ in  \eqref{eq:Lschematic2} should be altered with an additional factor $(|M_{pq}|\rho)^{-N^0_{pq}/3}$. Then the decoupling requirement \eqref{eq:check} is traded for
\begin{align}
\label{eq:check2}
\tilde C^{\rm UV}_I(\rho)\,(\rho M_{pq}^*)^{N^0_{pq}} \,(\rho |M_{pq}|)^{-N^0_{pq}/3}=\tilde C^{\rm IR}_I(\rho)+O\left(\frac{1}{\rho |M_{pq}|}\right),
\end{align}
which is indeed satisfied, as follows from equation \eqref{eq:Cmismatch}. We remind the reader that the insertions of $\rho M^*_{pq}$ corresponded to the determinant of the Dirac operator restricted to the $\Psi_{pq}$ Dirac fermion, and the extra power of $(|M_{pq}|\rho)^{-N^0_{pq}/3}$ is meant to capture contributions from eigenvalues beyond the lowest mode.

Aside from interactions that do not involve $\Psi_{pq}$, as we just considered, 
the $n_{\rm top} = 1$ contribution to the effective Lagrangian in the UV theory also includes terms of the form $\overline\Psi_{pq}P_L\Psi_{pq}$, which are proportional to the the fermion determinant times the fermion propagator.\footnote{ The term $\overline\Psi_{pq}P_R\Psi_{pq}$
is generated by the $n_{\rm top} = -1$ instanton,
which is necessary to make the effective Lagrangian
Hermitian (i.e.~$\overline\Psi_{pq} (P_L + P_R) \Psi_{pq} = \psi_p^\dagger \psi_q^\dagger +
\psi_q \psi_p$). 
}
These terms will also be modified for large $M_{pq}$. As seen before, the determinant piece will pick up a factor of $(|M_{pq}|\rho)^{-N^0_{pq}/3}$. On the other hand, the propagator is of the form \eqref{eq:Diracprop}; if $\rho| M_{pq}|\gtrsim1$, we expect all terms to contribute similarly, being of the same order as the contribution from the lowest mode. Thus, as we have a sum of terms --as opposed to a product in the determinant-- we don't expect a modification of the power of $\rho| M_{pq}|$ coming from the propagator. The normalisation of the product of determinant and propagator in the large $\rho| M_{pq}|$ regime is fixed by requiring that the modified effective Lagrangian matches the result of equation \eqref{eq:Lschematic2}, valid in the small $M_{pq}$ limit,  at $\rho=|M_{pq}|$. This is already achieved by the insertion of $(|M_{pq}|\rho)^{-N^0_{pq}/3}$  in the fermion determinant. Note that the full mass-dependent form-factors in equations \eqref{eq:FFF} and \eqref{eq:FFA} do implement as well some form of decoupling, as is clear from the large $\rho M$ expansions in equation \eqref{eq:Fexpansions}: for $\rho M\gg1$, the form factors go to zero, meaning that instantons of sizes much larger than the the inverse mass of the heavy fermions do not contribute to the interactions of the latter.

The previous results can also be extended to the integration of a heavy Majorana spinor $\Psi_{qq}$; all goes as before, but $N^0_{pq}$ should be substituted by $N^0_{qq}/2$. Thus we conclude that the effective Lagrangian \eqref{eq:Lschematic2} generalised to large $\rho| M_{pq}|$ is of the form
\begin{equation}
\label{eq:Lschematic3}\begin{aligned}
{\Delta L}\supset&\sum_{n_W,n_h} \int\frac{d\rho}{\rho^5}\,\tilde C_I(\rho)\,(-\sqrt{2}\pi^2 \rho^2 v h)^{n_h}\left(-\frac{4\pi^2\rho^2}{g}\eta_{a\mu\nu}\partial_\nu W^{a}_{\mu}\right)^{n_W}\\
 %%%%%
 &\times\prod_{[k,l]}\left\{(\rho|M_{kl}|)^{N^0_{kl}b_{kl}}\sum_{j=0}^{N^0_{kl}} \left({\cal F}_{kl}\right)^{j}(\psi_k\psi_l)^{j}(\rho M^*_{kl})^{N^0_{kl}-j}\right\}\\
%%%%%
&\times\prod_{m}\left\{(\rho|M_{mm}|)^{1/2N^0_{mm}b_{mm}}\sum_{i=0}^{1/2N^0_{mm}} \left({\cal F}_{mm}\right)^{i}(\psi_m\psi_m)^{i}(\rho M^*_{mm})^{1/2N^0_{mm}-i}\right\},\\
%%%%%
%%%%%%
b_{m n}=&\,\left\{\begin{array}{cc}
           0, &\rho |M_{m n}|<1,\\
           %%%%
           -1/3, &\rho |M_{m n}|\gtrsim1. %\quad\text{$\psi_m,\psi_n\in$ vector fermion.}
          \end{array}\right.
\end{aligned}\end{equation}
When $\rho$ crosses a fermion mass threshold, the behaviour of the interaction changes, but there is continuity at the threshold. As anticipated before, decoupling can be recovered by insertions of powers of $\rho |M|$ in the effective Lagrangian, maintaining compatibility with the selection rule in \eqref{eq:chiral_anomaly}. The fact that this works out is not trivial, as it requires to account for the nontrivial matching between $\theta$ angles in \eqref{eq:thetamatch}. \footnote{If the matching of $\theta$ were to be ignored, one would obtain that the gauge coupling in the low energy theory involves the phases of the heavy masses, which would violate the selection rule \eqref{eq:chiral_anomaly}.}
%%%%%%%%%%%%%%%%%%%%%%%%%%%%%%%%%%%%%%%%%%%%%%%%%%%%%%%
%%%%%%%%%%%%%%%%%%%%%%%%%%%%%%%%%%%%%%%%%%%%%%%%%%%%%%%
%%%%%%%%%%%%%%%%%%%%%%%%%%%%%%%%%%%%%%%%%%%%%%%%%%%%%%%
%%%%%%%%%%%%%%%%%%%%%%%%%%%%%%%%%%%%%%%%%%%%%%%%%%%%%%%
%%%%%%%%%%%%%%%%%%%%%%%%%%%%%%%%%%%%%%%%%%%%%%%%%%%%%%%
\section{Enhancement of the polynomial contributions to $B+L$ violating rates in BSM theories}
\label{sec:heavy_fermions}
%%%%%%%%%%%%%%%%%%%%%%%%%%%%%%%%%%%%%%%%%%%%%%%%%%%%%%%
%%%%%%%%%%%%%%%%%%%%%%%%%%%%%%%%%%%%%%%%%%%%%%%%%%%%%%%
%%%%%%%%%%%%%%%%%%%%%%%%%%%%%%%%%%%%%%%%%%%%%%%%%%%%%%%
%%%%%%%%%%%%%%%%%%%%%%%%%%%%%%%%%%%%%%%%%%%%%%%%%%%%%%%
%%%%%%%%%%%%%%%%%%%%%%%%%%%%%%%%%%%%%%%%%%%%%%%%%%%%%%%
With the effective Lagrangian in \eqref{eq:Lschematic3}  we are now ready to study how the 
rates of $B+L$ violating processes are affected by the presence of heavy fermions. Equation \eqref{eq:Lschematic3} ignores the details of spinor algebra (possible contractions, insertions of Pauli matrices, etc) and 
performed only approximately the integrations over rigid rotations, 
but it should suffice for order-of-magnitude estimates. Furthermore, we will opt for estimating ratios, which should be less affected by theoretical uncertainties. 

We will assume a two-quark initial state with a fixed centre-of-mass energy, and consider cross-sections for the multi-particle final states that follow from the effective Lagrangian \eqref{eq:Lschematic3}, which accounts for the effects of anomalies. As mentioned in the introduction, BSM scenarios with new electroweak, nonchiral fermions predict SM-like anomalous interactions involving 12 SM fermions, as in \eqref{eq:Lanom} --which give rise to processes $qq\rightarrow 7q + 3l$\,--\footnote{
 In this notation and what follows, 
we do not distinguish fermions and anti-fermions.
} plus additional interactions involving not only the SM fermions, but the exotic ones. For concreteness, we will focus on the following BSM scenarios:
\begin{itemize}
\item Scenario $F$: M plus a Dirac fermion in the fundamental representation of $SU(2)$, or equivalently, two Weyl spinors in the (anti) fundamental representations, $\psi_F$, $\tilde\psi_F$. 
In this case, one has interactions of the form of \eqref{eq:Lanomf} --which imply processes with 12 final states $qq\rightarrow 7q+3l+\psi_F\tilde\psi_F$.
\item Scenario $A$: SM supplemented with a Weyl spinor in the adjoint of $SU(2)$, $\psi_A$. 
In this case, the allowed interactions are of the form in \eqref{eq:LanomA} --giving a 14 final state process $qq\rightarrow 7q+3l+4 \psi_A$\,-- and of the form of \eqref{eq:LanomAp}, which gives processes with 12 final states, $qq\rightarrow 7q+3l+2\psi_A$.
\item SUSY scenarios. 
In these models, we have processes with $16,14$ and $12$ final fermionic states, of the form $qq\rightarrow7q+3l+2\Psi_H+4\lambda$, $qq\rightarrow7q+3l+4\lambda$,  $qq\rightarrow7q+3l+2\Psi_H$.
Given the vast number of parameters in SUSY models, we concentrate here on two simplified scenarios:
\begin{itemize}
\item Scenario $S$, inspired by Split-SUSY, in which all SUSY particles except for Higgsinos and gauginos are decoupled.
\item Scenario $MSSM$, a simplified degenerate SUSY setting in which all BSM particles are assumed to be approximately degenerate.
\end{itemize}
\end{itemize}
As was discussed in the introduction, the $B+L$-violating cross section in the SM is known to have the form of equation \eqref{eq:sigmaF}, involving a polynomial part fixed by a function $f(\hat{s})$, and an exponential part featuring the holy grail function $F(\hat{s})$. Fermionic interactions only affect the polynomial part, and thus by studying ratios of rates for processes that only involve the fermions listed above, we may estimate how $f(\hat{s})$ is affected by the presence of BSM fermions. Regarding the effect of gauge boson emission, which is encoded by the holy grail function, we remind the reader that, as mentioned in the introduction, leading-order instanton results are known to only capture the first energy-dependent term of the  expansion of $F[s]$ in $(\sqrt{\hat{s}}/E_0)$ (see \eqref{eq:holymoly}), which does not converge for energies above the sphaleron barrier. However, we will still compute the leading order instanton corrections for gauge-boson emission, in order to see how they are affected by the nonzero masses of the heavy fermions.

We are interested in ratios of cross-sections, which in a collider setting will translate into ratios of event rates. To fix the notation, we will denote event rates producing $n_F$ fermions, $n_W$ gauge bosons, and $n_h$ Higgses from a $qq$ initial state by $\Gamma^{n_F,n_W,n_H}_{SM/F/A/S/MSSM}$, the suffix depending on whether the rate is calculated in the SM or one of its extensions.  We will consider three types of ratios: 
\begin{itemize}
\item Ratios of SM-like rates with no boson emission: 
\begin{align}
\frac{\Gamma_{BSM}^{10,0,0}}{\Gamma_{SM}^{10,0,0}},
\end{align}
with ``$BSM$" denoting the models $F$,$A$,$S$, or $MSSM$.
We will confirm that the ratio quickly tends to one when the exotic fermions become heavy, as a consequence of decoupling.
\item Ratios of BSM rates with BSM fermions over SM-like BSM rates:
\begin{align}
\frac{\Gamma_{BSM}^{10+\delta_1(+\delta_2),0,0}}{\Gamma_{BSM}^{10,0,0}},
\end{align}
where $\delta_1$ counts the number of BSM fermions in the fundamental of $SU(2)$, and $\delta_2$ counts BSM fermions in the adjoint, if applicable.
These ratios allow us to determine whether $B+L$ violating rates will be dominated by processes involving exotic fermions, or by SM-like processes. Also, since $\Gamma_{BSM}^{ 10,0,0}\rightarrow \Gamma_{SM}^{10,0,0}$, the ratio will determine whether one can have faster $B+L$ violating rates in theories beyond the Standard Model.
\item Ratios of BSM rates with and without boson emission:
\begin{align}
\frac{\Gamma_{BSM}^{10+\delta_1(+\delta_2),n_W,n_h}}{\Gamma_{BSM}^{10+\delta_1(+\delta_2),0,0}}.
\end{align}
This allows us to infer whether the dominant $B+L$ violating processes are expected to involve the emission of gauge bosons, or not. This effect was studied in the SM in \cite{Ringwald:1989ee,Espinosa:1989qn}.
\item Ratio of BSM rates with boson emission and SM-like vertex with boson emission: 
\begin{align}
\frac{\Gamma_{BSM}^{10+\delta_1(+\delta_2),n_W,n_h}}{\Gamma_{BSM}^{10,n_W,n_h}}.
\end{align}

\end{itemize}

When computing the cross-sections, averaging over spins and polarizations, one ends up with lengthy traces over the gamma matrices. However, since our effective Lagrangian ignored details on the possible operators acting on the spinors, we will just estimate the traces as yielding a product of the energies of the spinors, as was done in  \cite{Ringwald:1989ee}  (see also \cite{Gibbs:1994cw}). This can be justified from dimensional reasons, as the modulus of the amplitude squared involves the product of two on-shell spinors per initial of final state, with each spinor having mass-dimension $1/2$.\footnote{This is clear from the completeness relations of Dirac spinors, $\sum_s u^s(p)\overline u^s(p)=\slashed{p}+m$.} Regarding the polarization sums over gauge bosons, the modulus of the amplitude square involves contractions of the form $\sum_{{\rm pol}}\eta_{a\mu\nu}\eta_{a\rho\sigma}\epsilon_\mu  k_\nu \epsilon^*_\rho  k^*_\sigma$. Note that in Euclidean space, the 4-momenta and polarization vectors are complex; with the Euclidean conventions in appendix \ref{app:Euclidean}, this yields 
\begin{align}
\label{eq:fW}
\sum_{{\rm pol}}\eta_{a\mu\nu}\eta_{a\rho\sigma}\epsilon_\mu k_\nu \epsilon^*_\rho  k^*_\sigma\equiv m^2_W f_W(k)=4E^2_W-m^2_W.
\end{align}

We note that from the point of view of our effective Lagrangian in \eqref{eq:Lschematic3} applied to $SU(2)$, the SM fermions are massless, as the mass parameters $M_{kl}$ are meant to couple Weyl fermions charged under the gauge group. However, the SM doublets only get masses by coupling to $SU(2)$ singlets. In our chosen scenarios the only relevant mass parameters are then those of the heavy BSM fermions.

For the first two scenarios, involving a Dirac fermion in the fundamental, or an Weyl spinor in the adjoint, there is a single  mass parameter which plays a role in the mass insertions in \eqref{eq:Lschematic3}, but also in the evolution of the gauge coupling $g_2(\rho)$. In the MSSM, there are more dimensionful parameters involved, corresponding to the $\mu$ and $M_2$ masses of the fundamental and adjoint spinors, as well as other thresholds that may affect the evolution of the running gauge coupling. In view of this, we will consider two simplifying scenarios. First, a degenerate MSSM scenario in which all SUSY mass parameters aside from the mass defining the lightest Higgs are of the same order $M$, which we will take as real. Here we have to implement the decoupling of the heavy scalars and fermions that are not charged under $SU(2)$, since our decoupling discussion in \ref{subsec:decoupling} only applied to fermions with weak interactions. Since the additional fields only enter the effective Lagrangian through their virtual effects in the running coupling $g(\rho)$, it suffices to consider the running generated by fields with masses $M$ such that $\rho M<1$. A second supersymmetric scenario to consider is a Split SUSY-like scenario, in which all BSM fields except those charged under $SU(2)$ (i.e., except Higgsinos and weak gauginos) are assumed to be decoupled. In this case decoupling is accounted for as in \ref{subsec:decoupling}.

Taking into account the above, the ratios of cross sections/rates can be captured for all scenarios with the following set of formulae:
\begin{equation}\label{eq:ratios}\begin{aligned}
\frac{\Gamma_{BSM/SM}^{10+\delta_1+\delta_2,n_W,n_h}}{\Gamma_{BSM/SM}^{10+\delta'_1+\delta'_2,n'_W,n'_h}}=&\,\frac{{\cal N}[10+\delta_1+\delta_2,n_W,n_h]PS[10+\delta_1+\delta_2,n_W,n_h]}{{\cal N}[10+\delta'_1+\delta'_2,n'_W,n'_h]PS[10+\delta'_1+\delta'_2,n'_W,n'_h]},\\
%%%%%%%
{\cal N}[10+\delta_1+\delta_2,n_W,n_h]=&\,2^{n_h+2n_W+12}\pi^{4(n_h+n_W)+24}v^{2(n_h+n_W)}\times\\
%%%
&\hskip-3.85cm\times\left[\int\frac{d\rho}{\rho^5}\,\overline C_{\rm SM}(\rho)
\rho^{18+2(n_h+n_W)}(\rho \,M)^{\Delta(1+3b[\rho])+\overline{N}(1/3+b[\rho])-\delta/2}\, ({\cal F}^F_M)^{\delta_1/2} ({\cal F}^A_M)^{\delta_2/2}\right]^2,\\
%%%%%%%%%%
&\overline C_{\rm SM}(\rho)=\left(\frac{8\pi^2}{g_{2,SM}^2(\rho)}\right)^4\exp\left[-\frac{8\pi^2}{g_{2,SM}^2(\rho)}-\pi^2v^2\rho^2\right],\\
%%%%
&b[\rho]=\,\left\{\begin{array}{cc}
           0, &\rho\, M<1,\\
           %%%%
           -1/3, &\rho M\gtrsim1,
           \end{array}\right.\\
%%%%%%
 PS[10+\delta,n_W,n_h]=&\,\int \left(\prod_f^{10+\delta} \frac{d^3p_f}{2(2\pi)^3}\right)
 \left(\prod_h^{n_h} \frac{d^3p_h}{2(2\pi)^3E_h}\right)\left(\prod_W^{n_W}
   \frac{d^3p_W}{2(2\pi)^3E_W}f_W({\bf p}_W)\right).
\end{aligned}\end{equation}
In the above equations, $\delta\equiv\delta_1+\delta_2$, $f_W(\bf p)$ is defined in equation \eqref{eq:fW}, and $g_{2,SM}(\rho)$ is the $SU(2)$ running coupling in the SM, with a one-loop beta function coefficient given by $b_{2,SM}=19/6$. The form factors ${\cal F}^F_M$ and ${\cal F}^A_M$ are given in equations \eqref{eq:FFF} and \eqref{eq:FFA}, respectively. We expressed the instanton density $C_I(\rho)$ in terms of the SM gauge coupling, choosing the parameterisation
\begin{align}
\label{eq:Deltab}
b_2=b_{2,SM}+\Delta-\frac{2}{3}\overline{N},
\end{align}
with 
\begin{align}
\overline{N}=\sum_{{\rm heavy}\,r_a}T_a.
\end{align}
When the heavy fermions are the only BSM particles, then $\Delta=0$, since the change of $b_2$ is just given by $-2/3 \overline{N}$, as in equation \eqref{eq:deltab} (see also \eqref{eq:bdef}). In the degenerate MSSM case, however, the additional scalar particles also modify the beta function, and $\Delta$ accounts for this effect. We have in summary that for our four  scenarios, 
\begin{equation}\label{eq:Deltas}\begin{aligned}
&\overline{N}_{\rm F}=1,&
&\overline{N}_{\rm A}=2,& 
&\overline{N}_{{\rm S}}=3,& 
&\overline{N}_{{\rm MSSM}}=3, \\
%%%%
&\Delta_{F}=0, &
&\Delta_{A}=0,& 
&\Delta_{S}=0,& 
&\Delta_{\rm MSSM}=-\frac{13}{6}.
\end{aligned}\end{equation}
Note that, under the assumption of a unique mass threshold, our estimates for the rates in the MSSM only depend on the number of BSM particles, no matter whether they are in the adjoint or the fundamental. Also, as $\delta$ appears with a positive coefficient in the exponents of the prefactor of ${\cal N}[10+\delta, n_W, n_h]$, we can expect enhancements of the rate for growing $\delta$, if the power-like enhancement is not compensated by either the phase space suppression inherent to the additional final states, or possible suppressions in the $\rho$ integral. Due to the $1/\rho^5$ factor, the $\rho$ integral is dominated by the small $\rho$ contributions, and for growing $\delta$ the powers of $\rho^\delta$ suppress the integrand in this region; the effect is however subleading with respect to the power-like enhancement of the prefactors. Similarly, as already known from the work in references \cite{Ringwald:1989ee,Espinosa:1989qn}, the constant prefactors in ${\cal N}[10+\delta,n_W, n_h]$ also grow as powers of $n_W,n_h$, with the added feature that anomalies impose no restriction on $n_W,n_h$, in contrast to the fermion case. The enhancement from bosonic emission will be dominated by $n_W$, not only due to the power of $4^{n_W}$ --as opposed to $2^{n_h}$ for Higgs emission, but also because the dimensionless factor $f_W$ accompanying the phase space integral of gauge bosons in \eqref{eq:ratios} can be large when they are created with a sizable boost. The rates will grow with $n_W$ until 
the phase-space suppression finally
thwarts the enhancement. The dominance of the corrections from gauge boson emission over those arising from Higgs bosons is known from the SM case: while both corrections exponentiate, giving rise to the holy-grail function contribution to the cross-section $f(\hat{s})$ in equations \eqref{eq:sigmaF} and \eqref{eq:holymoly}, Higgs bosons contribute to the Holy Grail function at second order and beyond in the expansion of equation \eqref{eq:holymoly} \cite{Khoze:1990bm}, and with a contribution that remains subleading with respect to that of gauge bosons \cite{Khlebnikov:1990ue,Arnold:1991dv,Mueller:1991fa}. Regarding the effect of the heavy fermions in gauge boson emission, we expect that in the presence of BSM fermions, the maximum rate will happen for lower values of $n_W$, since the additional fermions decrease the energy available for producing bosons.

In order to estimate the phase space integrals\footnote{
Note that $PS$ is not exactly the phase space
because it contains extra energy dependences 
coming from the wave function factors, e.g.~$f_W({\bf p}_W)$.
} $PS[10+\delta,n_W,n_h]$, we use {\tt RAMBO}
\cite{Platzer:2013esa}, which populates events with a probability that follows the measure $\prod_i d^3p_i/(2\pi)^3/(2E(p_i))$. The integrals are then evaluated by reweighting each event with the additional factors 
in $PS[10+\delta,n_W,n_h]$, that is a factor of $E_f$ for each fermion, and a factor of $f_W(p_W)$ for each gauge boson. (For other tools specifically designed for instanton-induced processes, see HERBVI \cite{Gibbs:1995bt} --for SU(2) instantons and $B+L$ violating processes-- and  QCDINS \cite{Ringwald:1999jb}), which accounts for QCD-instanton effects in deep inelastic scattering).
The results of our numerical calculations are described in the following subsections.

\subsection{Fermionic final states}
\paragraph{Scenario F:}
\begin{figure}[t] 
  \begin{subfigure}[b]{0.5\linewidth}
    \centering
    \includegraphics[width=0.95\linewidth]{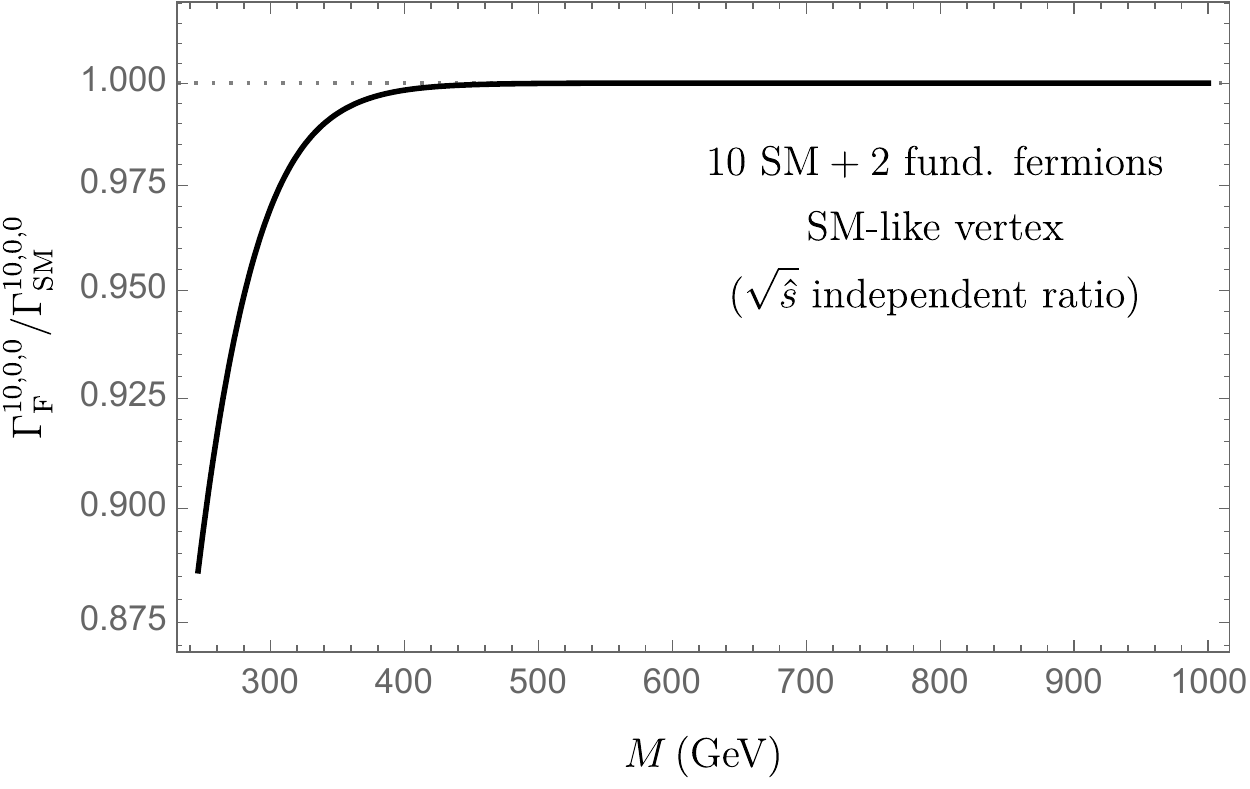} 
    \caption{SM-like/SM vertex} 
    \label{fig:heavy1} 
    \vspace{4ex}
  \end{subfigure}%% 
  \begin{subfigure}[b]{0.5\linewidth}
    \centering
    \includegraphics[width=0.95\linewidth]{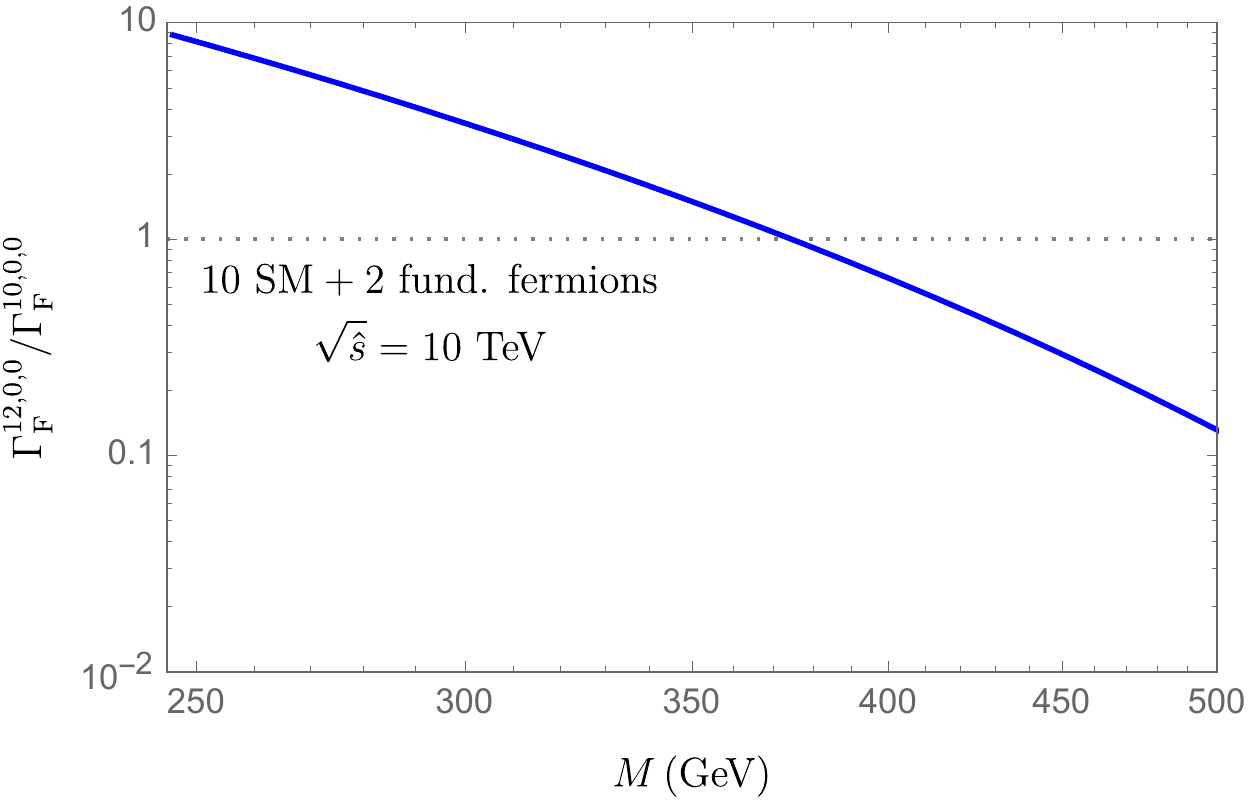} 
    \caption{$q q$ collision energy $\sqrt{\hat{s}}=10$~TeV} 
    \label{fig:heavy2} 
    \vspace{4ex}
  \end{subfigure} 
  \begin{subfigure}[b]{0.5\linewidth}
    \centering
    \includegraphics[width=0.95\linewidth]{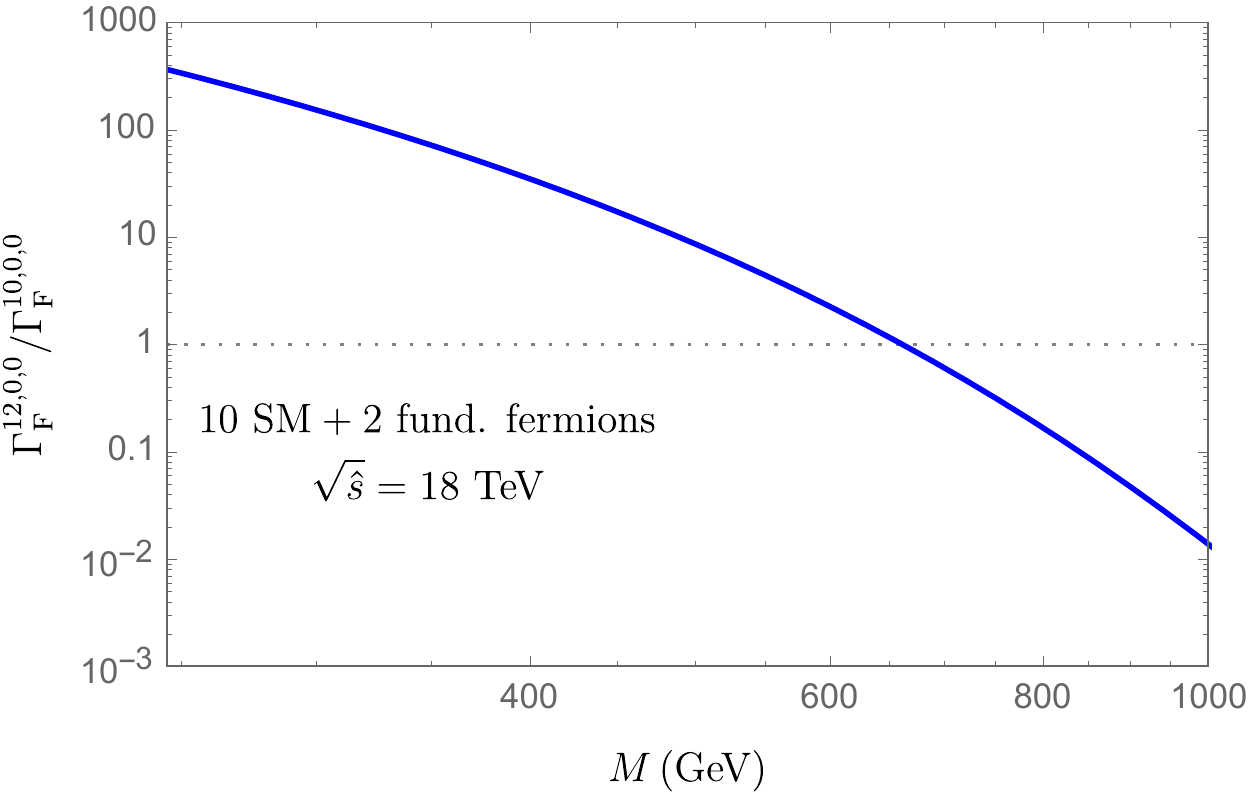} 
    \caption{$q q$ collision energy $\sqrt{\hat{s}}=18$~TeV} 
    \label{fig:heavy3} 
  \end{subfigure}%%
  \begin{subfigure}[b]{0.5\linewidth}
    \centering
    \includegraphics[width=0.95\linewidth]{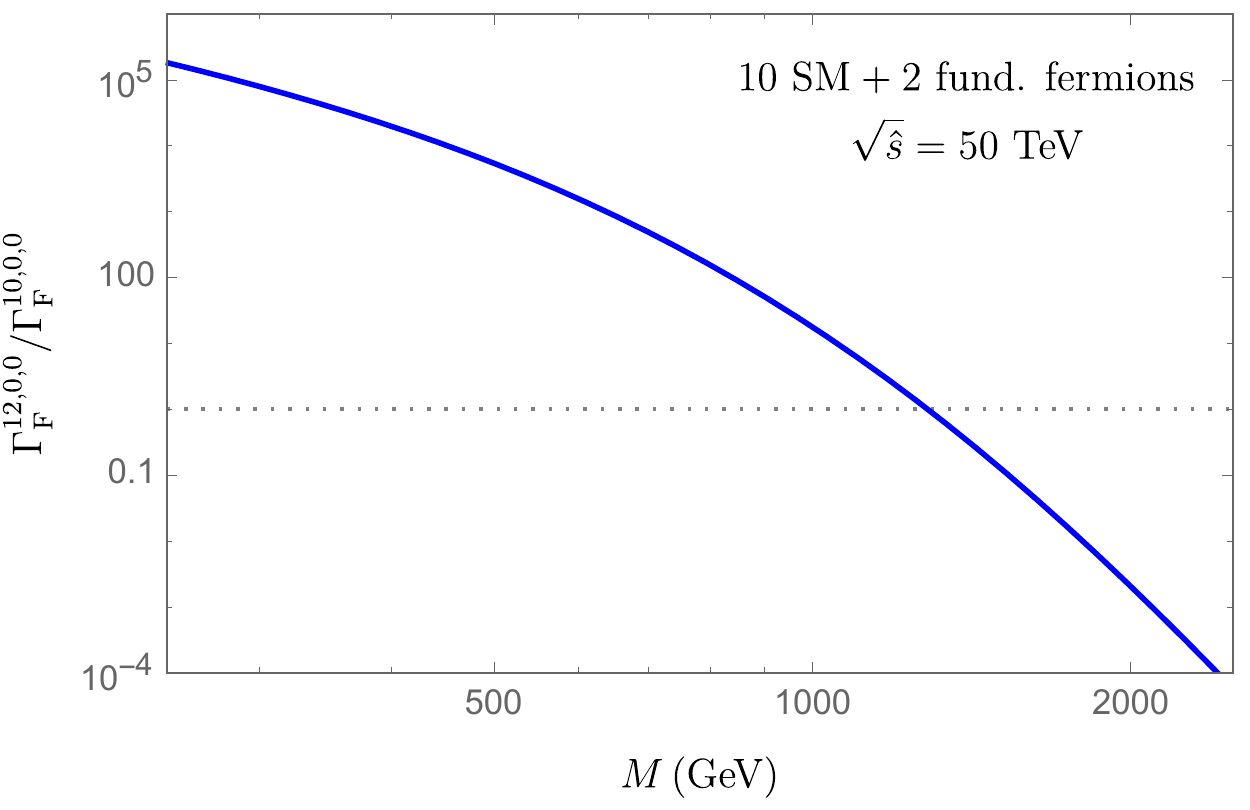} 
    \caption{$q q$ collision energy $\sqrt{\hat{s}}=50$~TeV} 
    \label{fig:heavy4} 
  \end{subfigure} 
  \caption{SM-like/SM (upper left) and BSM/SM-like rate ratios
          for two additional heavy mass fermions in the fundamental
          representation of $SU(2)$.}
  \label{fig:plots2f_high} 
\end{figure}
The results can be found in Fig.~\ref{fig:plots2f_high}.
The upper left plot shows the ratio of SM-like processes, $\Gamma_{F}^{10,0,0}/\Gamma_{SM}^{10,0,0}$.  In accordance with decoupling, the rates converge for large enough $M$, being essentially indistinguishable for masses $M\gtrsim400$ GeV. For lower masses, the ratio falls below one because the $\rho^{\overline{N}(1/3+b[\rho])}$ factor in the integrand stays less than one in the dominant small $\rho$ region, suppressing the rate. The other plots illustrate the ratio $\Gamma_{F}^{12,0,0}/\Gamma_{F}^{10,0,0}$ between the rate of the anomalous process involving a pair of the new fermions, and the rate of the SM-like process, both computed in the BSM theory. 
We have chosen  partonic centre of mass energies, $\sqrt{\hat s}$, of 10, 18, and 50 TeV. 
{Although in reality the $\sqrt{\hat s}$\, can be spread in fixed energy proton-proton collisions,
its distribution may be sharply peaked at an energy scale where the 
instanton-type exponential suppression is overcome (or minimised), 
which should be (well) above the sphaleron energy, $E_{\rm sph} \simeq$ 9 TeV.
This is because below such a scale the cross-section is exponentially suppressed and above it is also suppressed by the sharply falling parton distribution function.
The fixed $\sqrt{\hat s}$\, 
in our presentation 
therefore crudely represents this energy scale.\footnote{ Note also that one of our choices, 18 TeV, is originated from $E_0 \equiv \sqrt{6} \pi m_W / \alpha_W \simeq 18$ TeV.}}
Although the rates of the new processes decay with a growing $M$ --as expected from the reduction of phase space-- the results show that $B+L$ violating processes involving the new fermions can be dominating if the new particles are light enough, thanks to the power-like enhancement in the prefactor of the rate. For $\sqrt{\hat{s}}=10$ TeV, only slight enhancements are possible, for $M\lesssim 350$ GeV, but increasing the centre-of-mass energy has a dramatic effect, allowing for enhancements by one or two orders of magnitude for $\sqrt{\hat{s}}=18$ TeV, with the BSM processes dominating up to $M\sim650$ GeV, and with enhancements up to 5 orders of magnitude for $\sqrt{\hat{s}}=50$ TeV, with $\Gamma_{F}^{12,0,0}$ dominating for $M\lesssim 1$ TeV.

\paragraph{Scenario A:}
\begin{figure}[h!] 
  \begin{subfigure}[b]{0.5\linewidth}
    \centering
    \includegraphics[width=0.95\linewidth]{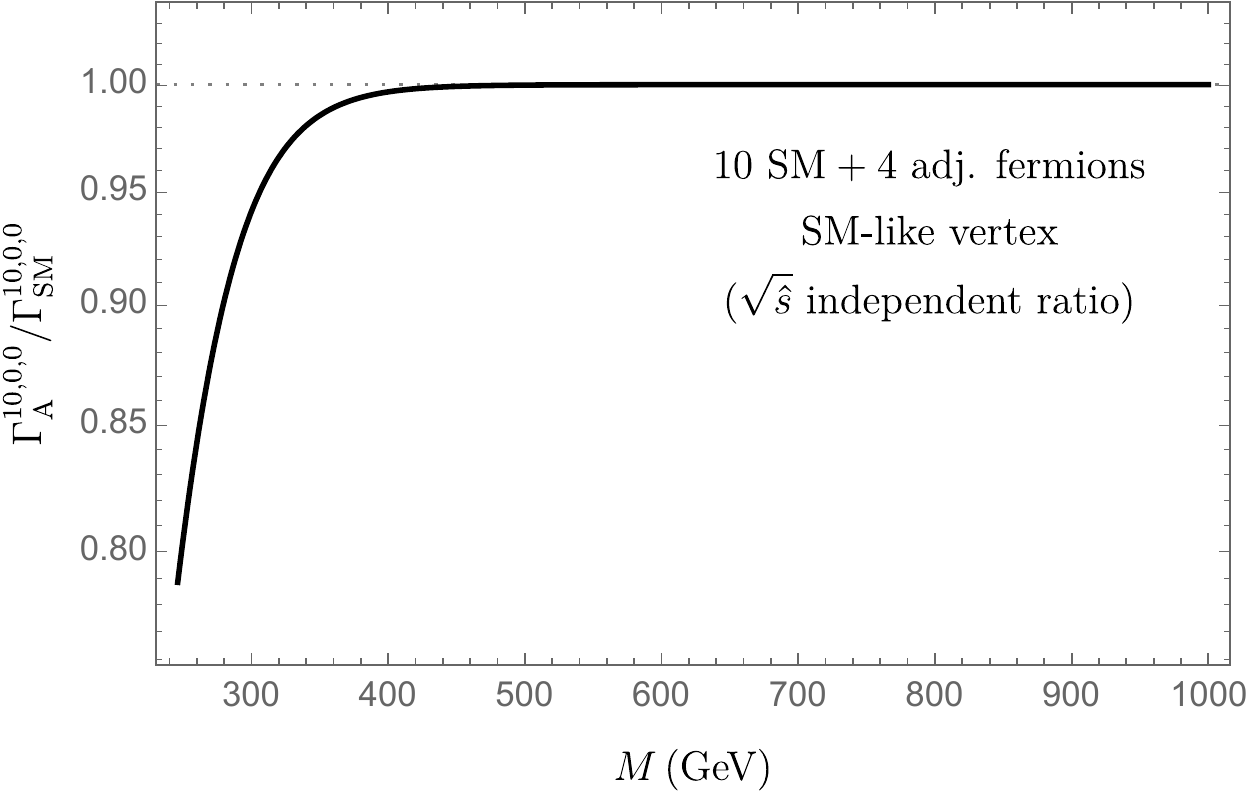} 
    \caption{SM-like/SM vertex} 
    \label{fig:heavyadj1} 
    \vspace{4ex}
  \end{subfigure}%% 
\begin{subfigure}[b]{0.5\linewidth}
    \centering
    \includegraphics[width=0.95\linewidth]{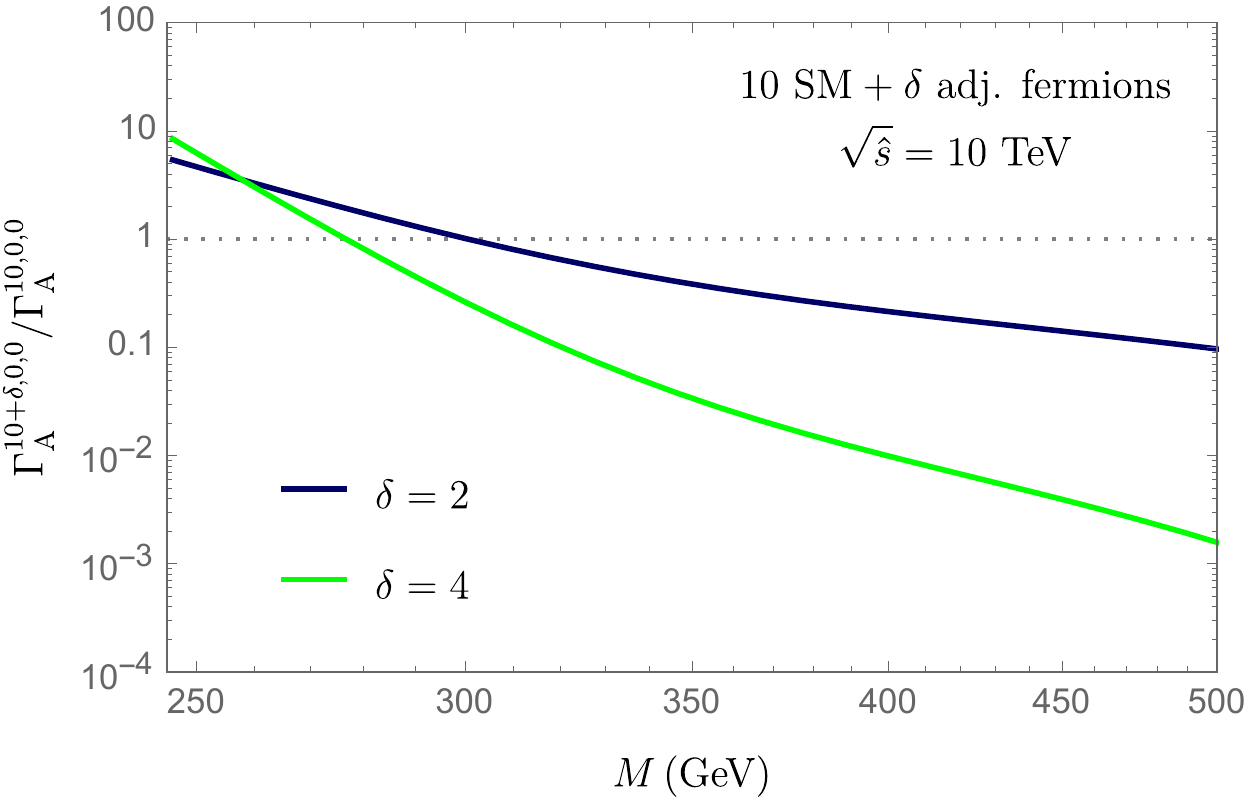} 
    \caption{$q q$ collision energy $\sqrt{\hat{s}}=10$~TeV} 
    \label{fig:heavyadj3}
    \vspace{4ex}
  \end{subfigure}%%
  
  \begin{subfigure}[b]{0.5\linewidth}
    \centering
    \includegraphics[width=0.95\linewidth]{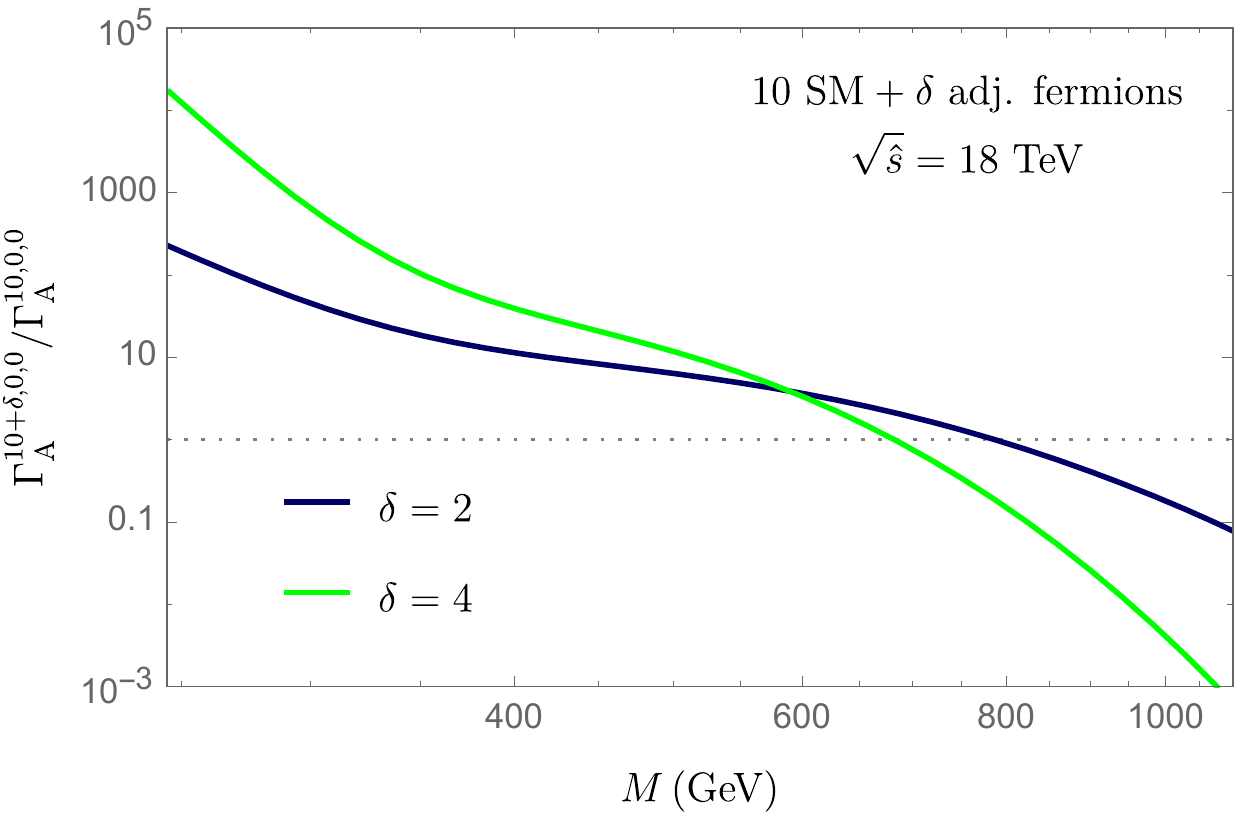} 
    \caption{$q q$ collision energy $\sqrt{\hat{s}}=18$~TeV} 
    \label{fig:heavyadj4}
  \end{subfigure}
  \begin{subfigure}[b]{0.5\textwidth}
    \centering
    \includegraphics[width=0.95\linewidth]{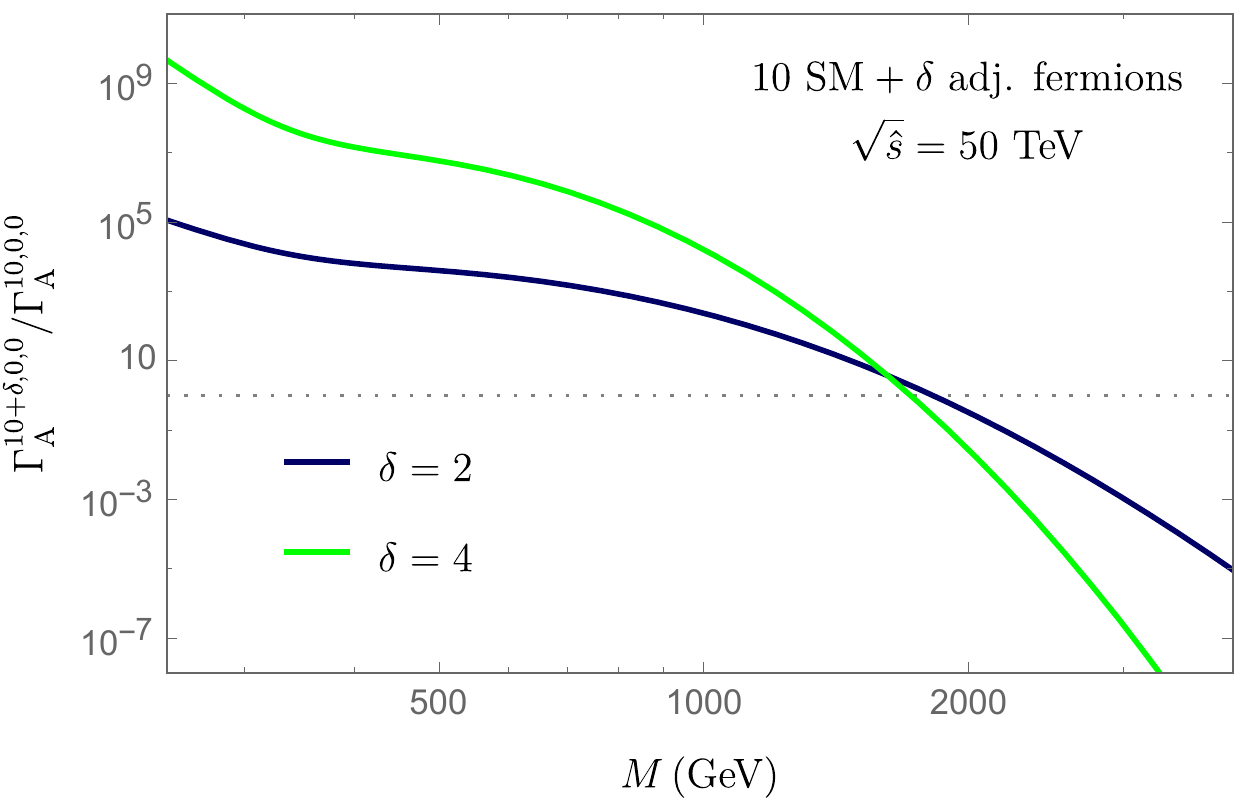} 
    \caption{$q q$ collision energy $\sqrt{\hat{s}}=50$~TeV} 
    \label{fig:heavyadj5} 
  \end{subfigure} 
  \caption{SM-like/SM (upper left) and BSM/SM-like rate ratios
          for up to four additional heavy mass fermions in the adjoint
          representation of $SU(2)$}
  \label{fig:plots4f_high} 
\end{figure}
The behaviour of the  ratios of rates in this case  is shown in Fig. ~\ref{fig:plots4f_high}. The results are similar to the ones in the theory with a new Dirac fermion in the fundamental. Again, decoupling is at work, although the
$\Gamma_{A}^{10,0,0}/\Gamma_{SM}^{10,0,0}$ ratio approaches unity for lower values of $M$ than before, while also dropping down to zero
more rapidly in the lower mass range due to the larger $\rho^{\overline{N}(1/3+b[\rho])}$ suppression in the $\rho$ integral which follows from a higher $\overline{N}$. As pertains to the ratios $\Gamma_{A}^{14,0,0}/\Gamma_{A}^{10,0,0}$ and $\Gamma_{A}^{12,0,0}/\Gamma_{A}^{10,0,0}$, one can have much larger enhancements than for a Dirac fermion in the fundamental, (due to the power-like enhancement of the prefactors with $\delta$) although the ratio is more sensitive to $M$ and decays faster as the mass grows. This effect is more accused for $\Gamma_{A}^{14,0,0}$ than for $\Gamma_{A}^{12,0,0}$, due to the larger phase space suppression with four heavy final states. Enhancements of 1 order of magnitude are already possible at $\sqrt{\hat{s}}=10$ TeV for $M<300$ GeV, and can reach $>10^8$ at a centre of mass energy of 50 TeV.

\begin{figure}[t] 
  \begin{subfigure}[b]{0.5\linewidth}
    \centering
    \includegraphics[width=0.95\linewidth]{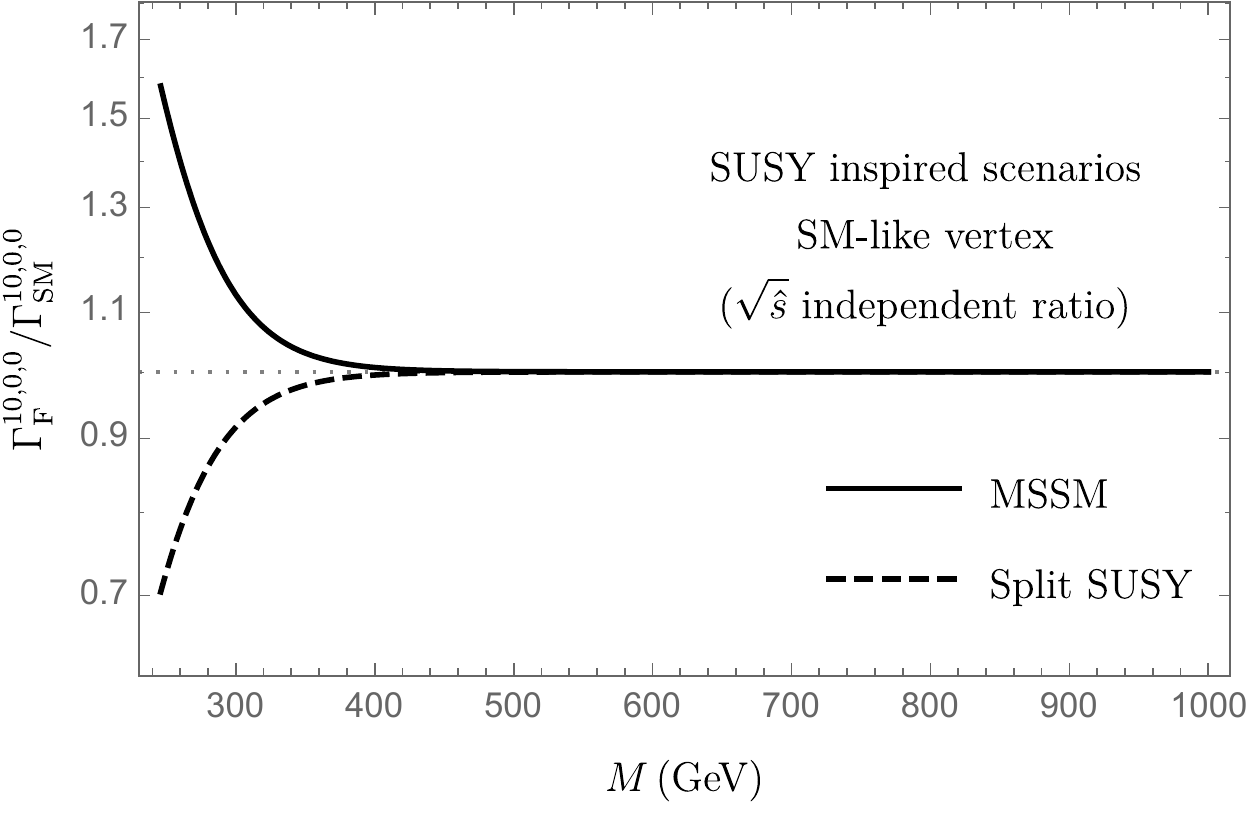} 
    \caption{SM-like/SM vertex} 
    \label{fig:heavyMSSM1} 
    \vspace{4ex}
  \end{subfigure}%% 
  \begin{subfigure}[b]{0.5\linewidth}
    \centering
    \includegraphics[width=0.95\linewidth]{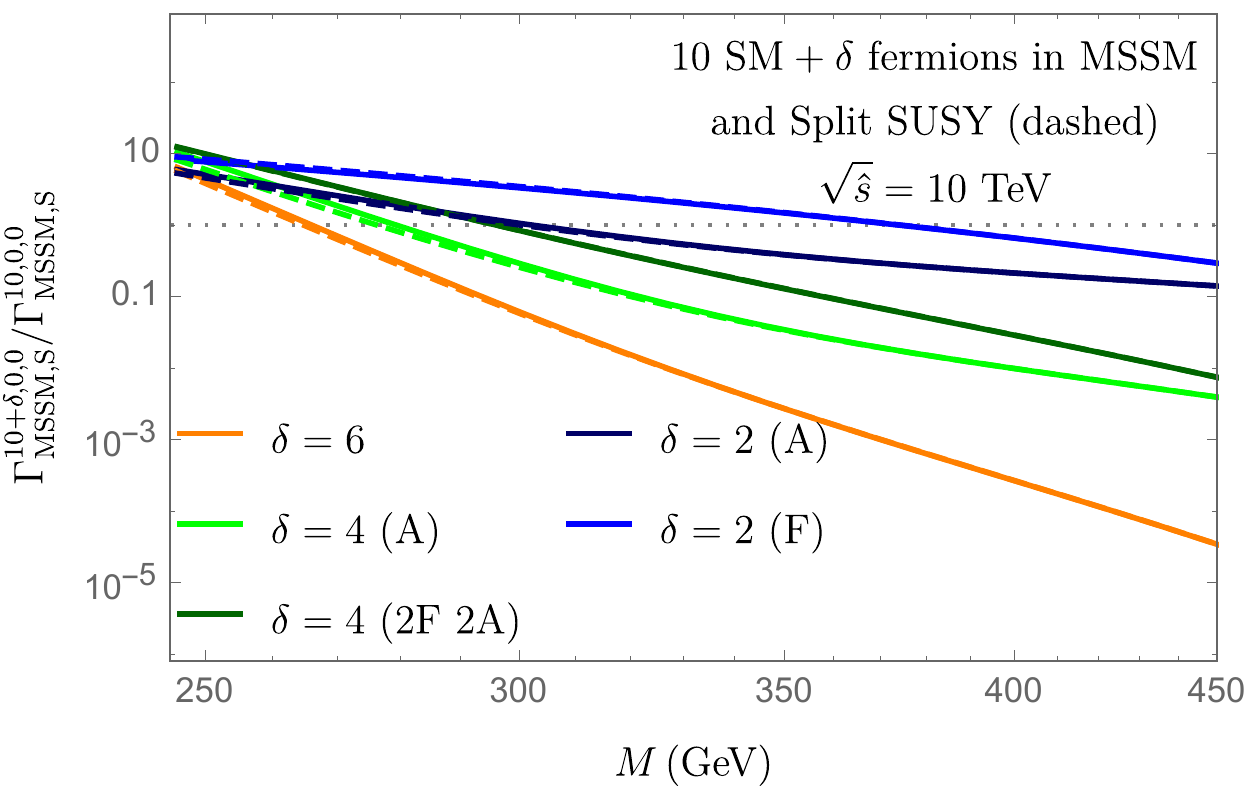} 
    \caption{$q q$ collision energy $\sqrt{\hat{s}}=10$~TeV} 
    \label{fig:heavyMSSM2} 
    \vspace{4ex}
  \end{subfigure} 
  \begin{subfigure}[b]{0.5\linewidth}
    \centering
    \includegraphics[width=0.95\linewidth]{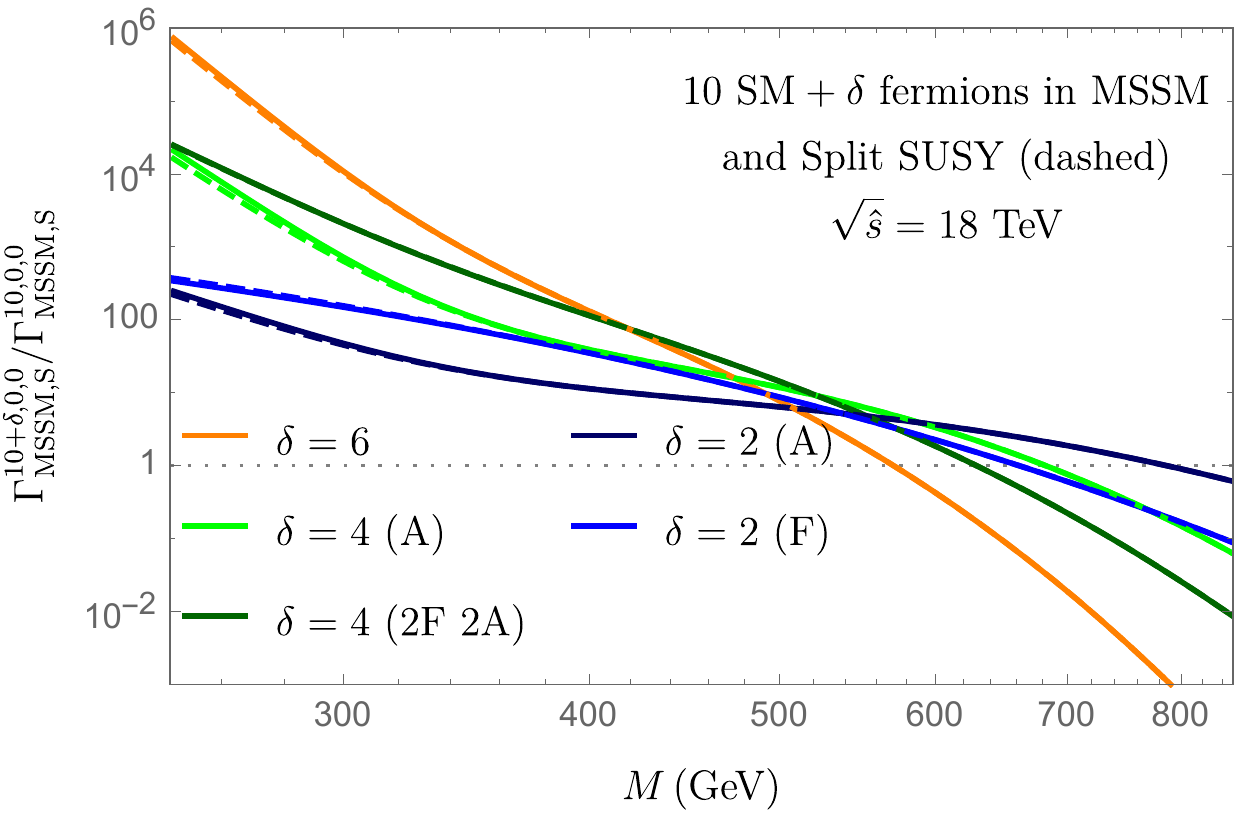} 
    \caption{$q q$ collision energy $\sqrt{\hat{s}}=18$~TeV} 
    \label{fig:heavyMSSM3} 
  \end{subfigure}%%
  \begin{subfigure}[b]{0.5\linewidth}
    \centering
    \includegraphics[width=0.95\linewidth]{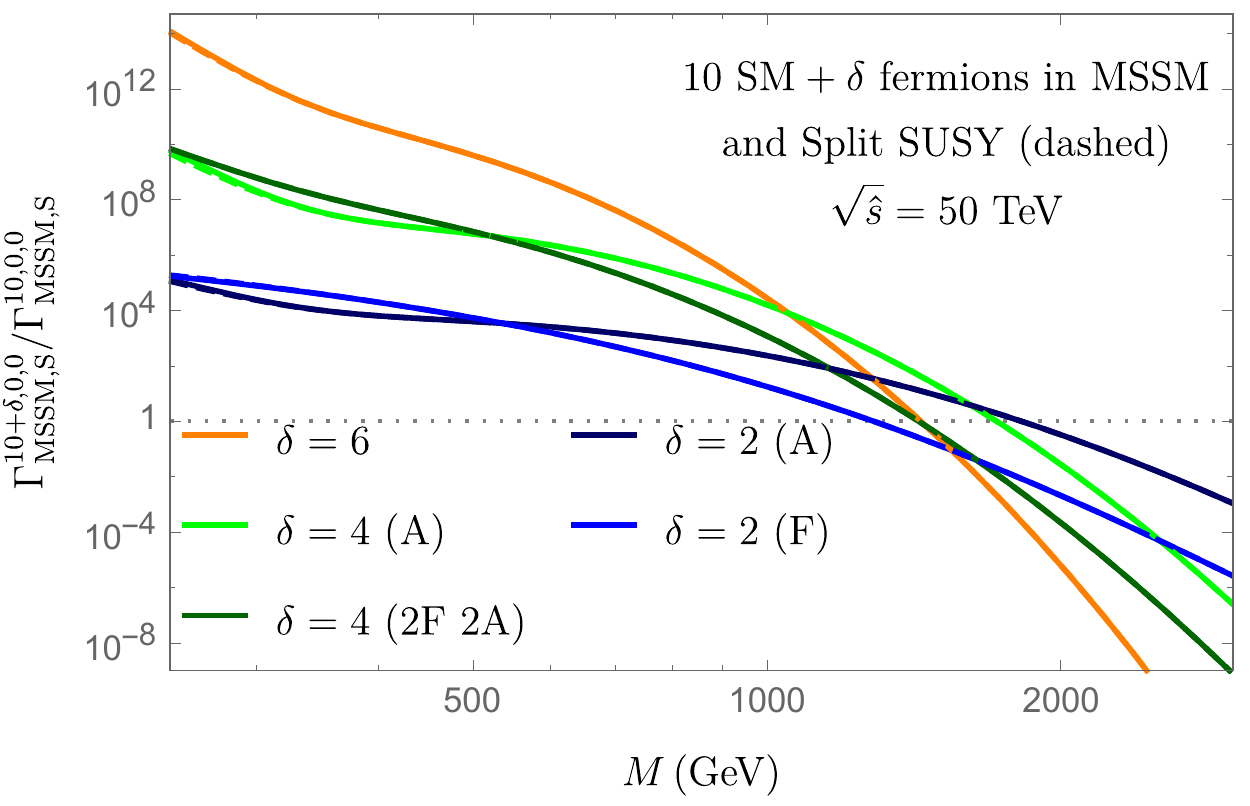} 
    \caption{$q q$ collision energy $\sqrt{\hat{s}}=50$~TeV} 
    \label{fig:heavyMSSM4} 
  \end{subfigure} 
  \caption{SM-like/SM (upper left) and BSM/SM-like rate ratios
          in the MSSM and Split SUSY.}
  \label{fig:plots6f_high} 
\end{figure}

{\bf Fermionic final states, SUSY inspired scenarios:}
{The results for SUSY-inspired scenarios are represented in Fig.~\ref{fig:plots6f_high}.} Results are similar for the degenerate MSSM and the Split-SUSY cases, with the largest difference coming from the rates of the SM-like interactions; both converge to the SM rate as before, for masses above $400$ GeV, but the SM rate is approached from above in the MSSM, as a consequence of the additional degrees of freedom that modify the running of the $g_2$ coupling. This leads to a nonzero $\Delta<0$ parameter  (see equations \eqref{eq:ratios} and \eqref{eq:Deltas}), which enhances the $\rho$ integral for small $\rho$. Regarding the rates for interactions involving the exotic fermions, we get again enhancements that grow with $\delta$ and the centre-of-mass-energy, reaching factors of $10^{12}$ for interactions with $\delta=6$ exotic fermions with masses of $300$ GeV at $\sqrt{\hat{s}}=50$ TeV.
A compilation of values of the enhancement factors for different scenarios is given in table  \ref{tab:enh}.

\begin{table}
\begin{align*}
\begin{array}{|c | c | c | c | } 
\hline
\delta  & M~[\text{GeV}]  & \sqrt{\hat s} \,\,{\text{(TeV)}} & \Gamma^{10+\delta,0,0}/{\Gamma^{10,0,0}}
\ \,
\\ \hline \hline
\rule{0pt}{3ex}   
2 (F)& 300 & 10  & 3.43 \\
     &     &  18 &   1.53\cdot 10^2 \\
     &     &  50 &  7.90 \cdot 10^4   \\\cline{2-4}
 & 600 & 10& 2.71\cdot 10^{-2}  \\
  &     &  18 &    2.26\\
     &     &  50 &   1.58 \cdot 10^3   \\\cline{2-4}
 & 1000 & 10& 5.08\cdot 10^{-5}\\
  &     &  18 &   1.38\cdot 10^{-2}  \\
     &     &  50 &   1.78  \\\cline{1-4}
4 (A) & 300 & 10& 2.62\cdot 10^{-1} \\
 &     &  18 &  6.46\cdot 10^{2}  \\
     &     &  50 &   1.89 \cdot 10^8  \\\cline{2-4}
 & 600 & 10& 1.65\cdot 10^{-4} \\
 &     &  18 &  3.34  \\
     &     &  50 &    2.28 \cdot 10^6  \\\cline{2-4}
 & 1000 & 10& 6.08\cdot 10^{-9} \\ 
 &     &  18 &    4.17\cdot 10^{-3} \\
     &     &  50 &    1.61 \cdot 10^4  \\\cline{1-4}
6 (MSSM) & 300 &10&  5.86\cdot 10^{-2} \\
 &     &  18 &   1.03\cdot 10^{4}  \\
     &     &  50 &   1.92 \cdot 10^{12}   \\\cline{2-4}
 & 600 & 10 & 5.23\cdot 10^{-8} \\ 
  &     &  18 &  4.27\cdot 10^{-1}   \\
     &     &  50 &    4.27 \cdot 10^8  \\\cline{2-4}
 & 1000 &10& 5.27\cdot 10^{-18}  \\
  &     &  18 &   7.75\cdot 10^{-7}  \\
     &     &  50 &     2.84\cdot 10^4 \\\cline{1-4}
6 (S) & 300 &10& 5.86\cdot 10^{-2} \\
 &     &  18 &    1.03\cdot 10^{4} \\
     &     &  50 &    1.92 \cdot 10^{12}   \\\cline{2-4}
 & 600 &10&  5.23\cdot 10^{-8}  \\
  &     &  18 &    4.27\cdot 10^{-1}   \\
     &     &  50 &     4.27 \cdot 10^8  \\\cline{2-4}
 & 1000 &10& 5.27\cdot 10^{-18}\\
  &     &  18 &    7.75\cdot 10^{-7}\\
     &     &  50 &    2.84\cdot 10^4 \\\cline{2-4}
\hline
\end{array}
\end{align*}
\caption{\label{tab:enh}Enhancement factors for example scenarios.} 
\end{table}

\newpage
\subsection{Processes with fermionic and bosonic final states}

Lastly, we consider $B+L$ violating processes accompanied by the emission of $W$ and Higgs bosons in the instanton background, as first analyzed in the SM context in \cite{Ringwald:1989ee,Espinosa:1989qn}. The leading instanton result is expected to give rise to an exponential enhancement involving  the first energy-dependent term in \eqref{eq:holymoly}, due to gauge boson emission; the  dominant higher-order corrections require more sophisticated methods \cite{Zakharov:1990dj,Porrati:1990rk,Khoze:1990bm,Khoze:1991mx,Ringwald:2002sw,Rubakov:1991fb,Tinyakov:1991fn,Arnold:1991dv,Mueller:1991fa,Bezrukov:2003er,Bezrukov:2003qm}.

\begin{figure}[t!]
  \begin{subfigure}[t]{\textwidth}
    \centering
    \includegraphics[width=1.\textwidth]{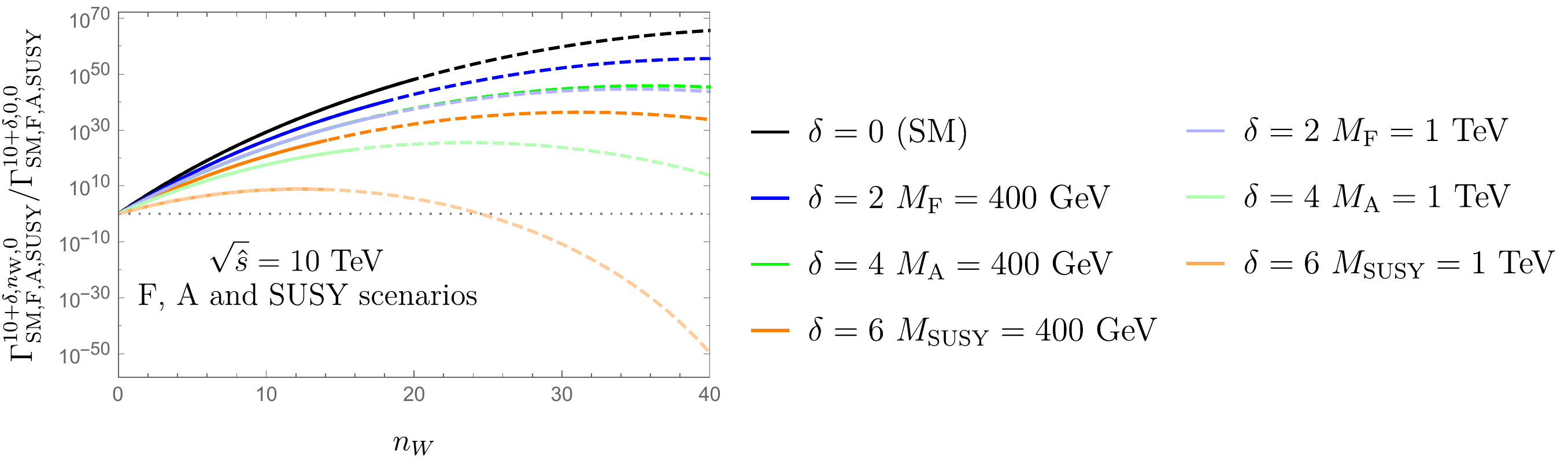} 
    \caption{$\sqrt{\hat{s}}=10$~TeV} 
    \label{fig:nW_nr} 
    \vspace{4ex}
  \end{subfigure}\\%% 
  \begin{subfigure}[t]{1.0\linewidth}
    \centering
    \includegraphics[width=1.0\linewidth]{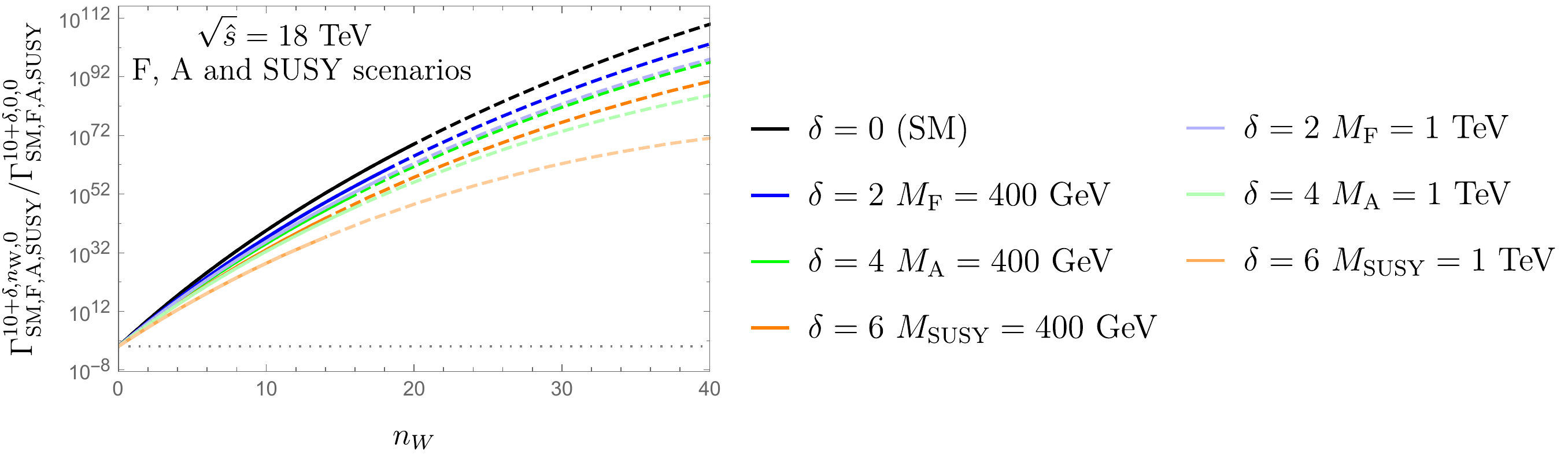} 
    \caption{$\sqrt{\hat{s}}=18$~TeV} 
    \label{fig:nW_nr_m} 
   \vspace{4ex}
  \end{subfigure} 
  \begin{subfigure}[b]{1.0\linewidth}
    \centering
    \includegraphics[width=1.0\linewidth]{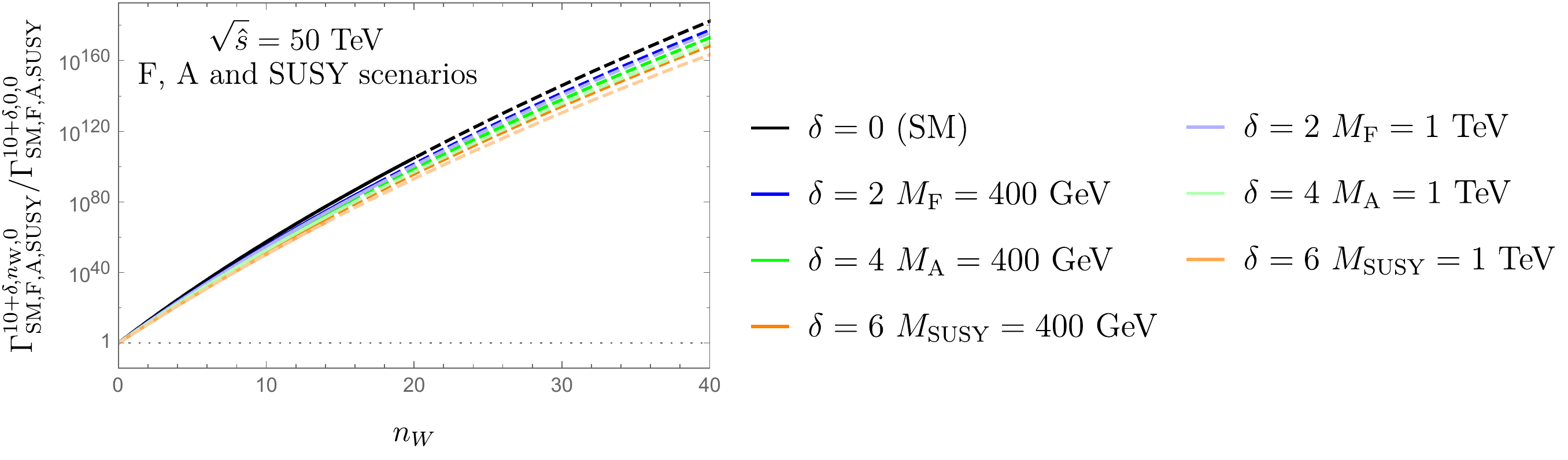} 
    \caption{$\sqrt{\hat{s}}=50$~TeV} 
    \label{fig:nW_E} 
  \end{subfigure}%%
  \caption{$n_W$ distributions for several $q q$ collision energies $\sqrt{\hat{s}}$ and new fermion masses $M$}
  \label{fig:nW} 
\end{figure}

%%%%%%%%%%%%%%%%%%%%
%%%%%%%%%%%%%%%%%%%%%

\begin{figure}[ht] 
  \begin{subfigure}[b]{1.0\linewidth}
    \centering
    \includegraphics[width=1.0\linewidth]{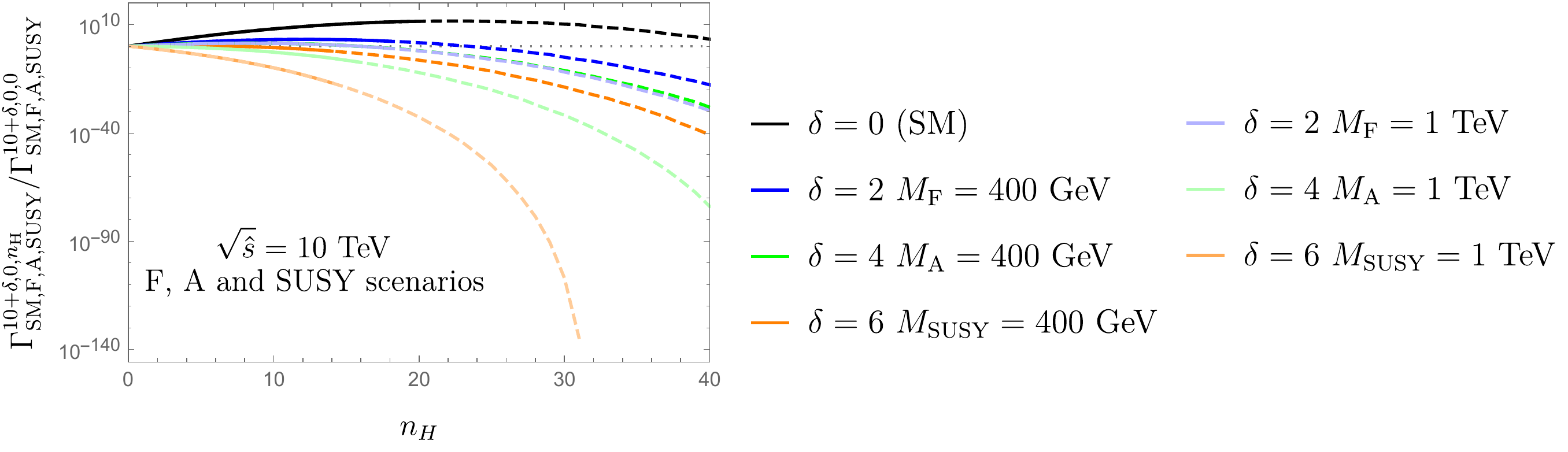} 
    \caption{$\sqrt{\hat{s}}=10$~TeV} 
    \label{fig:nH_10} 
    \vspace{4ex}
  \end{subfigure}\\%% 
  \begin{subfigure}[b]{1.0\linewidth}
    \centering
    \includegraphics[width=1.0\linewidth]{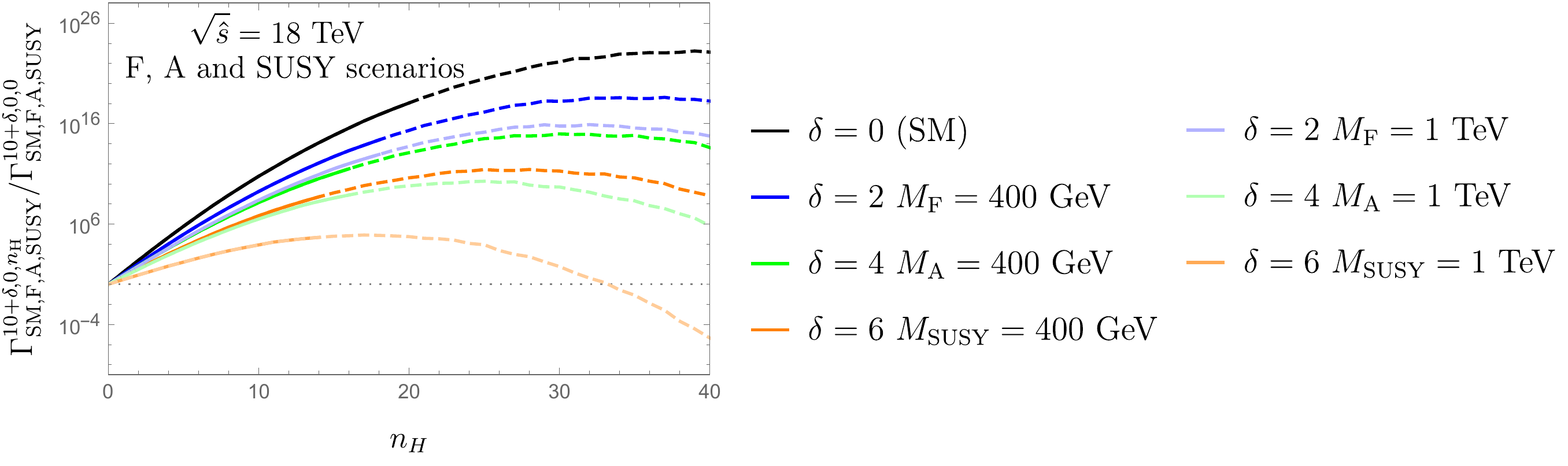} 
    \caption{$\sqrt{\hat{s}}=18$~TeV} 
    \label{fig:nH_18} 
    \vspace{4ex}
  \end{subfigure} 
  \begin{subfigure}[b]{1.0\linewidth}
    \centering
    \includegraphics[width=1.0\linewidth]{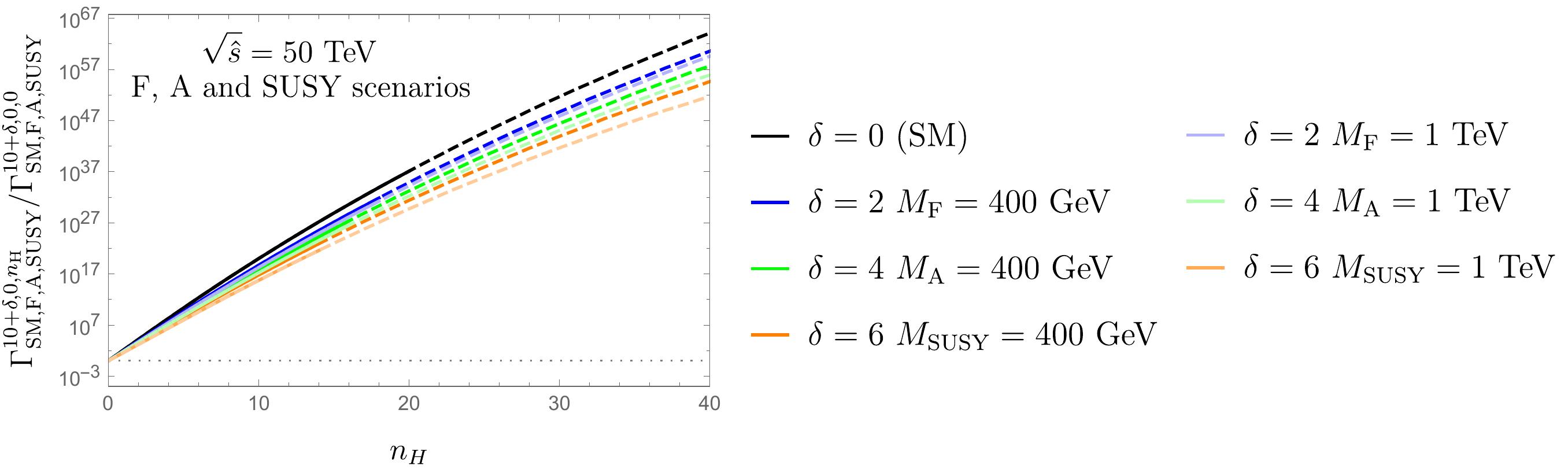} 
    \caption{$\sqrt{\hat{s}}=50$~TeV} 
    \label{fig:nH_50} 
  \end{subfigure}%%
  \caption{$n_H$ distributions for several $q q$ collision energies $\sqrt{\hat{s}}$ and new fermion masses $M$}
  \label{fig:nH} 
\end{figure}

Our earlier expectations are confirmed by our results  in Fig.~\ref{fig:nW}. In the (unreliable) leading-order calculation in the one-instanton background, additional gauge bosons increase the rate by many orders of magnitude, and when one considers processes with additional BSM fermions, the maximum enhancement  is shifted towards lower
values of of $n_W$, the effect being more pronounced for larger $M$. As we have commented earlier, these results are obtained from an instanton perturbative expansion that does not converge for $E>E_0$; additionally, the expansion is expected to break down when the total number of external lines exceeds $1/\alpha_W$, i.e. $(12+\delta+n_W+n_H)\alpha_W\geq1$, because diagrams with  propagator corrections in the external lines (implying insertions of $\alpha$) become comparable to the leading order diagrams \cite{Ringwald:1989ee}. We have indicated the values of $n_W$ in which the above inequality is safisfied by using dashed lines. If we include Higgs bosons in our vertex, we still get
an enhancement --unless the energies are low-- yet much weaker, as seen in Fig.~\ref{fig:nH} where we consider the same scenarios as in the $W$ boson case.

Although these estimates are very far from capturing the real effect of gauge boson emission, they may serve as a testing ground to understand how the holy grail function is affected by the presence of heavy fermions.  It turns out that the effect can be  understood in a very simple way: the enhancement from gauge boson emission in the leading-order instanton result is given approximately by the $\exp(\sqrt{\hat{s}}/E_0)^{4/3}$ contribution in the expansion \eqref{eq:holymoly} of the holy grail function, after substituting $\sqrt{\hat{s}}$ with the maximum energy available for gauge boson emission, 
\begin{align}
\label{eq:shift}
\sqrt{\hat{s}}\rightarrow\sqrt{\hat{s}}-\delta \cdot M\,.
\end{align}
To justify this quantitatively, we have calculated the enhancement of the cross section due to boson emission for different masses and centre-of-mass energies, and computed the ratios
\begin{align}
R_{BSM}[M,\delta_1,\delta_2]=\frac{\sum_{n_W}\Gamma^{10+\delta_1+\delta_2,n_W,0}_{BSM}}{\sum_{n_W'}\Gamma^{10,n_W',0}_{SM}}.
\end{align}
Values of $R_{BSM}$ different than one can be interpreted as a change in the holy-grail function. 
We expect the leading-order, one-instanton enhancement to be captured by the second term in the expansion of the holy grail function in \eqref{eq:holymoly}, which in turn is expected to be modified in the presence of massive fermions by shifting the energy as in \eqref{eq:shift}. We may then define a parameter $\kappa$ characterizing the deviation of $R_{BSM}$ from one as follows:
\begin{align}
R_{BSM}[M,\delta_1,\delta_2]\equiv\frac{\exp\left[\frac{4\pi}{\alpha_W}\frac{9}{8}\left(\frac{\sqrt{\hat{s}}-\kappa(\delta_1+\delta_2) M}{E_0}\right)^{4/3}\right]}{\exp\left[\frac{4\pi}{\alpha_W}\frac{9}{8}\left(\frac{\sqrt{\hat{s}}}{E_0}\right)^{4/3}\right]}.
\end{align}
For $\kappa=0$, one recovers the SM result, while $\kappa=1$ is compatible with the interpretation that the modified holy-grail function is simply obtained from the SM one after shifting the energy as in \eqref{eq:shift}.
We list results for $\kappa$ in table \ref{tab:kappa}, which shows that in our numerical estimates $\kappa$ is compatible with one within a 6\% accuracy at  energies above twice the sphaleron barrier.

\begin{table}[h]
\begin{align*}
\begin{array}{|c|c|c|c|c|}
\hline
\text{Model} & (\delta_1,\delta_2)	&  M\, \text{(GeV)} 	&    \sqrt{\hat{s}}\,\,({\text{TeV}}) &  \kappa									\\
  \hline
  \hline
   F & (2,0) & 400 & 10 & 1.3009 \\
     &       &     & 18 & 1.0646 \\
       &       &     & 20 & 1.0271 \\
  \cline{3-5}
    &  & 1000 & 10 & 1.2651\\
    &   &     &  18 & 1.0196\\
     &   &     &  20 & 0.9824\\
     \cline{2-5}
  \hline
  A & (0,4) & 400 & 10 & 1.3622 \\
    &       &     & 18 & 1.1000 \\
     &   &     &  20 & 1.0602\\
   \cline{3-5}
     & & 1000 & 10& 1.2598\\
      &       &      &  18& 0.9811\\
       &       &      &  20& 0.9427\\
     \cline{1-5}
  MSSM & (2,4) & 400 & 10 & 1.3814 \\
       &       &     &  18 & 1.1011 \\
       &       &     &  20 & 1.0596 \\
   \cline{3-5}
    &  & 1000 & 10& 1.4264\\
    &  &      & 18& 1.0251\\
    &  &      & 20& 0.9810\\
    \cline{1-5}
   S & (2,4) & 400 & 10 & 1.3814\\
   &        &     & 18 & 1.1011\\
      &       &     &  20 & 1.0596 \\
   \cline{3-5}
   &  & 1000 & 10&  1.4264\\
   &   &     & 18&  1.0251\\
    &  &      & 20& 0.9810\\
   \hline
 \end{array} 
 \end{align*}
 \caption{\label{tab:kappa} Values of $\kappa$, characterizing the deviation of the enhancement due to gauge boson emission from its SM value. $k=1$ is compatible with the following modification of the holy grail function: $F[\sqrt{\hat{s}}]\rightarrow F[\sqrt{\hat{s}}-\delta \cdot M]$. }
\end{table}

%%%%%%%%%%%%%%%%%%%%%%%%%%%%%%%%%%%%%%%%%%%%%%%%%%%%%%%
%%%%%%%%%%%%%%%%%%%%%%%%%%%%%%%%%%%%%%%%%%%%%%%%%%%%%%%
%%%%%%%%%%%%%%%%%%%%%%%%%%%%%%%%%%%%%%%%%%%%%%%%%%%%%%%
%%%%%%%%%%%%%%%%%%%%%%%%%%%%%%%%%%%%%%%%%%%%%%%%%%%%%%%
%%%%%%%%%%%%%%%%%%%%%%%%%%%%%%%%%%%%%%%%%%%%%%%%%%%%%%%
%\section{Summary and conclusions}
\section{ Discussion and conclusions}
%%%%%%%%%%%%%%%%%%%%%%%%%%%%%%%%%%%%%%%%%%%%%%%%%%%%%%%
%%%%%%%%%%%%%%%%%%%%%%%%%%%%%%%%%%%%%%%%%%%%%%%%%%%%%%%
%%%%%%%%%%%%%%%%%%%%%%%%%%%%%%%%%%%%%%%%%%%%%%%%%%%%%%%
%%%%%%%%%%%%%%%%%%%%%%%%%%%%%%%%%%%%%%%%%%%%%%%%%%%%%%%
%%%%%%%%%%%%%%%%%%%%%%%%%%%%%%%%%%%%%%%%%%%%%%%%%%%%%%%
In this paper, we have studied the impact that BSM fermions in nontrivial $SU(2)_L$ representations can have in the rates of $B+L$-violating interactions at colliders. These processes involve numbers of elementary fermions which are restricted by chiral $SU(2)_L$ anomalies. As a consequence of this, new fermions  charged under the weak gauge group allow for novel $B+L$-violating fermionic interactions in addition to the 12 fermion vertex in the SM.

In addition to fermion production, $B+L$ violating-rates can be accompanied by the production of as many bosons as the centre-of-mass energy allows. The ensuing cross-sections can be parametrised as in equation \eqref{eq:sigmaF}
by a function $f(\hat{s})$ that depends polynomically in the energy, and an exponential contribution involving the holy grail function $F(\hat{s})$. Fermion production only affects $f(\hat{s})$, while $F(\hat{s})$ incorporates the effects of massive gauge bosons. We have used leading-order instanton perturbation theory in the one-instanton background, modified to account for decoupling effects of heavy BSM fermions, to compute the effect on the latter on $f(\hat{s})$ and $F(\hat{s})$ in different BSM scenarios: a new Dirac fermion in the fundamental of $SU(2)_L$, a Weyl fermion in the adjoint, and SUSY-inspired scenarios including Higgsinos and an electroweakino.

The effect of BSM fermions in the polynomial function $f(\hat{s})$ can be substantial, leading to an enhancement with respect to the SM value (given in equation \eqref{eq:prefactor} \cite{Khoze:1990bm}) which, for a fixed BSM fermion mass $M$, grows with the number of BSM fermion fields and the centre-of-mass energy. The enhancement diminishes for growing $M$, but can still reach very large values for masses compatible with collider limits. In SUSY-like scenarios, which allow for $B+L$-violating interactions involving six BSM fermions, the enhancement can reach $10^{12}$ for $M=300$ GeV at a centre-of-mass energy of 50 TeV. Enhancement factors for different scenarios are given in table \ref{tab:enh}; the reader is also referred to figures \ref{fig:plots2f_high}, \ref{fig:plots4f_high} and \ref{fig:plots6f_high}.

Regarding the holy-grail function $F(\hat{s})$, it is known that leading-order instanton calculations can only capture its first energy-dependent contribution in an expansion in powers of the energy over the sphaleron barrier. As such, the results  for the rates of gauge boson production using instanton perturbation theory cannot be relied upon for collider predictions. Nevertheless, they might be used to infer how the full holy grail function  changes in the presence of heavy fermions. Our calculations show that for energies sufficiently above the sphaleron barrier, the one-instanton results in the presence of BSM fermions can be understood from the $(\sqrt{\hat{s}}/E_0)^{4/3}$ term in \eqref{eq:holymoly} by substituting $\sqrt{\hat{s}}$ with the maximum  energy available for gauge boson production, that is $\sqrt{\hat{s}}-\delta M$, where  $\delta$ is the number of $BSM$ fermions involved in a given $B+L$-violating interaction.

\begin{figure}[ht] 
    \centering
    \includegraphics[width=1.0\linewidth]{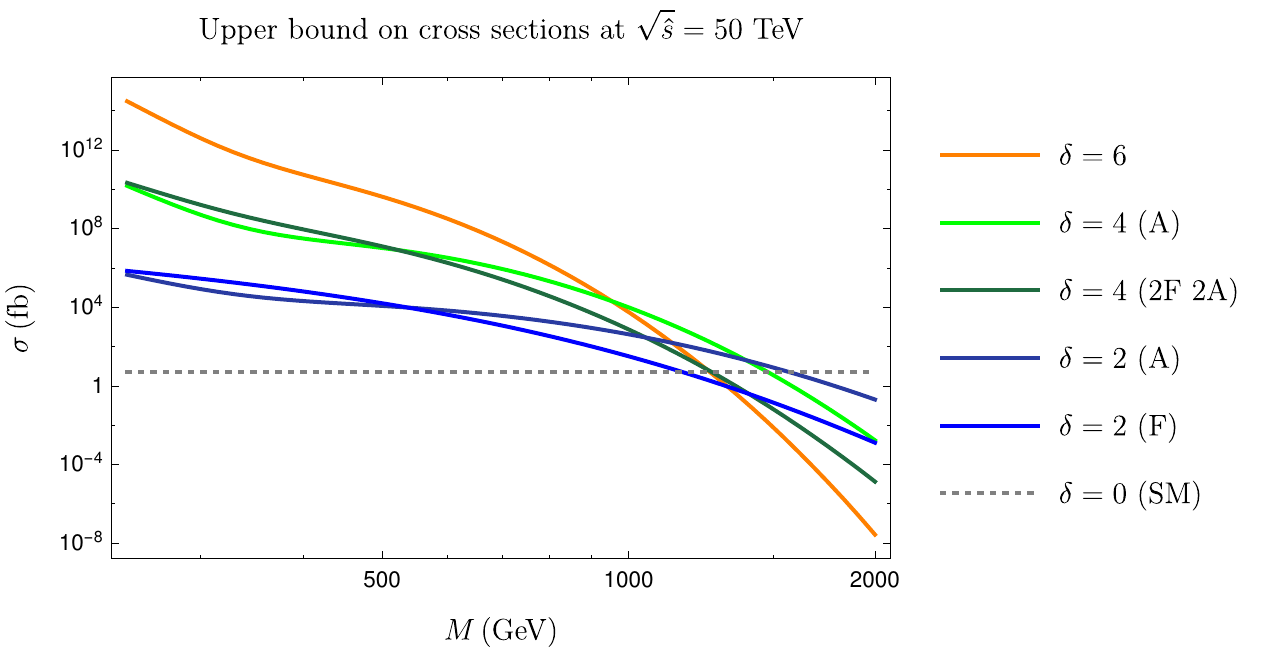} 
  \caption{Upper bounds on the cross sections in different models.}
  \label{fig:Sigmas} 
\end{figure}

We conjecture that this substitution might apply for the full holy-grail function. In this manner, starting from the SM value of $f(\hat{s})$ in equation \eqref{eq:prefactor}, the $B+L$-violating rate for an interaction involving $\delta$ BSM fermions will be of the form
\begin{align}
\label{eq:sigmaF2}
\sigma_{B+L}^{2\rightarrow {\rm any}}=\frac{ E(s,\delta,M)}{m^2_W}\left(\frac{2\pi}{\alpha_W}\right)^{7/2}e^{-\frac{4\pi}{\alpha_W}F[(\sqrt{\hat{s}}-\delta M)/E_{0}]}\ ,
\end{align}
where $ E(s,\delta,M)$ is the enhancement factor calculated in figures \ref{fig:plots2f_high}, \ref{fig:plots4f_high} and \ref{fig:plots6f_high}, and tabulated in table \ref{tab:enh}. One may then proceed as in \cite{Ringwald:2003ns} and use the lower bounds for $F(\hat{s})$ derived in \cite{Bezrukov:2003er,Bezrukov:2003qm} to estimate upper bounds for cross-sections at colliders. Taking for example a Split SUSY-like scenario at $\sqrt{\hat{s}}=50$ TeV with BSM fermions masses of 400 GeV --leading to an enhancement factor of $3.8\times 10^{10}$, as seen in figure \ref{fig:plots6f_high}-- the lower bound of $4\pi F(50-6\times0.4 \,{\rm TeV})\gtrsim 1.21$ in \cite{Bezrukov:2003er,Bezrukov:2003qm} implies a maximum cross-section of the order of 50 $\mu$barn. Taking $M=1$ TeV in the same scenarios gives in turn an enhancement factor of $10^{4.45}$, $4\pi F(44 \,{\rm TeV})\gtrsim 1.28$, and an upper bound on the cross section of around 6pb. These are to be compared with an SM cross section at $\sqrt{\hat{s}}=50$ TeV bounded by $\sim5$ fb. For more estimates, see figure \ref{fig:Sigmas}.

Our results indicate that, if $B+L$-violating interactions are ever detected at a collider, they could predominantly involve exotic particles and thus be tied to new physics. 
 This does not necessarily imply that such nonperturbative interactions could be a discovery channel for new particles, though this intriguing possibility is still open due to the 
 large backgrounds of traditional collider searches --reviewed below-- and the fact that, as advocated in \cite{Tye:2015tva}, the overall normalization of the rate for $B+L$ violating interactions may be 
 significantly larger than the results obtained with instanton methods. Note that our results concerning the enhancement of the rate in the presence of fermions is independent of the overall normalization of the rate, as our estimates were based on ratios of cross sections.
 The role of $B+L$ violating interactions as a discovery channel deserves further study, but regardless of the outcome, the nonperturbative processes analyzed in this paper are interesting on their own,
as they are direct probes of nonperturbative effects with connections to physics in the early universe that might be connected to baryogenesis.

Finally, we may comment on the current experimental limits on vector-like fermions charged under $SU(2)$.
The limit from LEP is as weak as $\sim 100$ GeV (see \cite{Egana-Ugrinovic:2018roi} for recent discussion)
for both doublet and triplet fermions.
The LHC limit strongly depends on the decay modes and mass splitting between charged and neutral components of the $SU(2)$ multiplet, $\Delta m_{\pm0}$.
In the minimal case where these multiplets do not mix with other multiplets, 
the mass splitting is generated radiatively, which is typically 
$\Delta m_{\pm0} \simeq 350$ \cite{Thomas:1998wy} and 165 \cite{Ibe:2012sx} MeV for doublet and triplet fermions, respectively.
For such a small mass splitting, the decay products of the charged state become very soft and easily lost in the background. 
The missing transverse energy also becomes small because the two neutral states are produced 
in a back-to-back configuration in the transverse plane and their missing momenta are cancelled.
For doublet fermions, the projected sensitivity has been estimated \cite{Low:2014cba} 
using a mono-jet channel and turns out to be 
$\sim (80 - 185$) GeV for ($5 - 2$) $\sigma$ at the high-luminosity 14 TeV LHC with 3 fb$^{-1}$.
This sensitivity is improved at a 100 TeV $pp$ collider with 3 fb$^{-1}$
to $\sim (285 - 870)$ GeV for ($5 - 2$) $\sigma$.
For triplet fermions, the mass splitting is small enough so that the charged state 
can become long-lived in terms of the collider scale.
These charged states can travel and decay inside the tracker system, leaving a distinctive disappearing charged track signature. 
Using the disappearing track channel, the current LHC data sets the limit on the mass of triplet fermions, $M \gtrsim 460$ GeV \cite{Aaboud:2017mpt,CMS:2014gxa}.
For a 100 TeV collider with 3 fb$^{-1}$, 
the projected sensitivity is estimated to be $\sim (2.2 - 2.9)$ TeV for $(5 - 2)$ $\sigma$ \cite{Low:2014cba}.
In summary, the current limit on the $SU(2)$ fermions is not very strong: $\gtrsim 100$ and 460 GeV 
for doublet and triplet fermions.
This justifies the fermion mass rage used in our numerical calculation.

As has been emphasized before, the overall normalization of the sphaleron production rate at colliders is still under debate.
If the resonant enhancement advocated in ref.~\cite{Tye:2015tva} is correct, then $B+L$ violating interactions could have an observable rate even at the LHC.
If this is the case, it is possible that these sphaleron processes could be observable immediately after 
the 100 TeV collider is turned on~\cite{Ellis:2016ast}. Given the weak mass limits expected at the LHC, this leaves open  the possibility that sphaleron interactions may be 
observed before the discovery of exotic fermions through perturbative production processes. 
The main result of this paper, enhancement of the sphaleron rate due to new $SU(2)$ fermions,
even encourages this very optimistic scenario.

%%%%%%%%%%%%%%%%%%%%%%%%%%%%%%%%%%%%%%%%%%%%%%%%%%%%%%%
%%%%%%%%%%%%%%%%%%%%%%%%%%%%%%%%%%%%%%%%%%%%%%%%%%%%%%%
%%%%%%%%%%%%%%%%%%%%%%%%%%%%%%%%%%%%%%%%%%%%%%%%%%%%%%%
%%%%%%%%%%%%%%%%%%%%%%%%%%%%%%%%%%%%%%%%%%%%%%%%%%%%%%%
%%%%%%%%%%%%%%%%%%%%%%%%%%%%%%%%%%%%%%%%%%%%%%%%%%%%%%%
\section*{Acknowledgements}

The authors wish to thank Andreas Ringwald and Valya Khoze for valuable discussions and feedback. D.G.C. is supported by the SFTC. P.R. is funded by the Graduiertenkolleg \textit{‘Particle physics beyond the Standard Model’} (GRK 1940) and acknowledges financial support by the ERASMUS programme for a research stay at the IPPP, Durham University. The work of K.S. was partially supported by the National Science Centre, Poland, under research grants DEC-2014/15/B/ST2/02157 and DEC-2015/18/M/ST2/00054, as well as by MEXT KAKENHI Grant-in-Aid for Scientific Research on Innovative Areas, Japan, Grant Number JP16K21730.  C.T. acknowledges support by the Collaborative Research Centre SFB1258 of the Deutsche Forschungsgemeinschaft (DFG). 
\vspace*{2cm}
%%%%%%%%%%%%%%%%%%%%%%%%%%%%%%%%%%%%%%%%%%%%%%%%%%%%%%%
%%%%%%%%%%%%%%%%%%%%%%%%%%%%%%%%%%%%%%%%%%%%%%%%%%%%%%%
%%%%%%%%%%%%%%%%%%%%%%%%%%%%%%%%%%%%%%%%%%%%%%%%%%%%%%%
%%%%%%%%%%%%%%%%%%%%%%%%%%%%%%%%%%%%%%%%%%%%%%%%%%%%%%%
%%%%%%%%%%%%%%%%%%%%%%%%%%%%%%%%%%%%%%%%%%%%%%%%%%%%%%%
%\clearpage
\appendix
\noindent{\bf \Large Appendix}

\section{\label{app:Euclidean}Euclidean conventions and identities. Group integration}

The Euclidean coordinates are $x_\mu=(x_1,x_2,x_3,x_4)=(x,y,z,it)$. We define the Euclidean antisymmetric tensor with the convention $\epsilon_{1234}=1$. The Euclidean gauge fields $A_\mu$ are related to their Minkowski counterparts $A^M_\mu$ as:
\begin{align}
A_i=A^M_i,\,i=1,\dots,3, \quad A_4= -i A^M_0.
\end{align}
A particle of mass $m$ with Minkowski momentum $k^{M}_{\mu}=(E,k_x,k_y,k_z)$ has a corresponding Euclidean momentum  $k_\mu=(k_x,k_y,k_z,-iE)$.
For spatial momentum aligned with the $z$ axis, the Euclidean polarization vectors of gauge bosons are:
\begin{align}
\label{eq:polarizations}
k_\mu=(0,0,k_z,-iE)\Rightarrow
\epsilon^{T1}_{\mu}=(1,0,0,0),\quad \epsilon^{T2}_{\mu}=(0,1,0,0),\quad \epsilon^{L}_{\mu}=\left(0,0,\frac{E_W}{m_W},-i\frac{k_z}{m_W}\right).
\end{align}
The 't Hoof symbols $\eta_{a\mu\nu}$ of equation \eqref{eq:etas} --and the analogous $\overline\eta_{a\mu\nu}$ symbols that satisfy  \eqref{eq:etas} but with the opposite sign in the $\delta$s-- have the following properties:
\begin{equation}\label{eq:etas2}\begin{aligned}
\eta_{a\mu\nu}\eta_{a\rho\sigma}=&\,\delta_{\mu\rho}\delta_{\nu\sigma}-\delta_{\mu\sigma}\delta_{\nu\rho}+\epsilon_{\mu\nu\rho\sigma},\\
%%%
\epsilon^{abc}\eta_{b\mu\nu}\eta_{c\rho\sigma}=&\,\delta_{\mu\rho}\eta_{a\nu\sigma}-\delta_{\mu\sigma}\eta_{a\nu\rho}-\delta_{\nu\rho}\eta_{a\mu\sigma}+\delta_{\nu\sigma}\eta_{a\mu\rho},\\
%%%
\overline\eta_{a\mu\nu}\overline\eta_{a\rho\sigma}=&\,\delta_{\mu\rho}\delta_{\nu\sigma}-\delta_{\mu\sigma}\delta_{\nu\rho}-\epsilon_{\mu\nu\rho\sigma},\\
%%%%%
\eta_{a\mu\nu}\overline\eta_{b\mu\nu}=&\,0.
\end{aligned}\end{equation}
Using the first identity above, together with \eqref{eq:polarizations}, one obtains \eqref{eq:fW}.

We define fermion fields and their partition functions through analytic continuation from Minkowski space \cite{Vainshtein:1982ic,Leutwyler:1992yt}, thus avoiding the problem of the non-existence of Majorana fermions compatible with the Euclidean SO(4) symmetry \cite{Ramond:1981pw}. We choose Euclidean gamma matrices, left and right projectors, as well as the Euclidean Dirac adjoint, as
\begin{equation}\begin{aligned}
\gamma_\mu=&\,\left[
\begin{array}{cc}
0 & \sigma_\mu\\
%%%
\overline\sigma_\mu & 0
\end{array}\right],\,\,  {\sigma}_\mu =(\vec{\sigma},i), \,\,\overline{\sigma}_\mu =(-\vec{\sigma},i),\\
%%%%
\gamma_5=&\,-\prod_\mu\gamma_\mu=\left[
\begin{array}{cc}
-\mathbb{I} &0\\
%%%
0 & \mathbb{I}
\end{array}\right],\,\,P_L=\frac{1-\gamma_5}{2},\,\,P_R=\frac{1+\gamma_5}{2},\\
%%%%%
\overline\Psi_{kl}=&\,-i\Psi_{kl}^\dagger{\gamma}_4.
\end{aligned}\end{equation}
In addition, we introduce the matrices $\gamma_{\mu\nu}$ as
\begin{align}
\gamma_{\mu\nu}=\frac{i}{4}[\gamma_\mu,\gamma_\nu],
\end{align}
which satisfy the following duality properties:
\begin{equation}\label{eq:gammamunudual}\begin{aligned}
P_R\,\gamma_{\mu\nu}=&\,\frac{1}{2}\epsilon_{\mu\nu\rho\sigma}P_R\,\gamma_{\mu\nu},\\
%%%
P_L\,\gamma_{\mu\nu}=&\,-\frac{1}{2}\epsilon_{\mu\nu\rho\sigma}P_L\,\gamma_{\mu\nu}.
\end{aligned}\end{equation}
One has
\begin{equation}
\label{eq:sigmamunu}
\gamma_{\mu\nu}=\left[
\begin{array}{cc}
\overline{\sigma}_{\mu\nu} &0\\
%%%
0 & {\sigma}_{\mu\nu}
\end{array}\right],
\end{equation}
with
\begin{equation}\label{eq:identities_sigma}
\begin{aligned}
\overline{\sigma}_{\mu\nu} =&\,\frac{1}{2}\overline\eta_{a\mu\nu}\sigma^a,\quad {\sigma}_{\mu\nu} =\frac{1}{2}\eta_{a\mu\nu}\sigma^a,\\
%%%%%%%
\sigma_{\mu\nu}\overline{\sigma}_\rho=&\,\frac{1}{2i}\,\delta_{\nu\rho}\,\overline\sigma_\mu-\frac{1}{2i}\,\delta_{\mu\rho}\,\overline\sigma_\nu-\frac{1}{2i}\epsilon_{\mu\nu\rho\delta}\,\overline\sigma_\delta.
\end{aligned}
\end{equation}
Finally, for the integration over rigid rotations $\tilde U$ in a given representation $r$, we use
\begin{align}
\int d\tilde U\tilde U_{r,mn}\tilde U^\dagger_{r,pq}=\frac{1}{\dim(r)}\,\delta_{np}\delta_{mq},
\end{align}
where $\tilde U_{r,mn}$ denotes the matrix with indices $m,n$ representing a given element $\tilde U$ of the group in the representation $r$. The normalisation of the above integral is such that  
\begin{align}\label{eq:group_integral}
\int d\tilde U (\tilde U \tilde U^\dagger)_{mq}=\int d\tilde U \mathbb{I}_{mq}=\delta_{mq}.
\end{align}
%%%%%%%%%%%%%%%%%%%%%%%%%%%%%%%%%%%%%%%%%%%%%%%%%%%%%%%%%%%%%%%%%%%%%%%%%
%%%%%%%%%%%%%%%%%%%%%%%%%%%%%%%%%%%%%%%%%%%%%%%%%%%%%%%%%%%%%%%%%%%%%%%%%
%%%%%%%%%%%%%%%%%%%%%%%%%%%%%%%%%%%%%%%%%%%%%%%%%%%%%%%%%%%%%%%%%%%%%%%%%
%%%%%%%%%%%%%%%%%%%%%%%%%%%%%%%%%%%%%%%%%%%%%%%%%%%%%%%%%%%%%%%%%%%%%%%%%
%%%%%%%%%%%%%%%%%%%%%%%%%%%%%%%%%%%%%%%%%%%%%%%%%%%%%%%%%%%%%%%%%%%%%%%%%
%%%%%%%%%%%%%%%%%%%%%%%%%%%%%%%%%%%%%%%%%%%%%%%%%%%%%%%%%%%%%%%%%%%%%%%%%
%%%%%%%%%%%%%%%%%%%%%%%%%%%%%%%%%%%%%%%%%%%%%%%%%%%%%%%%%%%%%%%%%%%%%%%%%

\section{\label{app:zeromodes}Fermion zero modes}

In this appendix we collect formulae for the fermionic zero modes of Dirac spinors in the fundamental and adjoint representations. The Fourier transforms of these modes, when evaluated on on-shell Euclidean momenta with $|p|=M$ ($M$ being a fermion mass), define the $\rho$-dependent form-factors that accompany instanton-induced fermion interactions.

\subsection{Fundamental representation}

A Dirac fermion in the fundamental representation has a single zero mode \cite{tHooft:1976rip}, given in the singular gauge by 
\begin{align}
 \Psi_{i m}^{0}(x)=\phi(r)\left[P_R(\xslash-\xslash_0)\right]_{ij}\tilde U_{mn}\tilde\epsilon_{j n},\quad \phi(r)=\frac{\rho}{\pi r(r^2+\rho^2)^{3/2}},
\end{align}
where $i=1,\dots,4$ and $m=1,2$ are Dirac and representation indices, respectively, $r=(\sum_i x^2_i)^{1/2}$, $\tilde U$ is the rigid group transformation in \eqref{eq:inst_sing} in the fundamental representation, and the matrix $\tilde\epsilon_{j m}$ is given by
\begin{align}
\label{eq:fundmode}
\tilde\epsilon=\left[
\begin{array}{cc}
0_{2\times2} & 0_{2\times2} \\
0_{2\times2} & \epsilon_{2\times2} 
\end{array}\right],
\end{align}
where $\epsilon_{2\times2}$ is the usual two-by-two antisymmetric matrix with $\epsilon_{12}=1$.

In momentum space one has
\begin{align}
\Psi_{i m}^{0}(p)=-\frac{i}{|p|}\,\phi(|p|)'[{P_R}\,\pslash]_{ij}\,\tilde U_{mn}\tilde\epsilon_{jn},
\end{align}
with $\phi(|p|)$ the Fourier transform of $\phi(x)$ in \eqref{eq:fundmode},
\begin{align}
\phi(|p|)=\int d^4 xe^{ipx}\phi(x)=2\pi\rho\left[I_0\left(\frac{|p|\rho}{2}\right)K_0\left(\frac{|p|\rho}{2}\right)-I_1\left(\frac{|p|\rho}{2}\right)K_1\left(\frac{|p|\rho}{2}\right)\right].
\end{align}
In the above expression, $I_i$ are modified Bessel functions
of the first kind, and $K_i$ denote modified Bessel functions of the second kind. The propagator in the instanton background can be approximated from the zero mode contribution (see the discussion around \eqref{eq:Diracprop}). A group-averaged propagator (calculated using  the integral in \eqref{eq:group_integral}), multiplied by a factor of $\rho$ times the mass --coming from the zero mode contribution to the fermion determinant-- and amputated with ordinary propagators, is  given in the on-shell limit by
\begin{align}
\nonumber
&\rho M\int d\tilde U(\pslash+M)\langle\Psi(p)\bar{\Psi}(q)\rangle(\qslash+M)|_{\rm o.s.}\sim \rho\int d\tilde U(\pslash+M)\Psi^0(p)\Psi^{0\dagger}(q)(\qslash+M)|_{\rm o.s.}\\
%%%%
\label{eq:correlatorfund}
&={2\rho(\phi'(M)M)^2}P_L\equiv {\cal F}^F_M(\rho)P_L.
\end{align}
Above, we ignored phases in the mass matrices, and ``o.s.'' refers to imposing ${\cal O}\qslash =M{\cal O}$, $\pslash{\cal O}=M{\cal O}$, where ${\cal O}$ designates an arbitrary operator. Such substitution is appropriate for the computation of matrix elements between physical states. Note how the result is proportional to the left-handed projector $P_L$, so that the on-shell effective Lagrangian reproducing the correlators \eqref{eq:correlatorfund} involves left-handed fermions, as expected from the chiral anomalies. For Green functions with more fermion insertions, we expect results looking like products of the above form factors, although with differences coming from the different group averaging and on-shell simplifications. For our estimates we will ignore these differences and approximate the results via products of the above correlators.  ${\cal F}^F_M(\rho)$ in \eqref{eq:correlatorfund} can be then viewed as an instanton form-factor for a massive fermion in the fundamental, whose behaviour at large and small $\rho M$ is given in \eqref{eq:Fexpansions}. For a massless fermion, the amputation is done with  massless propagators, and the result is
\begin{align}
\lim_{M\rightarrow0}\rho M\int d\tilde U\pslash\langle\Psi(p)\bar{\Psi}(q)\rangle\qslash|_{\rm on-shell}\sim 2\pi^2\rho^3P_L\equiv {\cal F}^F_0(\rho)P_L.
\end{align}

\subsection{Adjoint representation}

In the $n_{\rm top}=1$ instanton background, the Dirac operator has four zero modes, which can be understood as supersymmetric transformations of the instanton background \cite{Jackiw:1977pu}.\footnote{This is because in supersymmetric theory, an adjoint gauge field and an adjoint Majorana spinor belong to the same supersymmetric multiplet. Supersymmetric transformations map boson configurations to fermion configurations, while preserving the equations of motion. Hence a supersymmetric transformation of the instanton background is a fermion zero mode.}
The properly normalised modes are 
\begin{equation}\label{eq:adjointmodes}\begin{aligned}
\Psi^{0(i)}_a(x)=&\,\frac{g}{4\sqrt{2}\pi}\,\gamma_{\mu\nu}\,u^{(i)} \tilde U_{ab} F^b_{\mu\nu}(x),\quad i=1,2,\\
%%%%
\Psi^{0(i)}_a(x)=&\,\frac{g}{8\pi\rho}\,\gamma_{\mu\nu}\,\xslash\, v^{(i-2)} \tilde U_{ab} F^b_{\mu\nu}(x),\quad i=3,4.
\end{aligned}\end{equation}
In the above equation, $(i)$ labels the zero modes, and $a$ denotes the index of the adjoint representation; we have omitted Dirac indices.
$\tilde U_{ab}$ is the rigid rotation of equation \eqref{eq:inst_sing} in the adjoint representation, and $u^{(i)},v^{(i)},\,i=1,2$ are four constant spinors which can be chosen as
\begin{align}
u^{(1)}=[0,0,1,0]^\top,\quad u^{(2)}=[0,0,1,0]^\top, \quad v^{(1)}=[1,0,0,0]^\top,\quad v^{(2)}=[0,1,0,0]^\top.
\end{align}
The $u^{(i)},v^{(i)}$  satisfy the completeness relation
\begin{align}
\sum_i u^{(i)}_k u^{(i)\dagger}_l= [P_R]_{kl},\quad \sum_i v^{(i)}_k v^{(i)\dagger}_l= [P_L]_{kl}.
\end{align}
In equation \eqref{eq:adjointmodes}, $F^a_{\mu\nu}$ is the field strength in the singular gauge, which reads
\begin{align}
F^a_{\mu\nu}=-\frac{8}{g}\left[\frac{(x-x_0)_\mu(x-x_0)_\sigma}{(x-x_0)^2}-\frac{1}{4}\delta_{\mu\sigma}\right]\,\overline\eta_{a\nu\sigma}\,\frac{\rho^2}{[(x-x_0)^2+\rho^2]^2}-(\mu\leftrightarrow\nu).
\end{align}
The Fourier transforms of the modes \eqref{eq:adjointmodes} can be given again in terms of Bessel functions. Denoting $u\equiv |p|\rho$, and using the identities in  \eqref{eq:identities_sigma}, one can show that
\begin{align}
\nonumber\Psi^{0(i)}_a(p)=&\,-\frac{4\sqrt{2}\pi}{|p|^6\rho^2}\left[u(u(8+u^2)K_0(u)+4(4+u^2)K_1(u))-16\right]\,U^{ab}\,\overline{\eta}_{b\nu\rho}\,p_\mu p_\rho\,\gamma_{\mu\nu}\,u^{(i)} ,\\
%%%
\nonumber & i=1,2,\\
%%%%
\Psi^{0(i)}_a(p)=&\,\frac{2i\pi \rho}{|p|^2}\left[u K_1(u)-2K_0(u)\right]\,U^{ab}\,\overline{\eta}_{b\nu\rho}\,p_\rho\,\gamma_\nu \, v^{(i-2)} ,\quad i=3,4.
\end{align}
Finally, we can again estimate the propagator in the instanton background by summing over the zero mode contributions in \eqref{eq:Diracprop}. Integrating over the rigid rotations using \eqref{eq:group_integral}, the on-shell, group-averaged propagator multiplied by a factor of the mass and amputated with ordinary propagators, is given by
\begin{align}\nonumber
&\rho M\int d\tilde U(\pslash+M)\langle\Psi(p)\bar{\Psi}(q)\rangle(\qslash+M)|_{\rm o.s.}\sim \rho\sum_{i=1}^4\int d\tilde U(\pslash+M)\Psi^{0(i)}(p)\Psi^{0(i)\dagger}(q)(\qslash+M)|_{\rm o.s.}\\
%%%%
&=16\pi^2\rho^3[u K_1(u)-2K_0(u)]^2P_L\\
%%%
\nonumber&+\frac{128\pi^2}{3M^{10}\rho^3}[u(u(8+u^2)K_0(u)+4(4+u^2)K_1(u))-16]^2\left(\frac{3M^4}{4}+(p\cdot q)(M^2-p\cdot q)\right)P_L,
\end{align}
where now $u=\rho\, M$. The above equations where derived using the identities in \eqref{eq:etas2}, \eqref{eq:gammamunudual}, \eqref{eq:sigmamunu}, \eqref{eq:identities_sigma}, and standard properties of the Pauli matrices. Again, the correlators involve a $P_L$ projector, so that the effective on-shell Lagrangian reproducing the correlations only involves left-handed spinors. Note that when approximating fermionic Green functions with products of the above correlators, we now have a momentum-dependent form factor, due to the terms involving $(p\cdot q)$. However, these appear in contributions suppressed by higher orders in $\rho M$, and since the instanton integral is dominated by the contributions with small $\rho$, their effect will be subleading. Also, in the limit in which the fermions are emitted with small velocity, one has $p\sim q$ and $p\cdot q\sim M^2$, and the contributions proportional to $p\cdot q$ vanish. Using this approximation, we define then the adjoint form factor as
\begin{align}
{\cal F}^A_M(\rho)=&16\pi^2\rho^3[u K_1(u)-2K_0(u)]^2\\
%%%
\nonumber&+\frac{32\pi^2}{M^{6}\rho^3}[-16+u(u(8+u^2)K_0(u)+4(4+u^2)K_1(u))]^2,\quad u=\rho \,M.
\end{align}
The small and large $\rho M$ expansion of the form factor are given in \eqref{eq:Fexpansions}.
\bibliographystyle{JHEP-cerdeno}
\bibliography{library}
\end{document}